\documentclass[a4paper]{article} 
\usepackage{amssymb,amscd}
\usepackage[all]{xy}

\def\goth{\mathfrak}          
\def\double{\mathbb}

\def\cc{{\double C}}     
\def\nn{{\double N}}       
\def\zz{{\double Z}}

\def\rr{{\double R}}

\def\GG{{\double G}}
\def\TT{{\double T}}

\newtheorem{theorem}{Theorem}[section]
\newtheorem{lemma}[theorem]{Lemma}
\newtheorem{corollary}[theorem]{Corollary}
\newtheorem{definition}[theorem]{Definition}
\newtheorem{proposition}[theorem]{Proposition}
\newtheorem{remark}[theorem]{Remark}

\newtheorem{example}[theorem]{Example}

\newtheorem{exercise}[theorem]{Exercise}

\def\Kt{K^{\mathrm{top}}}
\def\Ka{K^{\mathrm{alg}}}
\def\Kr{K^{\mathrm{rel}}}
\def\Uc{{\cal U}}

\def\Pf{\mathop{\mathrm{Pf}}\limits_{t\to 0}}
\def\limind{\mathop{\mathrm{lim}}\limits_{\longrightarrow}}

\def\si{\sigma}

\def\cinf{C^{\infty}}

\newcommand{\be}{\begin{equation}}
\newcommand{\ee}{\end{equation}}
\newcommand{\beq}{\begin{eqnarray}}
\newcommand{\eeq}{\end{eqnarray}}
\newcommand{\om}{\omega}
\newcommand{\Om}{\Omega}
\newcommand{\al}{\alpha}
\def\nat{\natural}
\def\id{{\mathop{\mathrm{id}}}}
\def\Esf{{\sf E}}
\def\Asf{{\sf A}}
\def\Fsf{{\sf F}}
\newcommand{\La}{\Lambda}
\newcommand{\la}{\lambda}
\newcommand{\Ec}{{\cal E}}
\newcommand{\Vc}{{\cal V}}
\newcommand{\Lc}{{\cal L}}
\newcommand{\non}{\nonumber}
\newcommand{\eps}{\varepsilon}
\newcommand{\Sc}{{\cal S}}

\newcommand{\Rc}{{\cal R}}

\newcommand{\Ic}{{\cal I}}

\def\ch{\mathrm{ch}}
\def\cht{\mathrm{ch}^{\mathrm{top}}}
\def\chr{\mathrm{ch}^{\mathrm{rel}}}
\def\chb{\bar{\mathrm{ch}}}

\newcommand{\Tr}{{\mathop{\mathrm{Tr}}}}
\newcommand{\tr}{{\mathop{\mathrm{tr}}}}
\newcommand{\Ac}{{\cal A}}

\newcommand{\te}{\theta}

\newcommand{\cqfd}{\hfill\rule{1ex}{1ex}}

\def\d{\partial}
\def\dd{\,\mathrm{\bf d}}

\def\Hc{{\cal H}}
\def\Hcb{\bar{\cal H}}
\def\Cb{\bar{C}}
\def\hb{\bar{h}}
\def\HPb{\bar{H\!P}}

\def\Bc{{\cal B}}
\def\Cc{{\cal C}}

\def\ker{\mathop{\mathrm{Ker}}}

\def\Bb{\overline{B}}
\def\Xb{\overline{X}}
\def\Bbe{\overline{B}_{\epsilon}}
\def\bb{\overline{b}}

\def\hom{{\mathop{\mathrm{Hom}}}}

\def\End{{\mathop{\mathrm{End}}}}
\def\ev{\mathrm{ev}}
\def\hotimes{\hat{\otimes}}

\def\St{\widetilde{S}}
\def\Sd{S_{\delta}}

\def\Act{\widetilde{\cal A}}
\def\Tct{\widetilde{\cal T}}
\def\Bct{\widetilde{\cal B}}

\def\lg{{\goth k}}
\def\Hg{{\goth H}}
\def\gg{{\goth g}}
\def\Omt{\widetilde{\Omega}}
\def\Rct{\widetilde{\cal R}}
\def\xt{\widetilde{x}}
\def\Xt{\widetilde{X}}

\def\zt{\widetilde{z}}
\def\Omh{\widehat{\Omega}}

\def\rhoh{\widehat{\rho}}
\def\Dh{\widehat{D}}
\def\Ome{\Omega_{\epsilon}}
\def\Omte{\Omt_{\epsilon}}

\def\Oman{\Omega_{\mathrm{an}}}
\def\Omtan{\widetilde{\Omega}_{\mathrm{an}}}

\def\Tc{{\cal T}}

\def\Uc{{\cal U}}

\def\Acop{{\cal A^{\mathrm{op}}}}
\def\rlarrows{\textup{$\rightleftarrows$}}
\def\Lie{\mathrm{Lie}\,}
\def\mod{\ \mathrm{mod}\ }
\def\Icons{I_{\mathrm{cons}}}
\def\Icov{I_{\mathrm{cov}}}
\def\Qcons{Q_{\mathrm{cons}}}
\def\Qcov{Q_{\mathrm{cov}}}

\begin{document}

\begin{center}

{\large\bf CHERN CHARACTER, HOPF ALGEBRAS,\\[1mm] 
AND BRS COHOMOLOGY}
\vskip 1cm
{\bf Denis PERROT}
\vskip 0.5cm
SISSA, via Beirut 2-4, 34014 Trieste, Italy \\[2mm]
{\tt perrot@fm.sissa.it}\\[2mm]
\today 
\end{center}
\vskip 0.5cm
\begin{abstract} 
We present the construction of a Chern character in cyclic cohomology, involving an arbitrary number of associative algebras in contravariant or covariant position. This is a generalization of the bivariant Chern character for bornological algebras introduced in a previous paper \cite{P3}, based on Quillen superconnections and heat-kernel regularization. Then we adapt the formalism to the cyclic cohomology of Hopf algebras in the sense of Connes-Moscovici \cite{CM00}. This yields a Chern character for ``equivariant $K$-cycles'' over a bornological algebra $\Ac$, generalizing the Connes-Moscovici characteristic maps. In the case of equivariant spectral triples verifying some additional conditions, we also exhibit secondary characteristic classes. The latter are not related to topology but rather define characteristic maps for the higher algebraic $K$-theory of $\Ac$. In the classical case of spin manifolds, we compute and interpret these secondary classes in terms of BRS cohomology in Quantum Field Theory.
\end{abstract}

\vskip 0.5cm


\section{Introduction}

In this paper we pursue the construction, initiated in \cite{P3,P4}, of characteristic classes in cyclic cohomology, based on heat-kernel methods generalizing the well-known JLO cocycle \cite{JLO} to situations of arbitrary ``variance''. Our aim is to present the construction of a universal {\it generalized Chern character} involving an arbitrary number of associative algebras in contravariant or covariant positions. This allows in principle to obtain an explicit cocycle representing a multilinear transformation between the cyclic homologies of some algebras $\Ac_1,\ldots, \Ac_n$ and $\Bc_1,\ldots,\Bc_p$,
\be
\ch(A)\ :\ HC(\Ac_1)\times \ldots\times HC(\Ac_n)\to HC(\Bc_1)\hotimes\ldots\hotimes HC(\Bc_p)\ ,\label{map}
\ee
from an abstract {\it geometric datum} given by a Quillen superconnection $A$ living in a appropriate differential graded convolution algebra \cite{Q1}. This encompasses the bivariant Chern character constructed previously in \cite{P3,P4}. Here $HC$ is a generic symbol denoting any interesting kind of cyclic homology (periodic $HP$, entire $HE$, non-periodic $HC_n$, $n\in\nn$, negative $HC_n^-$, $n\in\zz$, etc...) depending on the category of algebras and the situation considered. For the sake of generality, we will formulate everything in the framework of {\it bornological} algebras \cite{HN,Me}. The idea of the construction is based on the following remark about a ``self-dual'' property of the cyclic homology of an algebra $\Ac$. The latter can be computed by the cyclic bicomplex of Connes \cite{C0, L}. Upon an observation of Quillen \cite{Q2}, the cyclic bicomplex is isomorphic to the $X$-complex of the differential graded {\it bar coalgebra} of $\Ac$. The second possibility comes from the Cuntz-Quillen formalism \cite{Cu2,CQ0,CQ1,CQ2}. Here the cyclic homology is computed by the $X$-complex of the {\it tensor algebra} of $\Ac$. Since the definitions of the $X$-complexes for coalgebras and algebras are dual to each other, we obtain the map (\ref{map}) by a straightforward generalization of Quillen's constructions using connections and curvatures \cite{Q2,Q3}. The rule is that for the algebras $\Ac_i$ in contravariant position, we describe their cyclic homology via the $X$-complex of the bar coalgebras $\Bbe(\Ac_i)$, whereas for the algebras $\Bc_j$ in covariant position, we use the $X$-complex of their tensor algebras $\Tc\Bc_j$. A generic Quillen superconnection $A$ is given by an odd element of the differential graded convolution algebra $\hom(\Bbe(\Ac_1)\hotimes\ldots\hotimes \Bbe(\Ac_n),\Tc\Bc_1\hotimes\ldots\hotimes\Tc\Bc_p)$. We obtain roughly the Chern character $\ch(A)$ by exponentiating the curvature of $A$. When the superconnection contains a Dirac operator (this appears for example in the unbounded description of bivariant $K$-theory \cite{Bl}), the corresponding Chern character includes naturally a ``heat-kernel'' used as a regulator of traces. One thus extends the usual index formulas \cite{BGV,B,Gi} to a very general non-commutative framework.\\
In practice the explicit computation of $\ch(A)$ becomes rapidly cumbersome when the number of algebras $\Ac_i$ and $\Bc_j$ increases. Since the most interesting examples certainly arise in the bivariant case, i.e. one contravariant algebra $\Ac$ and one covariant algebra $\Bc$, we will actually restrict our attention to this situation. However we formulate everything in such a way that the generalization to (\ref{map}) becomes clear. It is also extremely important to note that by dualizing the formulas in the obvious fashion, one gets a generalized Chern character involving cyclic (co)homology for {\it coalgebras}. Furthermore, incorporating the notion of {\it invariance} leads to the cyclic homology of Hopf algebras \cite{CM99,Cr}. One thus gets stirking generalizations of characteristic maps for Hopf actions in the sense of Connes-Moscovici \cite{CM98,CM00}. With a little addition of analysis, we will also exhibit secondary characteristic classes for equivariant spectral triples related to BRS cohomology appearing in Quantum Field Theory \cite{D,MSZ}. The point is that all these constructions, ranging from the bivariant Chern character to BRS cohomology, are specific {\it examples} of the universal construction (\ref{map}); the relevant information is simply encoded in the connection $A$. We hope this will be apparent in the subsequent pages.\\

Let us briefly outline the paper. Section \ref{sgen} is devoted to the explanation of the recursive method abutting to the generalized Chern character. We recall Quillen's definition of the $X$-complex for differential graded (DG) algebras and coalgebras \cite{Q2}, and the cocycle associated to a connection. Using some ideas of Cuntz-Quillen \cite{CQ1} concerning free products of algebras, we generalize this construction to an arbitrary number of DG algebras $\Rc_1,\ldots,\Rc_p$: given a connection $A$ in the DG algebra tensor product $\Rc_1\hotimes\ldots\hotimes\Rc_p$ (i.e. an odd element), we associate a cocycle $\ch(A)\in X(\Rc_1)\hotimes\ldots\hotimes X(\Rc_p)$, by exponentiating the curvature of $A$. By dualising all structure maps, we also incorporate DG coalgebras $\Cc_1,\ldots,\Cc_n$. A connection $A$ is now an odd element of the DG convolution algebra $\hom(\Cc_1\hotimes\ldots\hotimes\Cc_n,\Rc_1\hotimes\ldots\hotimes\Rc_p)$, and its Chern character is a cocycle
\be
\ch(A)\in \hom(X(\Cc_1)\hotimes\ldots\hotimes X(\Cc_n), X(\Rc_1)\hotimes\ldots\hotimes X(\Rc_p))\ .
\ee
We explicitly compute $\ch(A)$ in terms of its curvature in the ``bivariant'' cases $X(\Rc_1)\hotimes X(\Rc_2)$, $\hom(X(\Cc), X(\Rc))$ and $\hom(X(\Cc_1)\hotimes X(\Cc_2),\cc)$. The three expressions are (up to signs) formally identical. This construction is performed in a bornological language which will become crucial in the subsequent applications.\\

In section \ref{sbiv}, we deal with bivariant cyclic cohomology. If $\Ac$ is a complete bornological algebra, its bar coalgebra is completed into the {\it entire bar coalgebra} $\Bbe(\Ac)$ \cite{P3}, and its tensor algebra is also completed into the {\it analytic tensor algebra} $\Tc\Ac$ \cite{Me}. Then the entire cyclic homology of $\Ac$ is computed either by $X(\Bbe(\Ac))$ or $X(\Tc\Ac)$. If $\Bc$ is another complete bornological algebra, the complex of bounded linear maps $\hom(X(\Bbe(\Ac)),X(\Tc\Bc))$ computes the bivariant entire cyclic cohomology $HE_*(\Ac,\Bc)$. We introduce the set $\Psi(\Ac,\Bct)$ as an algebraic version of Kasparov's bivariant $K$-theory \cite{Bl}. We consider the unitalization $\Bct$ of $\Bc$ for technical reasons. A bivariant element $\Ec\in\Psi(\Ac,\Bct)$ corresponds to a quintuple $(\Lc,\ell^1,\tau,\rho,D)$, where $\Lc$ is a $\zz_2$-graded auxiliary algebra of ``abstract pseudodifferential operators'' or symbols. $\ell^1$ is a two-sided ideal in $\Lc$ provided with a graded trace of homogeneous degree $\tau:\ell^1\to \cc$. $\rho:\Ac\to\Lc\hotimes\Bct$ is an homomorphism and $D\in \Lc\hotimes\Bct$ is an odd element (``Dirac operator'') whose laplacian $D^2$ has a ``trace-class'' heat kernel. Using the universal construction of section \ref{sgen}, we form the Quillen superconnection $A=\rho+D$ in the DG convolution algebra $\hom(\Bbe(\Ac),\Lc\hotimes\Tct\Bc)$. Then under suitable $\te$-summability asumptions on $\Ec$, the cocycle $\tau\ch(A)$ defines the {\it bivariant Chern character}
\be
\ch(\Ec)\in HE_i(\Ac,\Bc)\ ,\quad i=\ \mbox{degree of}\ \tau\ .
\ee
It is homotopy invariant with respect to the homomorphism $\rho$ and the Dirac operator $D$. The exponential of the curvature of $A$ naturally includes the heat kernel of $D^2$ with its tracial properties. This is essentially equivalent to the construction presented in \cite{P3} for unbounded families of spectral triples.\\ 

Section \ref{shopf} incorporates Hopf algebras. We let $\Hc$ be a Hopf algebra provided with a character (algebra homomorphism) $\delta:\Hc\to\cc$. The convolution product of $\delta$ with the antipode of $\Hc$ defines the {\it twisted antipode} $\Sd$. Connes and Moscovici develop in \cite{CM98,CM99} a cyclic cohomology theory for $\Hc$ when the twisted antipode verifies the involution condition $\Sd^2=\id$. This cohomology is interpreted by Crainic \cite{Cr} as the cohomology of the cyclic bicomplex for the {\it coalgebra} $\Hc$, quotiented by the subspace spanned by coinvariant elements. This adapts immediately to the setting of our generalized Chern character by formal manipulations. All what we have to do is to check the invariance of the cyclic cocyles with respect to Hopf actions. For any complete bornological algebra $\Ac$, we introduce the set $\Psi^{\Hc}(\Ac,\cc)$ of $\Hc$-equivariant $K$-cycles over $\Ac$. They correspond to quintuples $\Ec=(\Lc,\ell^1,\tau,\rho,D)$, where the auxiliary algebra $\Lc$ and the two-sided ideal $\ell^1$ are endowed with an $\Hc$-action. $\tau:\ell^1\to\cc$ is a $\delta$-invariant graded trace, that is, $\tau(h(x))=\delta(h)\tau(x)$ for any $h\in\Hc$, $x\in\ell^1$. $\rho:\Ac\to\Lc$ is an homomorphism and $D\in\Lc$ is an odd element with trace-class heat kernel. We introduce a Quillen superconnection $A$ in a suitable convolution algebra. $A$ encodes simultaneously $\rho$, $D$, and {\it the action of } $\Hc$ on them. Upon $\te$-summability for $\Ec$, the universal Chern character of section \ref{sgen} yields a characteristic map
\be
\ch(\Ec): HE_*(\Ac)\to HP^{\delta}_{*+i}(\Hc)\ ,\quad i=\ \mbox{degree of}\ \tau\ ,
\ee
with target the periodic cyclic homology of $\Hc$ in the sense of Connes-Moscovici. Again, it is homotopy invariant with respect to $\rho$ and $D$ and naturally includes the heat kernel of $D^2$. In the purely algebraic situation $D=0$, one recovers the construction of \cite{CM98} as a particular case. \\

Section \ref{ssec} is more analytical in nature. We consider the situation of the preceding section, with the additional hypothesis that all algebras are represented as linear operators on a Hilbert space. This allows to exhibit much finer invariants, {\it a priori} not related to topology. Let $\Ac$ be a complete bornological algebra and $\Hc$ a Hopf algebra. For any integer $p$, we introduce the set of $p_+$-summable $\Hc$-equivariant spectral triples $\Esf^{\Hc}_p(\Ac,\cc)$. They are given by triples $\Ec=(\Ac,\Hg,D)$ in Connes' sense of non-commutative geometry \cite{C1}: $\Hg$ is a separable Hilbert space on which the algebra $\Ac$ acts by bounded operators, $D$ is a self-adjoint unbounded operator with compact resolvent satisfying certain finite summability assumptions, and the commutators $[D,a]$ are bounded for any $a\in\Ac$. Moreover, the Hopf algebra is also linearly represented as (not necessarily bounded) operators on $\Hg$, and ``almost commutes'' with the other data of the spectral triple in the sense that the adjoint action of $\Hc$ induces bounded perturbations of $D$, and $[h,a]=0=[h,[D,a]]$ for any $h\in\Hc$, $a\in\Ac$. If the antipode of $\Hc$ is involutive, then we can define the periodic cyclic homology $HP^{\eps}_*(\Hc)$ associated with the trivial character given by the counit $\eps$ of $\Hc$. Unfortunately, since $\Hc$ acts on the algebra of operators by the adjoint representation, the associated characteristic map $\ch(\Ec): HE_*(\Ac)\to HP^{\eps}_*(\Hc)$ retracts on the trivial factor $\cc\subset HP^{\eps}_*(\Hc)$ and does not detect any interesting class associated to the Hopf action. But this vanishing property makes way for {\it secondary characteristic classes}. The $p_+$-summable element $\Ec$ yields a collection of maps defined on the {\it non-periodic} cyclic homology of $\Ac$
\be
\psi(\Ec): HC_n(\Ac)\to HC^{\eps,+}_{n-p}(\Hc)\ ,\quad \forall n\ge 0\ ,\label{map2}
\ee
where $HC^{\eps,+}_{n-p}(\Hc)$ denotes a cyclic homology of the Hopf algebra $\Hc$ associated to the vertical filtration of its cyclic bicomplex. The good property of these characteristic maps is their homotopy invariance with respect to the Dirac operator $D$. However, they are {\it not} homotopy invariant with respect to the representation of $\Ac$ in the Hilbert space $\Hg$. If $\Ac$ is a unital Banach or Fr\'echet algebra, we may adopt the (non-standard) notation $\Ka_n(\Ac)$ for the kernel of the natural homomorphism $K_n(\Ac)\to \Kt_n(\Ac)$ from the higher algebraic $K$-theory \cite{R} to the topological $K$-theory of $\Ac$. Then the characteristic classes (\ref{map2}) induce a bilinear pairing
\be
\Ka_{n+1}(\Ac)\times\Esf^{\Hc}_p(\Ac,\cc)\to HC^{\eps,+}_{n-p}(\Hc)\ ,\quad \forall n\ge 0\ ,
\ee
defined at the level of homotopy classes of $p_+$-summable Dirac operators. These pairings do not detect topological invariants of $\Ac$ (the topological $K$-theory is dropped), but rather {\it geometric} invariants contained in algebraic $K$-theory. It should be noted that these secondary classes do not correspond to the higher regulators introduced by Connes-Karoubi in \cite{CK}.\\

In section \ref{sBRS}, we illustrate the nature of these secondary characteristic classes by an explicit computation in classical (commutative) geometry. If $\Ac=\cinf(M)$ is the Fr\'echet algebra of smooth functions over a compact riemannian spin manifold $M$, and $E\to M$ an hermitian vector bundle over $M$, we may form the ordinary Dirac spectral triple $\Ec=(\Ac,\Hg,D)$ with coefficients in $E$: $\Hg$ is the Hilbert space of square-integrable sections of the spinor bundle tensored with $E$, the algebra of functions $\Ac$ acts by pointwise multiplication and $D$ is the usual Dirac operator obtained from the metric on $M$ and a choice of Yang-Mills connection on $E$. The gauge group $\GG$ of unitary endomorphisms of $E$ is represented in $\Hg$, hence also the enveloping algebra $\Hc=\Uc(\Lie\GG)$ of its complexified Lie algebra. $\Hc$ is a cocommutative Hopf algebra, and $\Ec$ defines an element of $\Esf^{\Hc}_p(\Ac,\cc)$, where $p$ is the dimension of the manifold. Then we compute and interpret the secondary characteristic classes (\ref{map2}) in terms of BRS cohomology \cite{D,MSZ}. These are {\it invariants} associated to the $\Hc$-equivariant Dirac operator in the sense that they do not depend on the metric on $M$ nor on the Yang-Mills connection. They are {\it local} on $M$ in the sense that they are independent of global issues such as the topological $K$-theory of $\Ac$, but are detected by the finer geometric objects $\Ka_n(\Ac)=\ker(K_n(\Ac)\to\Kt_n(\Ac))$. It follows that higher algebraic $K$-theory detects the local BRS cohomology class of the (consistent) chiral anomaly and Schwinger term of Quantum Field Theory \cite{MSZ}, which is a priori not related to the {\it topological} anomaly of \cite{P1,P2}. This interpretation should hold in the case of non-commutative Yang-Mills theories as well. \\

By convention, we work in the formulation of cyclic (co)homology for {\it non-unital} algebras. This may appear quite inadapted to the description of cyclic cohomology for Hopf algebras (we will be forced to consider a Hopf algebra as an augmented algebra), but in general it is not possible to stay in the unital category. For example in the case of bivariant $K$-theory, the homomorphism $\rho:\Ac\to\Lc\hotimes\Bct$ cannot be unital in general. Also, the isomorphism between the $X$-complex of the bar coalgebra and the cyclic bicomplex of non-commutative differential forms only holds in the non-unital setting, whereas it is only a quasi-isomorphism in the unital case. This motivates our choice.

\section{Generalized Chern character}\label{sgen}

The aim of this section is to explain the basic material used throughout the paper. We recall Quillen's definition of the $X$-complex of a differential graded (DG) algebra $\Rc$ \cite{Q2,Q3} and the construction of the Chern character cocycle associated to a connection form $A\in\Rc$:
\be
\ch(A)\in \Xt(\Rc)\ .
\ee
Roughly, this cocycle is obtained by exponentiating the curvature of $A$. Here $\Xt(\Rc)$ denotes the $X$-complex of $\Rc$ augmented by adjoining a unit to $\Rc$. We should also mention that our sign conventions differ slightly from \cite{Q2,Q3}. Then we present a recursive method allowing to generalize the Chern character construction to an arbitrary number of DG algebras. If $A$ is a connection form in a tensor product of augmented DG algebras $\Rct_1\hotimes\ldots\hotimes \Rct_n$, we get a cocycle in the corresponding tensor product of $X$-complexes
\be
\ch(A)\in \Xt(\Rc_1)\hotimes \ldots\hotimes  \Xt(\Rc_n)\ .
\ee
In particular for two algebras $\Rc_1$ and $\Rc_2$, we compute explicitly the cocycle $\ch(A)$ in terms of $A$ and its curvature. Passing to (counital) DG coalgebras by formally dualising all structure maps, we also obtain hybrid cocycles in $\hom$-complexes involving coalgebras in contravariant position and algebras in covariant position. The fundamental example which will be used for the construction of the bivariant Chern character (section \ref{sbiv}) is based on the following general situation. Let $\Cc$ be a DG coalgebra and $\Rc$ a DG algebra. Then to any connection form $A$ in the DG convolution algebra $\hom(\Cc,\Rct)$, we associate a cocycle
\be
\ch(A)\in \hom(X(\Cc),\Xt(\Rc))\ .
\ee
Another related example uses two DG coalgebras $\Cc_1$ and $\Cc_2$. Then any connection form $A\in\hom(\Cc_1\hotimes\Cc_2,\cc)$ implies a cocycle
\be
\ch(A)\in \hom(X(\Cc_1)\hotimes X(\Cc_2),\cc)\ .
\ee
The characteristic maps of Hopf algebras (section \ref{shopf} and \ref{ssec}) will be based on the latter cocycle.\\
Concerning the specific applications of these universal constructions, especially for the bivariant Chern character (section \ref{sbiv}), we cannot deal with purely algebraic objects and the spaces or algebras must posses some extra structure related to analysis. It turns out that the correct setting where our constructions make sense is the category of {\it bornological} vector spaces \cite{HN}. This is also the most general framework for entire cyclic cohomology \cite{Me}. We recall that a (convex) bornology on a vector space is given by a collection of subsets satisfying certain axioms inspired by the behaviour of the bounded sets of a normed space. Thus an element of the bornology is called a {\it bounded} or {\it small} subset. There is a notion of bornological {\it completion}, completed tensor products (denoted $\hotimes$), bounded linear maps between bornological spaces (denoted $\hom(\cdot,\cdot)$), etc... This material has already been used extensively in our previous papers on the bivariant Chern character \cite{P3,P4}, so that these notions will not be recalled here. The interested reader is referred to the fundational works of Hogbe-Nlend \cite{HN} and the thesis of Meyer \cite{Me} for further details. A concrete example of complete bornological space is given by a complete locally convex space with bornology consisting in all bounded subsets with respect to the family of seminorms defining the topology. In the case of Fr\'echet spaces, completed (bornological) tensor products correspond to ordinary projective tensor products, and bounded maps correspond to continuous maps. A complete bornological algebra $\Rc$ is complete as a vector space and such that the product is a bilinear bounded map $\Rc\times\Rc\to\Rc$.\\

Let $\Rc$ be an associative $\zz_2$-graded complete bornological algebra. We suppose that $\Rc$ is endowed with a bounded differential $\delta:\Rc\to\Rc$ of odd degree, that is,
\be
\delta (xy)=\delta(x) y +(-)^{|x|}x\delta(y)\quad \forall x,y\in\Rc\ ,\quad \delta^2=0\ ,
\ee
where $|x|=0,1$ denotes the degree of $x$. Thus $(\Rc,\delta)$ is a differential graded (DG) bornological algebra. The space of universal one-forms over $\Rc$ is the completed tensor product
\be
\Om^1\Rc=\Rct\hotimes\Rc\ ,
\ee
where $\Rct=\cc\oplus \Rc$ is the unitalization of $\Rc$. We denote by $1$ the unit of $\Rct$ considered as an element of degree zero (which differs from the unit of $\Rc$, if it exists), and extend the derivation $\delta$ to $\Rct$ by setting $\delta 1=0$. This implies that $(\Rct,\delta)$ is a unital DG bornological algebra. For $\widetilde{x}\in\Rct$ and $y\in\Rc$, the element $\xt\otimes y\in \Om^1\Rc$ will be denoted as usual by the one-form $\xt\dd y$. We endow $\Om^1\Rc$ with the unique $\zz_2$-graduation such that the degree of $\xt\dd y$ reads
\be
|\xt\dd y|= |\xt| + (|y|+1)\ .
\ee
$\Om^1\Rc$ is naturally a bornological $\Rct$-bimodule: the left multiplication $\Rct\hotimes \Om^1\Rc\to \Om^1\Rc$ given by $\zt\cdot (\xt\dd y)= (\zt\xt)\dd y$ is bounded as well as the right multiplication $\Om^1\Rc\hotimes \Rct\to \Om^1\Rc$ coming from the leibniz rule
\be
(\xt\dd y)\cdot \zt = \xt \dd(y\zt) -(-)^{|y|}\xt y\dd\zt\ .
\ee
Then extend $\delta$ as an odd degree derivation of $\Om^1\Rc$ viewed as an $\Rct$-bimodule, by setting
\be
\delta(\xt\dd y)= (\delta\xt)\dd y +(-)^{|x|+1}\xt\dd(\delta y)\ .
\ee
Note that the symbol $\dd$ always appears to have degree 1. With these sign conventions, the universal derivation
\be
\dd : \Rct \to \Om^1\Rc\ ,\quad \xt\mapsto \dd \xt\ ,\quad \dd 1=0
\ee
anticommutes with the derivation $\delta $ acting respectively on $\Rct$ and $\Om^1\Rc$. One defines also the Hochschild operator with appropriate signs
\be
b: \Om^1\Rc\to \Rc\ ,\qquad \xt\dd y\mapsto (-)^{|\xt|}[\xt,y]\ ,
\ee
so that $b$ is an odd degree bounded map anticommuting with $\delta$. Here the commutator $[\xt,y]=\xt y -(-)^{|\xt||y|}y\xt$ is {\it graded}. Let now $\Om^1\Rc_{\nat}$ be the quotient
\be
\Om^1\Rc_{\nat}=\Om^1\Rc/[\Rc,\Om^1\Rc]
\ee
by the subspace of graded commutators, using the bimodule structure of universal one-forms. Then $\Om^1\Rc_{\nat}$ inherits the grading of $\Om^1\Rc$ and the bounded map $\delta$ of odd degree. We denote by $\nat$ the projection $\Om^1\Rc\to\Om^1\Rc_{\nat}$, and $\nat \xt\dd y$ is the class of $\xt\dd y$. Since the Hochschild operator vanishes on the subspace of graded commutators $[\Rc,\Om^1\Rc]$, it passes to a bounded map of odd degree on the quotient:
\be
\bb:\Om^1\Rc_{\nat}\to \Rc\ ,\qquad \nat\xt\dd y\mapsto (-)^{|\xt|}[\xt,y]\ .
\ee
Of course $\bb$ anticommutes with $\delta$. One checks easily that the compositions of maps $\nat\dd\circ\bb:\Om^1\Rc_{\nat}\to \Om^1\Rc_{\nat}$ and $\bb\circ \nat\dd:\Rc\to\Rc$ vanish. Therefore, we get a periodic complex 
\be
X(\Rc):\ \Rc\ \xymatrix@1{\ar@<0.5ex>[r]^{\nat \dd} &  \ar@<0.5ex>[l]^{\bb}}\ \Om^1\Rc_{\nat}\ ,
\ee
where $\Rc$ and $\Om^1\Rc_{\nat}$ are themselves complexes for the boundary map $\delta$. The above $X$-complex of the DG algebra $\Rc$ is thus actually a bicomplex. In the following, we will also use the augmented $X$-complex
\be
\Xt(\Rc):\ \Rct\ \xymatrix@1{\ar@<0.5ex>[r]^{\nat\dd} &  \ar@<0.5ex>[l]^{\bb}}\ \Om^1\Rc_{\nat}\ ,
\ee
which obviously splits into the direct sum of $X(\Rc)$ and $X(\cc):\cc\rlarrows 0$. \\

We now recall Quillen's construction \cite{Q2} of cocycles in the total complex $\Xt(\Rc)$ endowed with the boundary $(\nat\dd\oplus\bb)+\delta$. These cocycles are analogues of the Chern character form in differential geometry: one chooses a ``connection'' $A\in \Rc_-$ in the odd part of the differential algebra $R$, with curvature
\be
F=\delta A+ A^2\ \in \Rc_+\ .
\ee
Then the Chern character is basically obtained by exponentiating $F$ through the formal power series
\be
e^{-F}=\sum_{i=0}^{\infty} \frac{(-)^n}{n!}\, F^n\ .
\ee
In fact we don't want to deal with formal power series, and in the sequel we will eventually assume that the exponential actually converges with respect to an appropriate topology (see remark \ref{rconv} below). One has the following result of Quillen:
\begin{proposition}[\cite{Q2}]\label{pqui}
Let $(R,\delta)$ be a complete bornological DG algebra, and $A\in \Rc_-$ a connection form. Then the Chern character $\ch(A)\in \Xt(\Rc)$ whose components are defined by 
\be
\ch^0(A)=e^{-F}\in \Rct\ ,\quad \ch^1(A)=-\nat\, e^{-F}\dd A\in \Om^1\Rc_{\nat}\ ,
\ee
is a cocycle in the total complex $(\Xt(\Rc),(\nat\dd\oplus\bb)+\delta)$, that is,
\be
\delta\ch^0(A)+\bb\ch^1(A)=0\ ,\quad \delta\ch^1(A)+\nat\dd\ch^0(A)=0\ .
\ee
\end{proposition}
{\it Proof:} It is a direct consequence of the Bianchi identity $\delta F+[A,F]=0$. One has
$$
\delta e^{-F}=[e^{-F},A]=\bb\, \nat e^{-F}\dd A
$$
because $e^{-F}$ has even degree. On the other hand
\beq
\delta\,\nat e^{-F}\dd A&=& \nat [e^{-F}, A]\dd A - \nat e^{-F}\dd\delta A\non\\
&=& \nat [e^{-F}, A]\dd A + \nat e^{-F}\dd(A^2)-\nat e^{-F}\dd F\non\\
&=& -\nat e^{-F}\dd F\ =\ \nat\dd(e^{-F})\ ,\non
\eeq
and the conclusion follows. \cqfd\\

We now discuss the homotopy invariance of the Chern character. Let $\cinf[0,1]$ be the commutative algebra of smooth complex-valued functions over the interval $[0,1]$. It is a commutative Fr\'echet algebra, hence a complete bornological algebra for the bornology given by bounded sets. The completed tensor product $\Rc\hotimes \cinf[0,1]$ is the algebra of smooth functions over $[0,1]$ with values in $\Rc$. The evaluation map $\ev_t:\cinf[0,1]\to\cc$ which sends a function $f$ to its value at $t\in[0,1]$ is bounded, hence extends to an evaluation map $\ev_t:\Rc\hotimes \cinf[0,1]\to \Rc$.
\begin{proposition}\label{phomo}
Let $A\in \Rc_-\hotimes \cinf[0,1]$ be a smooth family of connections forms in $\Rc$. For any $t\in [0,1]$, let $A_t=\ev_t A\in \Rc_-$ be the evaluation of $A$ at $t$. Then the cohomology class of the Chern character $\ch(A_t)\in \Xt(\Rc)$ is independent of $t$.
\end{proposition}
{\it Proof:} Let $\Om[0,1]$ be the graded commutative Fr\'echet algebra of differential forms over the interval $[0,1]$ parametrized by $t$. It is endowed with the de Rham differential $d_t$. We slightly generalize proposition \ref{pqui} with $\Xt(\Rc)$ replaced by the (graded) tensor product of complexes $\Xt(\Rc)\hotimes\Om[0,1]$ with total differential $(\nat\dd\oplus\bb)+\delta+d_t$. The family of connections $A$ is an element of $(\Rc\hotimes\Om[0,1])_-$ and its curvature $F$ reads
$$
F=(\delta+d_t)A+A^2\ \in (\Rc\hotimes\Om[0,1])_+\ .
$$
Then adapting the proof of proposition \ref{pqui} to the total complex $\Xt(\Rc)\hotimes\Om[0,1]$, one has with $\ch^0(A)=e^{-F}$ and $\ch^1(A)=-\nat e^{-F}\dd A\in \Om^1\Rc_\nat\hotimes\Om[0,1]$:
\be
(\delta+d_t) \ch^0(A)+\bb \ch^1(A)=0\ ,\quad (\delta+d_t)\ch^1(A)+\nat\dd \ch^0(A)=0\ .\label{eq}
\ee
Here we used the fact that $\Om[0,1]$ is a graded commutative differential algebra. We define $\chi(A)$ and $cs(A)$ as the projections of $\ch(A)$ onto $\Xt(\Rc)\hotimes \cinf[0,1]$ and $\Xt(\Rc)\hotimes\Om^1[0,1]$ respectively:
\beq
\chi^0(A)=e^{-F}|_{\Rct\hotimes\cinf[0,1]}\ ,\quad \chi^1(A)=-\nat e^{-F}\dd A|_{\Om^1\Rc\hotimes\cinf[0,1]}\ ,\non\\
cs^0(A)=e^{-F}|_{\Rct\hotimes\Om^1[0,1]}\ ,\quad cs^1(A)=-\nat e^{-F}\dd A|_{\Om^1\Rc\hotimes\Om^1[0,1]}\ .\non
\eeq
Then equations (\ref{eq}) imply
$$
d_t\chi^0(A)=-(\bb cs^1(A)+\delta cs^0(A))\ ,\quad d_t\chi^1(A)=-(\nat\dd cs^0(A)+\delta cs^1(A))\ .
$$
Upon integration over $t$, this shows that the evaluations $\ch(A_t)=\ev_t\chi(A)$ for different values of $t$ differ by a coboundary.\cqfd\\
\begin{remark}\label{rconv}\textup{
Taking $A_t=tA$ for $t\in[0,1]$ and $A\in\Rc_-$ shows that the cocycle $\ch(A)\in \Xt(\Rc)$ retracts onto the unit $1\in \Rct$ in the sense of formal power series. This does not need to be true if we impose that $e^{-F}$ actually converges towards an element of $\Rct$ with respect to an appropriate topology, because the transgressions $cs$ appearing in the proof of the proposition above may not converge for arbitrary values of $t$. This property will ensure that the subsequent cocycles (for example the bivariant Chern character of section \ref{sbiv}) are not trivial in general. }
\end{remark}

We now generalize Quillen's construction to a pair of DG algebras $(\Rc_1,\delta_1)$ and $(\Rc_2,\delta_2)$. We shall obtain a cocycle in the tensor product of total complexes $\Xt(\Rc_1)\hotimes\Xt(\Rc_2)$ from a connection form $A\in (\Rct_1\hotimes\Rct_2)_-$ and its curvature
\be
F=(\delta_1+\delta_2)A+A^2\ \in (\Rct_1\hotimes\Rct_2)_+\ .
\ee
For this we adapt some ideas from \cite{CQ1} and work with the free product $\Rc_1 *\Rc_2$. Recall that the free product is the non-unital associative algebra generated by $\Rc_1$ and $\Rc_2$. We can write
\be
\Rc_1*\Rc_2 = \Rc_1\oplus\Rc_2 \oplus (\Rc_1\hotimes\Rc_2)\oplus (\Rc_2\hotimes\Rc_1)\oplus (\Rc_1\hotimes\Rc_2\hotimes\Rc_1)\oplus (\Rc_2\hotimes\Rc_1\hotimes\Rc_2)\oplus\ldots\ ,
\ee
where only alternate tensor products of $\Rc_1$ and $\Rc_2$ appear. $\Rc_1*\Rc_2$ is endowed with the direct sum bornology, and the product is such that whenever two elements of $\Rc_1$ meet, we take the multiplication in $\Rc_1$ instead of the tensor product, and similarly for the elements of $\Rc_2$. The free product naturally inherits from $\Rc_1$ and $\Rc_2$ a $\zz_2$-grading (take the sum of all degrees in a tensor product $\Rc_1\hotimes\Rc_2\hotimes\ldots$), as well as two anticommuting differentials $\delta_1$ and $\delta_2$. Thus $\Rc_1*\Rc_2$ endowed with the total differential $\delta=\delta_1+\delta_2$ is a DG algebra, provided we take care of the signs. For example, if we denote by $*$ the product in $\Rc_1*\Rc_2$, one has
\be
\delta(x_1*x_2*y_1)=\delta_1(x_1)* x_2*y_1 +(-)^{|x_1|}x_1*\delta_2(x_2)*y_1 +(-)^{|x_1|+|x_2|}x_1*x_2*\delta_1(y_1)\ ,
\ee
for any $x_1,y_1\in \Rc_1$ and $x_2\in\Rc_2$. Note that there are two canonical inclusions $\iota_1:\Rc_1\to \Rc_1*\Rc_2$ and $\iota_2:\Rc_2\to\Rc_1*\Rc_2$ which are bounded DG algebra homomorphisms. The free product enjoys the following universal property: given two DG algebra homomorphisms $\al:\Rc_1\to \Sc$ and $\beta:\Rc_2\to\Sc$ with target a DG algebra $\Sc$, then there is a unique DG algebra homomorphism $\al*\beta: \Rc_1*\Rc_2\to \Sc$ such that $\al=(\al*\beta)\circ \iota_1$ and $\beta=(\al*\beta)\circ \iota_2$.\\
We shall use the universal property of $\Rc=\Rc_1*\Rc_2$ with respect to the unital DG algebra $\Sc=\Rct_1\hotimes\Rct_2$. Indeed, there are two canonical injections $\al:\Rc_1\to \Sc$ and $\beta:\Rc_2\to \Sc$ given by
\be
\al(x_1)=x_1\otimes 1\quad\forall x_1\in\Rc_1\ ,\quad \beta(x_2)=1\otimes x_2\quad \forall x_2\in\Rc_2\ .
\ee
We denote by $\mu$ the folding map $\al*\beta: \Rc\to\Sc$, and extend it to a unital DG homomorphism over $\Rct$ by setting $\mu(1_{\Rct})=1_{\Sc}$. For example (note the sign)
\be
\mu(x_1*x_2*y_1 *y_2)=(-)^{|x_2||y_1|}x_1y_1\otimes x_2y_2\quad \forall x_1,y_1\in\Rc_1\ ,\ x_2,y_2\in \Rc_2\ .
\ee
Consider now the $\zz_2$-graded tensor product $\Om^1\Rc_1\hotimes\Om^1\Rc_2$. It is endowed with the action of the total differential $\delta=\delta_1+\delta_2$. Since $\Om^1\Rc_1$ and $\Om^1\Rc_2$ are bimodules respectively over $\Rct_1$ and $\Rct_2$, the tensor product itself has a structure of $\Sc$-bimodule compatible with the action of $\delta$, provided we insert signs whenever elements of $\Om^1\Rc_1$ and $\Rct_2$ (or $\Om^1\Rc_2$ and $\Rct_1$) cross over. We want to build a bounded map
\be
\phi: \Rct\to \Om^1\Rc_1\hotimes\Om^1\Rc_2\ ,
\ee
commuting with $\delta$, and verifying the following property:
\be
\phi(X*Y)=\phi(X)\mu(Y) + \mu(X)\phi(Y)+(-)^{|X|}\dd_2\mu(X)\dd_1\mu(Y)\label{mul}
\ee
for any $X,Y\in\Rct$, where $\dd_1:\Sc\to \Om^1\Rc_1\hotimes \Rct_2$ is the universal differential for $\Rc_1$, and $\dd_2:\Sc\to \Rct_1\hotimes \Om^1\Rc_2$ is the universal differential for $\Rc_2$. By definition, $\Rct$ is linearly generated by the unit $1$ and arbitrary products of elements of $\Rc_1$ and $\Rc_2$, subject to the rules $x_1*y_1=x_1y_1$ for any $x_1,y_1\in\Rc_1$, and $x_2*y_2=x_2y_2$ for any $x_2,y_2\in\Rc_2$. Therefore, once we know the value of $\phi$ on 1, $\Rc_1$ and $\Rc_2$, equation (\ref{mul}) implies its values on $\Rct$ inductively. We try the following:
\be
\phi(1)=0\ ,\quad \phi(x_1)=0\quad \forall x_1\in \Rc_1\ ,\quad \phi(x_2)=0\quad \forall x_2\in \Rc_2\ .
\ee
Then (\ref{mul}) yields, in degree two, 
\beq
\phi(x_1*x_2)&=& (-)^{|x_1|}\dd_2\mu(x_1)\dd_1\mu(x_2)=0\ ,\label{bebert}\\
\phi(x_2*x_1)&=& (-)^{|x_2|}\dd_2\mu(x_2)\dd_1\mu(x_1)=(-)^{|x_2|}\dd_2x_2\dd_1x_1 \non
\eeq
for any $x_1\in\Rc_1$ and $x_2\in\Rc_2$, because $\mu(x_1)=x_1\otimes 1$ and $\mu(x_2)=1\otimes x_2$. One proceeds similarly for the products of three elements, and so on... It is easy to check that this process is compatible with the relations $x_1*y_1=x_1y_1$ and $x_2*y_2=x_2y_2$, hence $\phi$ is well-defined on $\Rct$. Moreover, it obviously commutes with $\delta$.\\

We are now ready to introduce a chain map of total complexes $\Xt(\Rc)\to \Xt(\Rc_1)\hotimes\Xt(\Rc_2)$, where $\Xt(\Rc)$, $\Xt(\Rc_1)$ and $\Xt(\Rc_2)$ are respectively endowed with the boundaries $(\nat\dd\oplus\bb)+\delta$, $(\nat\dd_1\oplus\bb_1)+\delta_1$ and $(\nat\dd_2\oplus\bb_2)+\delta_2$. We let $\nat_1$, $\nat_2$ and $\nat_{12}$ denote the projections
\beq
\nat_1 &:& {\Om^1\Rc_1}\hotimes\Rct_2 \to {\Om^1\Rc_1}_{\nat}\hotimes\Rct_2\ ,\quad \nat_2 \ :\ \Rct_1\hotimes {\Om^1\Rc_2}\to \Rct_1\hotimes {\Om^1\Rc_2}_{\nat}\non\\
\nat_{12} &:& {\Om^1\Rc_1}\hotimes {\Om^1\Rc_2} \to {\Om^1\Rc_1}_{\nat}\hotimes {\Om^1\Rc_2}_{\nat}\ .
\eeq
\begin{proposition}\label{ppsi}
Let $(\Rc_1,\delta_1)$ and $(\Rc_2,\delta_2)$ be complete bornological DG algebras, and $\Rc=\Rc_1*\Rc_2$ the free product DG algebra. The bounded linear map $\Psi: \Xt(\Rc)\to \Xt(\Rc_1)\hotimes\Xt(\Rc_2)$ given by
\be
\Psi(X) = \mu(X)+ \nat_{12}\phi(X)\ \in\ \Rct_1\hotimes\Rct_2\ \oplus\ {\Om^1\Rc_1}_{\nat}\hotimes {\Om^1\Rc_2}_{\nat}\ ,
\ee
\beq
{\left. \begin{array}{cr}
	\Psi(\nat X\dd Y)= & (-)^{|X|(|Y|+1)}\nat_1 \dd_1 \mu(Y)\mu(X) \\[1mm]
	& +(-)^{|X|}\bb_2 \nat_{12} \mu(X)\phi(Y)\end{array} \right\}} &\in& {\Om^1\Rc_1}_{\nat}\hotimes\Rct_2 \non\\
 {\left. \begin{array}{l}
	+ \nat_2\mu(X)\dd_2\mu(Y)\\[1mm]
	+(-)^{|X|}\bb_1 \nat_{12} \mu(X)\phi(Y)\end{array} \right\}} &\in& \Rct_1\hotimes {\Om^1\Rc_2}_{\nat}
\eeq
is well-defined for any $X\in \Rct$ and $Y\in\Rc$. It is a morphism of even degree from the total complex $\Xt(\Rc)$ to the tensor product of total complexes $\Xt(\Rc_1)\hotimes\Xt(\Rc_2)$.
\end{proposition}
{\it Proof:} For notational simplicity, we will check the statement only when the elements $X,Y$ have even degree. The general case is exactly analogous, provided one takes care of the signs occuring when odd elements are permuted.\\
Thus let $X\in\Rct$ and $Y\in\Rc$, both of even degree. We first show that $\Psi$ is well-defined on the quotient $\Om^1\Rc_{\nat}$, i.e. it vanishes on the commutator subspace $[\Rc,\Om^1\Rc]$. In other words, $X\dd Y$ and $\dd Y X=\dd(YX)-Y\dd X$ must have the same image under $\Psi$. Set $\mu(X)=x$, $\mu(Y)=y$. One has
\beq
\lefteqn{\Psi\,\nat(\dd(YX)-Y\dd X)= \nat_1\dd_1(yx)+\nat_2\dd_2(yx) +(\bb_1+\bb_2)\nat_{12}\phi(YX)} \non\\
&& -\nat_1(\dd_1x) y-\nat_2 y\dd_2x- (\bb_1+\bb_2)\nat_{12} y\phi(Y) \non\\
&=& \nat_1\dd_1(yx)+\nat_2 \dd_2(yx) +(\bb_1+\bb_2)\nat_{12}(\phi(Y)x+y\phi(X)+\dd_2y\dd_1x) \non\\
&& -\nat_1(\dd_1x)y -\nat_2 y\dd_2x -(\bb_1+\bb_2)\nat_{12} y\phi(x)\non
\eeq
But $(\bb_1+\bb_2)\nat_{12}(\dd_2 y\dd_1x)= -\nat_2[\dd_2 y,x]+ \nat_1[\dd_1x,y]$, so that
\beq
\Psi\,\nat(\dd(YX)-Y\dd X) &=& \nat_1(\dd_1y)x+\nat_1 y\dd_1x+\nat_2(\dd_2y)x +\nat_2 y\dd_2x\non\\
&&-\nat_2[\dd_2 y,x]+ \nat_1[\dd_1x,y] + (\bb_1+\bb_2)\nat_{12} \phi(Y)x\non\\
&&-\nat_1(\dd_1x)y -\nat_2y\dd_2x\non\\
&=& \nat_1(\dd_1y)x + \nat_2x\dd_2y +(\bb_1+\bb_2)\nat_{12} x\phi(Y)\non\\
&=& \Psi(\nat X\dd Y)\ .\non
\eeq
We used the fact that $\nat_{12}$ is a trace on the $\Rct_1\hotimes\Rct_2$-bimodule $\Om^1\Rc_1\hotimes\Om^1\Rc_2$, but we were not allowed to set $-\nat_2[\dd_2 y,x]+ \nat_1[\dd_1x,y]$ equal to zero because $\nat_2$ and $\nat_1$ are not traces. Thus $\Psi$ is well-defined.\\
Let us now show that $\Psi$ is a morphism of total complexes. In fact it is a morphism of bicomplexes. First, the differential $\delta$ on $\Xt(\Rc)$ corresponds to the sum of differentials $\delta_1+\delta_2$ on $\Xt(\Rc_1)\hotimes\Xt(\Rc_2)$:
$$
(\delta_1+\delta_2)\Psi= \Psi \delta\ .
$$
This follows easily from the fact that $\mu$ is a DG algebra homomorphism and $\phi$ commutes with $\delta$. Second, we denote collectively by $\d$ the $X$-complex differentials $(\nat\dd\oplus\bb)$ on $\Xt(\Rc)$ and $(\nat_1\dd_1\oplus\bb_1)+(\nat_2\dd_2\oplus\bb_2)$ on $\Xt(\Rc_1)\hotimes\Xt(\Rc_2)$. We want to show $\d\Psi=\Psi\d$. For $X\in\Rct$ and $x=\mu(X)$, one has
$$
\Psi(\d X)=\Psi(\nat\dd X)=\nat_1\dd_1 x+\nat_2\dd_2x +(\bb_1+\bb_2)\nat_{12}\phi(X)=\d(x+\nat_{12}\phi(X))=\d\Psi(X)\ .
$$
For $X\in\Rct$ and $Y\in\Rc$ of even degree, one has
\beq
\lefteqn{\Psi\d(\nat X\dd Y)=\Psi([X,Y])\ =\ [x,y] + \nat_{12}\phi([X,Y])}\non\\
&=& [x,y] +\nat_{12}(\phi(X)y+x\phi(Y)+\dd_2x\dd_1y-\phi(Y)x-y\phi(X)-\dd_2y\dd_1x)\non\\
&=& [x,y]+ \nat_{12}(\dd_2x\dd_1y-\dd_2y\dd_1x)\non
\eeq
because $\nat_{12}$ is a trace. On the other hand,
\beq
\d\Psi(\nat X\dd Y)&=& \d(\nat_1(\dd_1y)x+\nat_2 x\dd_2y +(\bb_1+\bb_2)\nat_{12} x\phi(Y))\non\\
&=& \bb_1(\nat_1(\dd_1y)x)+\nat_2\dd_2(\nat_1(\dd_1y)x) + \bb_2(\nat_2 x\dd_2y) +\nat_1\dd_1(\nat_2 x\dd_2y)\ .\non
\eeq
One has $\nat_2\dd_2(\nat_1(\dd_1y)x)+\nat_1\dd_1(\nat_2 x\dd_2y)= \nat_{12}(\dd_2((\dd_1y)x)+\dd_1(x\dd_2y))=\nat_{12}(-\dd_1y\dd_2x+\dd_1x\dd_2y)$. The terms in $\bb$ are more complicated to compute. We decompose $x$ and $y$ respectively into sums of tensor products like $x_1\otimes x_2$ and $y_1\otimes y_2$, involving elements $x_1,y_1\in \Rct_1$ and $x_2,y_2\in\Rct_2$ whose degrees verify $|x_1|+|x_2|=|y_1|+|y_2|\equiv 0$ mod 2. Then
\beq
\lefteqn{\bb_1(\nat_1(\dd_1y)x)+ \bb_2(\nat_2 x\dd_2y) =}\non\\
&& (-)^{|x_1||y_2|}\bb_1(\nat_1 \dd_1y_1x_1 \otimes y_2x_2) + (-)^{|y_1|(|x_2|+1)}\bb_2(x_1y_1\otimes \nat_2 x_2\dd_2y_2)\non\\
&&= [x_1,y_1]\otimes y_2x_2 + (-)^{|x_2||y_1|} x_1y_1\otimes [x_2,y_2]\non\\
&&= [x,y]\ .\non
\eeq
Hence $\d\Psi(\nat X\dd Y)=\Psi\d(\nat X\dd Y)$ and $\d$ is a map of complexes. One proceeds similarly when $X$ and/or $Y$ are odd. \cqfd\\

The map $\Psi$ allows to obtain easily the generalized Chern character of a connection form $A\in (\Rct_1\hotimes\Rct_2)_-$ in the tensor product $\Xt(\Rc_1)\hotimes \Xt(\Rc_2)$. First note there exists an obvious {\it linear} inclusion of 
\be
\Rct_1\hotimes\Rct_2 = \cc\oplus \Rc_1\oplus \Rc_2 \oplus (\Rc_1\hotimes\Rc_2)
\ee
into the unital free product $\Rct=\widetilde{\Rc_1*\Rc_2}$. We use this inclusion to consider $A$ as an odd element $\Asf\in\Rc$, together with its curvature
\be
\Fsf=\delta \Asf +\Asf*\Asf\ \in \Rc_+\ .
\ee
Here $*$ is the product in $\Rc$ and $\delta=\delta_1+\delta_2$ the differential. The DG algebra map $\mu:\Rct\to\Rct_1\hotimes\Rct_2$ shows that $\Asf$ and $\Fsf$ are canonical lifts of $A$ and its curvature $F=(\delta_1+\delta_2) A+A^2$ in the following sense:
\be
\mu(\Asf)=A\ ,\qquad \mu(\Fsf)=F\ .
\ee
Now, from proposition \ref{pqui} the Chern character $\ch(\Asf)$ is a cocycle in the total complex $\Xt(\Rc)$. By applying the morphism $\Psi$, we get a cocycle $\ch(A)\in\Xt(\Rc_1)\hotimes \Xt(\Rc_2)$, which is by definition the Chern character of the connection $A$. The following proposition computes $\ch(A)$ explicitly in terms of the curvature $F$.
\begin{proposition}\label{pper}
Let $(\Rc_1,\delta_1)$ and $(\Rc_2,\delta_2)$ be complete bornological DG algebras, and $\Rc=\Rc_1*\Rc_2$ their free product DG algebra, with differential $\delta =\delta_1+\delta_2$. Let $A\in (\Rct_1\hotimes\Rct_2)_-$ be a connection form, and $\Asf$ its natural lift to $\Rc_-$. Then the image of the Chern character $\ch(\Asf)\in\Xt(\Rc)$ under the map $\Psi$ of proposition \ref{ppsi} is by definition the generalized Chern character $\ch(A)\in\Xt(\Rc_1)\hotimes \Xt(\Rc_2)$, whose components read
\beq
\Rct_1\hotimes\Rct_2 &\ni& \ch^{0,0}(A) =e^{-F}\ ,\non\\
{\Om^1\Rc_1}_{\nat}\hotimes\Rct_2 &\ni& \ch^{1,0}(A) =-\nat_1 \dd_1Ae^{-F}\ ,\non\\
\Rct_1\hotimes{\Om^1\Rc_2}_{\nat} &\ni& \ch^{0,1}(A) = -\nat_2 e^{-F}\dd_2A\ ,\\
{\Om^1\Rc_1}_{\nat}\hotimes{\Om^1\Rc_2}_{\nat} &\ni& \ch^{1,1}(A)= \nat_{12} \dd_1A e^{-F}\dd_2A + \non\\
&+&\int_{\Delta_2}ds_1ds_2\,\nat_{12} e^{-s_0F}\dd_2Fe^{-s_1F}\dd_1Fe^{-s_2F}\ ,\non
\eeq
where $F=(\delta_1+\delta_2)A+A^2 \in\Rct_1\hotimes\Rct_2$ is the curvature of $A$, and $\Delta_2=\{(s_0,s_1,s_2)\in [0,1]^3\ |\ \sum_is_i=1\}$ is the standard 2-simplex. The cocycle property of $\ch(A)$ means concretely, with $\delta=\delta_1+\delta_2$:
\beq
\delta\ch^{0,0}(A)+ \bb_1\ch^{1,0}(A)+\bb_2\ch^{0,1}(A) &=& 0\ ,\non\\
\delta\ch^{1,0}(A)+ \nat_1\dd_1\ch^{0,0}(A)+\bb_2\ch^{1,1}(A) &=& 0\ ,\non\\ 
\delta\ch^{0,1}(A)+ \bb_1\ch^{1,1}(A)+\nat_2\dd_2\ch^{0,0}(A) &=& 0\ ,\\
\delta\ch^{1,1}(A)+ \nat_1\dd_1\ch^{0,1}(A)+\nat_2\dd_2\ch^{1,0}(A) &=& 0\ .\non
\eeq
Moreover, the cohomology class of $\ch(A)$ is invariant with respect to smooth homotopies of $A$.
\end{proposition}
{\it Proof:} One has $F=(\delta_1+\delta_2)A+A^2$, $\Fsf=\delta\Asf+\Asf*\Asf$, and $\mu(\Asf)=A$, $\mu(\Fsf)=F$. Since $\Asf\in \Rc_1\oplus \Rc_2 \oplus (\Rc_1\hotimes\Rc_2)$, equation (\ref{bebert}) implies $\phi(\Asf)=0$, and similarly $\phi(\delta\Asf)=0$. Thus one has from (\ref{mul})
$$
\phi(\Fsf)=\phi(\Asf*\Asf)=-\dd_2A\dd_1A\ ,
$$
because $\mu(\Asf)=A$ is odd.\\
Recall now that the two components of $\ch(\Asf)$ in $\Xt(\Rc)$ are given by
$$
\ch^0(\Asf)=e^{-\Fsf}\ ,\quad \ch^1(\Asf)=-\nat e^{-\Fsf}\dd \Asf\ .
$$
Since the morphism $\Psi$ has even degree, one deduces
$$
\Psi(\ch^0(\Asf))=\ch^{0,0}(A)+\ch^{1,1}(A)\ ,\quad \Psi(\ch^1(\Asf))=\ch^{1,0}(A)+\ch^{0,1}(A)\ .
$$
From the definition of $\Psi$ we see that
$$
\Psi(e^{-\Fsf})=e^{-F}+\nat_{12}\phi(e^{-\Fsf})\ ,
$$
so we get immediately $\ch^{0,0}(A)=e^{-F}$. We now want to compute $\phi(\exp(-\Fsf))$. For this, we have to solve a differential equation for the function $t\mapsto \phi(\exp(-t\Fsf))$, $t\in[0,1]$:
\beq
\frac{d}{dt}\phi(e^{-t\Fsf})&=& -\phi(\Fsf*e^{-t\Fsf})\ =\ -\phi(\Fsf)e^{-tF}- F\phi(e^{-t\Fsf})-\dd_2F \dd_1e^{-tF}\non\\
&=& \dd_2A\dd_1A e^{-tF} -F\phi(e^{-t\Fsf})-\dd_2F \dd_1e^{-tF}\ ,\non
\eeq
or equivalently,
$$
(\frac{d}{dt}+F)\phi(e^{-t\Fsf})= \dd_2A\dd_1A e^{-tF}-\dd_2F \dd_1e^{-tF}\ .
$$
For $t=0$, one has by definition $\phi(1)=0$. Introduce the function
$$
U(t)=e^{tF}\phi(e^{-t\Fsf})\ ,
$$
then $U$ fulfills the differential equation
$$
\frac{d}{dt}U(t)= e^{tF}\dd_2A\dd_1A e^{-tF}-e^{tF}\dd_2F \dd_1e^{-tF}\ .
$$
Since $U(0)=0$, the solution reads
$$
U(t)=\int_0^t ds\, (e^{sF}\dd_2A\dd_1A e^{-sF}-e^{sF}\dd_2F \dd_1e^{-sF})\ ,
$$
therefore
$$
\phi(e^{-\Fsf})=\int_0^1 ds\, (e^{-(1-s)F}\dd_2A\dd_1A e^{-sF}-e^{-(1-s)F}\dd_2F \dd_1e^{-sF})\ .
$$
Projecting the latter expression onto ${\Om^1\Rc_1}_{\nat}\hotimes {\Om^1\Rc_2}_{\nat}$ through the trace $\nat_{12}$, one finds
$$
\nat_{12} \phi(e^{-\Fsf})= \nat_{12} e^{-F}\dd_2A\dd_1A +\int_{\Delta_2}ds_1ds_2\,\nat_{12} e^{-s_0F}\dd_2Fe^{-s_1F}\dd_1Fe^{-s_2F}\ .
$$
This is the claimed value for $\ch^{1,1}(A)$.\\
One the other hand, the definition of $\Psi$ for odd components yields
$$
\Psi(\nat e^{-\Fsf}\dd\Asf)= \nat_1 \dd_1A e^{-F}+ \nat_2 e^{-F}\dd_2A\ ,
$$
whence the values of $\ch^{1,0}(A)$ and $\ch^{0,1}(A)$. The homotopy invariance of $\ch(A)$ is a consequence of the same property for $\ch(\Asf)$. \cqfd\\
\begin{remark}\label{raux}\textup{
Let $(\Rc,\delta)$ be a complete bornological DG algebra. If $(\Lc,d)$ is an auxiliary bornological DG algebra and $\tau:\Lc\to \Vc$ a bounded closed graded trace with values in a bornological DG vector space $(\Vc,d)$, that is, $\tau\circ d=\pm d\circ\tau$, then the construction generalizes in the following way. We take a connection $A\in (\Rct\hotimes\Lc)_-$ and its curvature $F=(\delta+d)A+A^2\in (\Rct\otimes\Lc)_+$. Then the cochain $\tau\ch(A)$ given by
\be
\tau\ch^0(A)=\tau e^{-F} \in \Rct\hotimes\Vc\ ,\quad \tau\ch^1(A)=\tau\nat e^{-F}\dd A\in \Om^1\Rc_{\nat}\hotimes\Vc\ ,
\ee
defines a cocycle in the graded tensor product of complexes $\Xt(\Rc)\hotimes \Vc$ endowed with the total differential $(\nat\dd\oplus\bb)+\delta+d$. This was implicitly used in the proof of the homotopy invariance (proposition \ref{phomo}), with $\Lc$ the graded commutative algebra of differential forms over the interval $[0,1]$ and $\Vc=\Lc$, $\tau=\id$. More generally, for two DG algebras $(\Rc_1,\delta_1)$ and $(\Rc_2,\delta_2)$ and an auxiliary $(\Lc,d)$ provided with a closed graded trace as before, any connection $A\in \Rct_1\hotimes\Rct_2\hotimes\Lc$ yields a cocycle $\tau\ch(A)$ in the total complex $\Xt(\Rc_1)\hotimes\Xt(\Rc_2)\hotimes\Vc$.}
\end{remark}
\begin{remark}\label{rrec}\textup{
The same method using free products can be generalized to an arbitrary number of DG algebras by induction. For example in the case of three DG algebras $(\Rc_1,\delta_1)$, $(\Rc_2,\delta_2)$, $(\Rc_3,\delta_3)$, one has a composition of morphisms
\be
\Xt(\Rc_1*(\Rc_2*\Rc_3))\to \Xt(\Rc_1)\hotimes \Xt(\Rc_2*\Rc_3)\to \Xt(\Rc_1)\hotimes \Xt(\Rc_2)\hotimes\Xt(\Rc_3)\ .
\ee
One thus gets a Chern character $\ch(A)$ for any connection $A\in \Rct_1\hotimes\Rct_2\hotimes\Rct_3$ with curvature $F=(\delta_1+\delta_2+\delta_3)A+A^2$, using its canonical lift $\Asf\in \Rc_1*(\Rc_2*\Rc_3)$. This however leads rapidly to complicated expressions when the number of algebras increases.}
\end{remark}

Let us now pass to coalgebras. The whole construction of universal one-forms, $X$-complex, and Chern character can be dualized by replacing the former unital DG algebra $\Rct$ with a counital DG coalgebra and transposing formally all structure maps. This is done by Quillen in \cite{Q2}, but since our sign conventions are different we have to set up carefully the definitions.\\
Let $(\Cc,\delta)$ be a (complete bornological) counital coassociative DG coalgebra. This means that $\Cc$ is a $\zz_2$-graded bornological vector space endowed with bounded coproduct and counit
\be
\Delta: \Cc\to \Cc\hotimes \Cc\ ,\quad \eta: \Cc\to \cc\ ,
\ee
verifying $(\Delta\otimes\id)\circ \Delta=(\id \otimes\Delta)\circ \Delta$ (coassociativity) and $(\eta\otimes\id)\circ\Delta=(\id\otimes\eta)\circ\Delta=\id$. We use Sweedler's notation \cite{Sw} for the coproduct (the sum may be infinite):
\be
\Delta(x)=\sum x_{(0)}\otimes x_{(1)}\ ,\quad \forall x\in \Cc\ .
\ee
The differential $\delta: \Cc\to \Cc$, $\delta^2=0$, is an odd degree coderivation:
\be
\Delta\circ \delta=(\delta\otimes\id \pm \id\otimes\delta)\circ\Delta\ ,
\ee
where the sign $\pm$ depends on the degree of the element in the first factor of $\Cc\hotimes \Cc$. Moreover $\delta$ is compatible with the counit in the sense that $\eta\circ\delta=0$. We also impose some additional conditions on $\Cc$: consider that $\Cc$ splits linearly into the direct sum $\cc\oplus\ker \eta$, such that the coproduct $\Delta$ carries the factor $\cc$ to $\cc\hotimes\Cc+\Cc\hotimes\cc\subset \Cc\hotimes\Cc$, and $\delta\cc=0$, $\delta\ker\eta\subset\ker\eta$. These requirements imply that the (bornological) dual space of $(\Cc,\delta)$ is an augmented DG algebra. Indeed, set $\Rc=\hom(\ker\eta,\cc)$, the space of bounded linear maps with its natural grading. It is an associative DG algebra for the convolution product and differential
\be
(fg)(x)=\sum (-)^{|g||x_{(0)}|}f(x_{(0)})g(x_{(1)})\ ,\quad (\delta f)(x)=(-)^{|f|}f(\delta x)\ ,
\ee
for any $f,g\in\Rc$ and $x\in\ker\eta$. The unitalization $\Rct$ identifies with $\hom(\Cc,\cc)$, the unit corresponding to the counit $\eta:\Cc\to\cc$ and $\delta\eta=0$. With this dual identification, we derive straightforwardly the construction of the $X$-complex for DG coalgebras. We first introduce the graded space of one-coforms $\Om_1\Cc$:
\be
\Om_1\Cc= \Cc\hotimes\ker\eta\ .
\ee
The degree of an homogeneous element $x\otimes y$ is $|x|+|y|+1$ (the symbol $\otimes$ has degree 1). $\Om_1\Cc$ is a bicomodule over the coalgebra $\Cc$, with left and right coactions
\beq
\Delta_l : \Om_1 \Cc\to \Cc\hotimes\Om_1 \Cc\ ,\quad \Delta_l(x\otimes y) &=& \sum x_{(0)}\otimes (x_{(1)}\otimes y)\ ,\non\\
\Delta_r : \Om_1 \Cc\to \Om_1 \Cc\hotimes \Cc\ ,\quad \Delta_r(x\otimes y) &=&\sum\ (x\otimes y_{(0)})\otimes y_{(1)} -\\
&& (-)^{|x_{(1)}|} (x_{(0)}\otimes x_{(1)})\otimes y\ .\non
\eeq
The differential $\delta$ is extended to $\Om_1\Cc$ by $\delta(x\otimes y)=\delta x\otimes y +(-)^{|x|+1}x\otimes\delta y$ and is compatible with the bicomodule structure in an obvious sense. The universal derivation $\dd:\Rct\to \Om^1\Rc$ can be dualized into a coderivation $\d: \Om_1 \Cc\to \Cc$. It is given by
\be
\d(x\otimes y)=\eta(x)y\ ,\quad \forall x\in\Cc,\ y\in\ker\eta\ .
\ee
In the same way, the Hochschild operator $b:\Om^1\Rc\to\Rc$ is dualized into a map $\beta: \Cc\to \Om_1 \Cc$ corresponding to
\be
\beta(x)=\sum \left( -(-)^{|x_{(0)}|}x_{(0)}\otimes x_{(1)}+ (-)^{|x_{(1)}|(|x_{(0)}|+1)}x_{(1)}\otimes x_{(0)}\right)\ \in\Cc\otimes\ker\eta\ ,
\ee
where $x_{(i)}$ is understood to be projected on the direct factor $\ker\eta$ when necessary, using the decomposition $\Cc=\cc\oplus\ker\eta$. The bicomodule structure of $\Om_1 \Cc$ allows to define the cocommutator subspace
\be
\Om_1 \Cc^{\nat}=\ker (\Delta_l-\si\Delta_r)\subset \Om_1 \Cc\ ,
\ee
where $\si:\Om_1 \Cc\hotimes \Cc\to \Cc\hotimes \Om_1 \Cc$ is the graded flip. It plays a role dual to the commutator quotient space $\Om^1\Rc_{\nat}=\Om^1\Rc/[\Rc,\Om^1\Rc]$. We denote by $\nat:\Om_1 \Cc^{\nat}\to\Om_1 \Cc$ the natural inclusion. The cocommutator subspace is stable by $\delta$ and contains the image of the map $\beta$. One has $\beta \circ \d\nat=0$ on $\Om_1 \Cc^{\nat}$ and $\d\nat\circ\beta=0$ on $\Cc$, and $\delta$ anticommutes with $\beta$ and $\d\nat$. The $X$-complex of the DG coalgebra $\Cc$ is thus the periodic complex
\be
X(\Cc): \Cc\ \xymatrix@1{\ar@<0.5ex>[r]^{\beta} &  \ar@<0.5ex>[l]^{\d\nat}}\ \Om_1\Cc^{\nat}\ ,
\ee
where the factors are themselves complexes for the differential $\delta$. At a formal level, $X(\Cc)$ plays a role dual to the augmented $X$-complex $\Xt(\Rc)$ (it is of course not the dual complex in a strict sense). The conditions $\Delta$ carries $\cc$ into $\cc\hotimes\Cc+\Cc\hotimes\cc$ and $\delta\ker\eta\subset \ker\eta$ imply that $X(\Cc)$ splits into the direct sum of the trivial complex $X(\cc):\cc\rlarrows 0$ and the {\it reduced} $X$-complex (formally dual to $X(\Rc)$)
\be
\Xb(\Cc): \ker \eta\ \xymatrix@1{\ar@<0.5ex>[r]^{\beta} &  \ar@<0.5ex>[l]^{\d\nat}}\ \Om_1\Cc^{\nat}\ .
\ee

The Chern character associated to a connection $A\in\Rc$ can be translated into this dual setting without difficulty. We simply replace $\Rct\hotimes\cdot$ by $\hom(\Cc,\cdot)$ everywhere, with sign rules adapted to the natural grading of the objects involved. Recall that the space of bounded linear maps $\hom(\Cc,\cc)$ is naturally a unital DG algebra, for the convolution product $fg=(f\otimes g)\circ\Delta$ and the differential
\be
\delta f= (-)^{|f|}f\circ \delta\ ,\quad \forall f\in\hom(\Cc,\cc)\ .
\ee
We introduce the derivation 
\be
\dd:\hom(\Cc,\cc)\to \hom(\Om_1 \Cc,\cc)\ ,\quad \dd f=(-)^{|f|}f\circ \d\ ,
\ee
verifying the Leibniz rule $\dd(fg)=(\dd f)g+(-)^{|f|}f(\dd g)$, where $\hom(\Om_1 \Cc,\cc)$ is viewed as a bimodule over the algebra $\hom(\Cc,\cc)$. We also define a trace
\be
\nat: \hom(\Om_1 \Cc,\cc)\to\hom(\Om_1 \Cc^{\nat},\cc)\ ,\quad \nat\gamma=\gamma\circ\nat\ ,
\ee
for $\gamma\in\hom(\Om_1\Cc,\cc)$. The connection and curvature construction now works in the complex $\hom(X(\Cc),\cc)$ endowed with the total differential $(\nat\dd\oplus\bb)+\delta$ transposed of the differential $(\beta\oplus\d\nat)+\delta$ on $X(\Cc)$:
\beq
\delta f\ =\ (-)^{|f|}f\circ\delta\ ,\quad \nat\dd f &=& (-)^{|f|}f\circ\d\nat\ ,\quad \forall f\in\hom(\Cc,\cc)\ ,\\
\delta \gamma\ =\ (-)^{|\gamma|}\gamma\circ\delta\ ,\quad \bb \gamma &=& (-)^{|\gamma|}\gamma\circ \beta\ ,\quad\forall \gamma\in\hom(\Om_1\Cc^{\nat},\cc)\ .\non
\eeq
Given any connection $A\in\hom(\Cc,\cc)_-$ with curvature $F=\delta A+A^2$, we define the Chern character $\ch(A)=\ch^0(A)+\ch^1(A)$ exactly as in proposition \ref{pqui},
\be
\ch^0(A)=e^{-F}\ \in \hom(\Cc,\cc)\ ,\quad \ch^1(A)=-\nat e^{-F}\dd A\ \in\hom(\Om_1 \Cc^{\nat},\cc)\ ,
\ee
which yields a cocycle in $\hom(X(\Cc),\cc)$. This method was used by Quillen in \cite{Q2}. It can be generalized including an arbitrary number of DG algebras or coalgebras, using the formulas we derived before. For example, the bivariant Chern character of section \ref{sbiv} will be based on the following hybrid cocycle, involving a DG coalgebra $(\Cc,\delta_1)$ with coproduct $\Delta$ and a DG algebra $(\Rc,\delta_2)$ with product $m$: form the DG algebra $\hom(\Cc,\Rct)$, provided with the convolution product $fg=m\circ(f\otimes g)\circ \Delta$, the unit $1\eta$ and two differentials $\delta_1$, $\delta_2$ induced by
\be
\delta_1 f=(-)^{|f|}f\circ\delta_1\ ,\quad \delta_2f=\delta_2\circ f\ ,\quad \forall f\in \hom(\Cc,\Rct)\ .
\ee
We have two derivations of bimodules over $\hom(\Cc,\Rct)$,
\beq
\dd_1 &:& \hom(\Cc,\Rct)\to \hom(\Om_1\Cc,\Rct)\ ,\quad \dd_1f=(-)^{|f|}f\circ\d\ ,\non\\
\dd_2 &:& \hom(\Cc,\Rct)\to \hom(\Cc,\Om^1\Rc)\ ,\quad \dd_2f=\dd\circ f\ ,
\eeq
where $\dd:\Rct\to\Om^1\Rc$ is the universal derivation for $\Rct$ and $\d:\Om_1\Cc\to\Cc$ is the universal coderivation. Using the natural trace $\nat:\Om^1\Rc\to\Om^1\Rc_{\nat}$ and cotrace $\nat:\Om_1\Cc^{\nat}\to\Om_1\Cc$, we define the partial traces 
\beq
\nat_1 &:& \hom(\Om_1\Cc,\Rct)\to \hom(\Om_1\Cc^{\nat},\Rct)\ ,\non\\
\nat_2 &:& \hom(\Cc,\Om^1\Rc)\to \hom(\Cc,\Om^1\Rc_{\nat})\ ,\\
\nat_{12} &:& \hom(\Om_1\Cc,\Om^1\Rc)\to \hom(\Om_1\Cc^{\nat},\Om^1\Rc_{\nat})\ .\non
\eeq
In fact $\nat_{12}$ is really a trace of $\hom(\Cc,\Rct)$-bimodule, but not $\nat_1$ and $\nat_2$. We consider the $\zz_2$-graded complex $\hom(X(\Cc),\Xt(\Rc))$, with differentials induced by $(X(\Cc),(\beta\oplus\d\nat)+\delta_1)$ and $(\Xt(\Rc),(\nat\dd\oplus\bb)+\delta_2)$. One has
\beq
\delta_1\gamma &=& (-)^{|\gamma|}\gamma\circ \delta_1\ ,\qquad \delta_2\gamma\ =\ \delta_2\circ\gamma\ ,\non\\
\bb_1\gamma &=& (-)^{|\gamma|}\gamma\circ \beta\ ,\qquad \bb_2\gamma\ =\ \bb\circ\gamma\ ,\\
\nat_1\dd_1\gamma &=& (-)^{|\gamma|} \gamma\circ \d\nat\ ,\qquad \nat_2\dd_2\gamma\ =\ \nat\dd\circ\gamma\ ,\non
\eeq
for any $\gamma$ in the appropriate component of $\hom(X(\Cc),\Xt(\Rc))$. Since it is a $\hom$-complex, the cocycles must correspond to chain maps between $X(\Cc)$ and $\Xt(\Rc)$. Given a connection $A\in\hom(\Cc,\Rct)_-$, we can associate a chain map $\ch(A)\in\hom(X(\Cc),\Xt(\Rc))$ by using exactly the same formulas as in the previous situation with $\Xt(\Rc_1)\hotimes\Xt(\Rc_2)$. The only difference is that we have to change the signs in front of the derivations $\delta_2$ and $\dd_2$ in order to get a chain map with right signs. Thus define the curvature $F=(\delta_1-\delta_2)A+A^2$, and the expression of the Chern character $\ch(A)\in \hom(X(\Cc),\Xt(\Rc))$ can literally be copied from proposition \ref{pper}, up to the sign of $\dd_2$:
\beq
\hom(\Cc,\Rct) &\ni& \ch_0^0(A) =e^{-F}\ ,\non\\
\hom(\Om_1\Cc^{\nat},\Rct) &\ni& \ch_1^0(A) =-\nat_1 \dd_1Ae^{-F}\ ,\non\\
\hom(\Cc,\Om^1\Rc_{\nat}) &\ni& \ch_0^1(A) = \nat_2 e^{-F}\dd_2A\ ,\\
\hom(\Om_1\Cc^{\nat},\Om^1\Rc_{\nat}) &\ni& \ch_1^1(A)= -\nat_{12} \dd_1A e^{-F}\dd_2A - \non\\
&-&\int_{\Delta_2}ds_1ds_2\,\nat_{12} e^{-s_0F}\dd_2Fe^{-s_1F}\dd_1Fe^{-s_2F}\ .\non
\eeq
The cocycle property of $\ch(A)\in\hom(X(\Cc),\Xt(\Rc))$ then reads
\beq
\delta_1\ch_0^0(A)+ \bb_1\ch_1^0(A) &=& \delta_2\ch_0^0(A)+\bb_2\ch_0^1(A)\ ,\non\\
\delta_1\ch_1^0(A)+ \nat_1\dd_1\ch_0^0(A) &=& \delta_2\ch_1^0(A)+\bb_2\ch_1^1(A)\ ,\\ 
\delta_1\ch_0^1(A)+ \bb_1\ch_1^1(A) &=& \delta_2\ch_0^1(A)+\nat_2\dd_2\ch_0^0(A)\ ,\non\\
\delta_1\ch_1^1(A)+ \nat_1\dd_1\ch_0^1(A) &=& \delta_2\ch_1^1(A)+\nat_2\dd_2\ch_1^0(A)\ .\non
\eeq
As usual the cohomology class of $\ch(A)$ is homotopy invariant. More generally, if $\Cc_1,\ldots,\Cc_n$ are DG coalgebras with counit and $\Rc_1,\ldots,\Rc_p$ are DG algebras, a connection form is an odd element of the DG convolution algebra $\hom(\Cc_1\hotimes\ldots\hotimes\Cc_n,\Rct_1\hotimes\ldots\hotimes\Rct_p)$. Then the adaptation of remark \ref{rrec} to coalgebras yields a cocycle
\be
\ch(A)\in \hom(X(\Cc_1)\hotimes\ldots\hotimes X(\Cc_n),\Xt(\Rc_1)\hotimes\ldots\hotimes\Xt(\Rc_p))\ .
\ee
Remark \ref{raux} generalizes also in a straightforward way.

\section{Bivariant cyclic cohomology}\label{sbiv}

The universal cocycles introduced in the previous section will now be used for the construction of a bivariant Chern character in the realm of bornological algebras. Our remark is that there are two equivalent ways of bescribing the cyclic homology of an algebra $\Ac$. The first one is the original cyclic bicomplex of Connes \cite{C0, L}. As observed by Quillen \cite{Q2}, the cyclic bicomplex is isomorphic to the (reduced) $X$-complex of the {\it bar coalgebra} $\Bb(\Ac)$. This led Quillen to build cyclic cocycles on $\Ac$ by means of the connection and curvature method in the complex $\hom(X(\Bb(\Ac)),\cc)$. The second description of cyclic homology is provided by the Cuntz-Quillen theory \cite{CQ1}. The latter involves the $X$-complex of the {\it tensor algebra} $T\Ac$, with its natural filtrations given by the powers of the ideal $\ker(T\Ac\to\Ac)$. If $\Ac$ is a complete bornological algebra, its bar coalgebra and tensor algebra are completed with respect to an {\it entire bornology}. This yields the entire bar coalgebra $\Bbe(\Ac)$ and the analytic tensor algebra $\Tc\Ac$. The {\it entire cyclic homology} of $\Ac$ is computed either by $X(\Bbe(\Ac))$ or $X(\Tc\Ac)$. This ``self-dual'' property of cyclic homology can be exploited for the description of bivariant cyclic cohomology. If $\Ac$ and $\Bc$ are complete bornological algebras, the complex $\hom(X(\Bbe(\Ac)),X(\Tc\Bc))$ computes the bivariant entire cyclic cohomology $HE_*(\Ac,\Bc)$.\\
Consider the unitalization $\Bct$ of $\Bc$. We introduce the set of bivariant elements $\Psi(\Ac,\Bct)$ given by quintuples $\Ec=(\Lc,\ell^1,\tau,\rho,D)$. They are intended to give an algebraic description of Kasparov bivariant $K$-theory \cite{Bl}. $\Lc$ is an auxiliary $\zz_2$-graded complete bornological algebra containing an ideal $\ell^1$ provided with a graded trace of homogeneous degree $\tau:\ell^1\to\cc$. $\rho:\Ac\to\Lc\hotimes\Bct$ is an homomorphism and $D\in\Lc\hotimes\Bct$ is a an odd element (``Dirac operator''). The laplacian of $D$ is assumed to have a trace-class heat kernel $\exp(-tD^2)\in\ell^1\hotimes\Bct$ for any $t>0$. We form the Quillen superconnection $A=\rho+D$ viewed as an element of the DG convolution algebra $\hom(\Bbe(\Ac),\Lc\hotimes\Tct\Bc)$. Then $\Ec$ is called $\te$-summable if the exponential of the curvature of $A$ converges to a bounded map $\Bbe(\Ac)\to \ell^1\hotimes\Tct\Bc$. If this condition is realized, the construction of section \ref{sgen} yields a cocycle
\be
\tau\ch(A)\in \hom(X(\Bbe(\Ac)),X(\Tc\Bc))
\ee
after composition with the trace $\tau$. Its cyclic cohomology class is homotopy invariant and defines the bivariant Chern character $\ch(\Ec)\in HE_*(\Ac,\Bc)$. The exponential of the curvature of $A$ incorporates the heat kernel $\exp(-tD^2)$ as a regulator of traces. The cocycle $\tau\ch(A)$ is a bivariant generalization of the JLO formula \cite{JLO}, and in fact is equivalent to our previous construction of the entire Chern character for unbounded Kasparov bimodules \cite{P3}. In \cite{P4}, we showed by a process analogous to \cite{CM93} that under some finite-summability assumptions this bivariant Chern character retracts on periodic cocycles related to the universal algebras of Cuntz and Zekri \cite{Cu1,Z}, and yields a bivariant Chern character adapted to the description of $KK$-theory via bounded Fredholm bimodules (these formulas are close to those of Nistor \cite{Ni1,Ni2}). The same method may be generalised by incorporating more than two algebras, as explained in the previous section. For example, judicious choices of connections $A$ lead to cup or shuffle products in periodic or entire cyclic homology. \\

Let $\Ac$ be a trivially graded complete bornological algebra. The {\it bar construction} of $\Ac$ is the $\zz_2$-graded vector space
\be
\Bb(\Ac)=\bigoplus_{n\ge 0}\Bb_n(\Ac)\ ,\quad \Bb_n(\Ac)=\Ac^{\hotimes n}\ ,\quad \Bb_0(\Ac)=\cc\ ,
\ee
with grading given by the parity of $n$ in $\Bb_n(\Ac)$. The bar construction endowed with the direct sum bornology is a complete bornological coassociative coalgebra, with coproduct $\Delta:\Bb(\Ac)\to \Bb(\Ac)\hotimes\Bb(\Ac)$ given by
\be
\Delta(a_1\otimes\ldots\otimes a_n)=\sum_{i=0}^n(a_1\otimes\ldots\otimes a_i)\otimes(a_{i+1}\otimes\ldots\otimes a_n)\ .
\ee
It has a counit $\eta:\Bb(\Ac)\to \cc$ corresponding to the natural projection onto $\Bb_0(\Ac)$. One defines a differential $b':\Bb_n(\Ac)\to \Bb_{n-1}(\Ac)$ of odd degree:
\be
b'(a_1\otimes\ldots\otimes a_n)=\sum_{i=1}^{n-1}(-)^{i-1}a_1\otimes\ldots\otimes a_ia_{i+1}\otimes \ldots\otimes a_n\ ,
\ee
with $b'=0$ for $n=0,1$. Then one has ${b'}^2=0$, and the coproduct and counit are morphisms of graded complexes: $\Delta \circ b'=(b'\otimes 1\pm 1\otimes b')\circ \Delta$ and $\eta\circ b'=0$. This turns $\Bb(\Ac)$ into a $\zz_2$-graded DG coalgebra. In the sequel, we will consider that $\Bb(\Ac)$ is endowed with the differential $\delta=-b'$ in order to fit well with the bicomplex of cyclic homology.\\
The associated universal bicomodule $\Om_1\Bb(\Ac)=\Bb(\Ac)\hotimes\ker\eta$ (section \ref{sgen}) identifies with the graded space
\be
\Om_1\Bb(\Ac)= \Bb(\Ac)\hotimes\Ac\hotimes\Bb(\Ac)\ ,
\ee
where the degree of the element $(a_1\otimes\ldots\otimes a_{i-1})\otimes a_i \otimes (a_{i+1}\otimes\ldots\otimes a_n)$ is equal to the parity of $n-1$. The rule is that the middle element $a_i$ always has {\it even degree}, whereas the other $a$'s appearing in the left and right factors $\Bb(\Ac)$ are odd. The isomorphism $\Bb(\Ac)\hotimes\Ac\hotimes\Bb(\Ac)\cong \Bb(\Ac)\hotimes\ker\eta$ is explicitly given by
\beq
\lefteqn{(a_1\otimes\ldots\otimes a_{i-1})\otimes a_i \otimes (a_{i+1}\otimes\ldots\otimes a_n)\longleftrightarrow}\\ 
&& \sum_{j=1}^{i}(-)^{j+1} (a_1\otimes\ldots\otimes a_{i-j})\otimes(a_{i-j+1}\otimes\ldots\otimes a_n)\ .\non
\eeq
The identification $\Om_1\Bb(\Ac)=\Bb(\Ac)\hotimes\Ac\hotimes\Bb(\Ac)$ simplifies the descriptions of the left and right bicomodule coactions of $\Bb(\Ac)$ on $\Om_1\Bb(\Ac)$. They correspond respectively to
\beq
\Delta_l=\Delta\otimes \id\otimes\id &:& \Om_1\Bb(\Ac)\to \Bb(\Ac)\hotimes\Om_1\Bb(\Ac)\ ,\non\\
\Delta_r=\id \otimes\id \otimes \Delta &:& \Om_1\Bb(\Ac)\to \Om_1\Bb(\Ac)\hotimes\Bb(\Ac)\ .
\eeq
The universal coderivation of odd degree $\d:\Om_1\Bb(\Ac)\to\Bb(\Ac)$ is given by
\be
\d((a_1\otimes\ldots\otimes a_{i-1})\otimes a_i \otimes (a_{i+1}\otimes\ldots\otimes a_n))=(-)^{i-1}(a_1\otimes\ldots\otimes a_n)\ .
\ee
The sign $(-)^{i-1}$ morally comes from the crossing of $\d$ over $(a_1\otimes\ldots\otimes a_{i-1})$ until the position of the element $a_i$, which becomes odd. The cocommutator subspace $\Om_1\Bb(\Ac)^{\nat}=\ker(\Delta_l-\si\Delta_r)$ identifies with
\be
\Om_1\Bb(\Ac)^{\nat}=\Ac\hotimes \Bb(\Ac)\ ,
\ee
where the element $a_0\otimes a_1\otimes\ldots\otimes a_n\in \Ac\hotimes\Bb_n(\Ac)$ has degree $n$ mod 2 ($a_0$ is even whereas the other $a_i$'s are odd). The natural inclusion $\nat:\Om_1\Bb(\Ac)^{\nat}\to \Om_1\Bb(\Ac)$ is therefore a bounded map of even degree, explicitly given by
\be
\nat(a_0\otimes a_1\otimes\ldots\otimes a_n)=\sum_{i=0}^n(-)^{i(n-1)}(a_{i+1}\otimes\ldots\otimes a_n)\otimes a_0 \otimes (a_1\otimes\ldots\otimes a_i)\ .
\ee
Under these identifications, the differential $\delta:\Om_1\Bb(\Ac)^{\nat}\to\Om_1\Bb(\Ac)^{\nat}$ induced by the differential $-b'$ on $\Bb(\Ac)$ corresponds to the Hochschild operator 
\be
b(a_0\otimes a_1\otimes\ldots\otimes a_n)=\sum_{i=0}^{n-1}(-)^ia_0\otimes\ldots\otimes a_ia_{i+1}\ldots\otimes a_n + (-)^na_na_0\otimes a_1\ldots \otimes a_n\ .
\ee
We are now in a position to identify the $X$-complex of the bar DG coalgebra (section \ref{sgen}). Since $\Bb(\Ac)$ and $\Om_1\Bb(\Ac)^{\nat}$ are both given by direct sums of tensor products $\Ac^{\hotimes n}$, it is useful to introduce the (signed) backward cyclic permutation
\be
\la(a_1\otimes\ldots\otimes a_n)=(-)^{n-1}a_n\otimes a_1\otimes \ldots\otimes a_{n-1}\ ,
\ee
thus acting on $\Bb(\Ac)$ and $\Om_1\Bb(\Ac)^{\nat}$ with the same sign convention. Consider also the norm operator
\be
N=\sum_{i=0}^{n-1}\la^i\quad \mbox{on}\ \Ac^{\hotimes n}\ .
\ee
Then the $X$-complex of the bar DG coalgebra is given by the supercomplex
\be
X(\Bb(\Ac)): \Bb(\Ac)\ \xymatrix@1{\ar@<0.5ex>[r]^{\beta=1-\la} &  \ar@<0.5ex>[l]^{\d\nat=N}}\ \Om_1\Bb(\Ac)^{\nat}\ ,
\ee
where $\Bb(\Ac)$ and $\Om_1\Bb(\Ac)^{\nat}$ are themselves complexes for the differential $\delta$, which anticommutes with $1-\la$ and $N$. Remark that $\Bb(\Ac)$ splits linearly as the direct sum of vector spaces $\cc\oplus\ker \eta$, where $\cc=\Bb_0(\Ac)$. Both $\cc$ and $\ker\eta$ are stable by $\delta$, and $\Delta$ maps $\cc$ to $\cc\hotimes\cc$. It follows that $X(\Bb(\Ac))$ splits into the direct sum of the reduced $X$-complex
\be
\Xb(\Bb(\Ac)): \ker\eta\ \xymatrix@1{\ar@<0.5ex>[r]^{1-\la} &  \ar@<0.5ex>[l]^{N}}\ \Om_1\Bb(\Ac)^{\nat}
\ee
and the trivial complex $\cc\rlarrows 0$. In this way, we obtain Quillen's fundamental result \cite{Q2} that the reduced $X$-complex of $\Bb(\Ac)$ identifies with the (two-sided) cyclic bicomplex \cite{L}:
\be
\xymatrix{
  & \vdots \ar[d]_{b} & \vdots \ar[d]_{-b'} & \vdots \ar[d]_{b} & \vdots \ar[d]_{-b'} &   \\
 \ldots &  \ar[l]_{N} \Ac^{\hotimes 3} \ar[d]_{b} & \ar[l]_{1-\la} \Ac^{\hotimes 3} \ar[d]_{-b'} & \ar[l]_{N} \Ac^{\hotimes 3} \ar[d]_{b} & \ar[l]_{1-\la} \Ac^{\hotimes 3} \ar[d]_{-b'} & \ar[l]_{N} \ldots \\
\ldots &  \ar[l]_{N} \Ac^{\hotimes 2} \ar[d]_{b} & \ar[l]_{1-\la} \Ac^{\hotimes 2} \ar[d]_{-b'} & \ar[l]_{N} \Ac^{\hotimes 2} \ar[d]_{b} & \ar[l]_{1-\la} \Ac^{\hotimes 2} \ar[d]_{-b'} &  \ar[l]_{N} \ldots \\
\ldots &  \ar[l]_{N} \Ac & \ar[l]_{1-\la} \Ac & \ar[l]_{N} \Ac & \ar[l]_{1-\la} \Ac & \ar[l]_{N} \ldots }\label{bcp}
\ee
The columns with $-b'$ (resp. $b$) correspond to $\ker\eta\subset \Bb(\Ac)$ (resp. $\Om_1\Bb(\Ac)^{\nat}$). It if often more convenient to reexpress the cyclic bicomplex in terms of noncommutative differential forms \cite{C1}. Let $\Om\Ac$ be the $\zz_2$-graded vector space 
\be
\Om\Ac=\bigoplus_{n\ge 0} \Om^n\Ac\ ,\quad \Om^n\Ac=\Act\hotimes\Ac^{\hotimes n}\ ,\quad \Om^0\Ac=\Ac\ ,
\ee
where $\Act=\cc\oplus \Ac$ is the unitalization of $\Ac$. The degree of $\Om^n\Ac$ is the parity of $n$. Since $\Om^n\Ac$ decomposes into the direct sum $\Ac^{\hotimes (n+1)}\oplus \Ac^{\hotimes n}$ for $n>0$, the corresponding elements are denoted by
\beq
a_0da_1\ldots da_n &\in& \Ac^{\hotimes (n+1)}\subset\Om^n\Ac\ ,\non\\
da_1\ldots da_n &\in& \Ac^{\hotimes n}\subset\Om^n\Ac\ ,
\eeq
for $a_i\in\Ac$. Then $\Om\Ac$ is a non-unital DG algebra, for the differential $d:\Om^n\Ac\to\Om^{n+1}\Ac$
\be
d(a_0da_1\ldots da_n)=da_0da_1\ldots da_n\ ,\quad d(da_1\ldots da_n)=0, \quad d^2=0\ ,
\ee
and the usual associative product of differential forms verifying the graded Leibniz rule
\be
d(\om_1\om_2)=d\om_1\om_2+(-)^{|\om_1|}\om_1d\om_2\qquad \forall \om_{1,2}\in\Om\Ac\ .
\ee
Introduce the Hochschild operator $b:\Om^n\Ac\to \Om^{n-1}\Ac$
\be
b(\om da)=(-)^{n-1}[\om,a]\ ,\quad b(a)=0\quad \forall \om\in\Om^{n-1}\Ac\ ,\ a\in\Ac\ ,
\ee
and Connes' boundary map $B:\Om^n\Ac\to \Om^{n+1}\Ac$
\be
B(a_0da_1\ldots da_n)=(1+\kappa+\ldots+ \kappa^n)(da_0da_1\ldots da_n)\ ,
\ee
where $\kappa$ is the Karoubi operator defined by $1-\kappa=db+bd$. One has $\kappa(\om da)=(-)^{|\om|}da\,\om$, $\forall \om\in\Om\Ac$, $a\in\Ac$. It turns out that $b^2=B^2=bB+Bb=0$ and $\kappa B=B=B\kappa$, hence $\Om\Ac$ is actually a bicomplex. Now remark that there is an isomorphism of graded vector spaces
\be
\Xb(\Bb(\Ac))\cong \Om\Ac\ ,\label{iso}
\ee
given by the following correspondence:
\beq
\Bb_n(\Ac) \ni a_1\otimes\ldots\otimes a_n &\leftrightarrow& da_1\ldots da_n \in \Om^n\Ac\\
\Om_1\Bb(\Ac)^{\nat}\ni a_0\otimes a_1\otimes\ldots\otimes a_n &\leftrightarrow& a_0da_1\ldots da_n \in \Om^n\Ac\ .\non
\eeq
Under this identification, it is easy to show that the operator $N:\Om_1\Bb(\Ac)^{\nat}\to\Bb(\Ac)$ corresponds to Connes' boundary map $B$. Similarly, the operators $-b'$, $(1-\la)$ and $b$ on $\Xb(\Bb(\Ac))$ collectively correspond to the Hochschild operator $b$ on $\Om\Ac$. Therefore, the isomorphism (\ref{iso}) is in fact an isomorphism of total complexes.\\

We now deal with cyclic homology. The latter is the homology of some completion of the cyclic bicomplex. There are different kinds of cyclic homologies, depending on the completion chosen. Since we are mainly interested in Chern characters of $\te$-summable modules over Banach or Fr\'echet algebras, the suitable version of cyclic homology is the entire one \cite{C2}. The general formulation of entire cyclic homology for bornological algebras, including the bivariant theory, appears in \cite{Me}. We already adapted these concepts to Quillen's formalism of algebra cochains in a previous paper \cite{P3} and constructed a bivariant Chern character along these lines. We will follow the same steps here.\\
First consider the {\it entire bornology} on $\Om\Ac$, as the bornology generated by the sets \cite{Me}
\be
\bigcup_{n\ge 0}[n/2]!\, \St(dS)^n\subset \Om\Ac\ ,
\ee
for any bounded set $S$ in the bornology of $\Ac$. Here $[n/2]=k$ if $n=2k$ or $n=2k+1$, and $\St=\cc+ S$. Let $\Ome\Ac$ be the completion of $\Om\Ac$ with respect to the entire bornology. One shows that the operators $(b,B)$ of $\Om\Ac$ are bounded and extend to this completion \cite{Me}. By definition, $(\Ome\Ac,b,B)$ is the cyclic bicomplex of entire chains over the bornological algebra $\Ac$. Using the isomorphism (\ref{iso}), one has of course a corresponding formulation using the $X$-complex of the entire completion of the bar DG coalgebra \cite{P3}. Endow $\Bb(\Ac)$ with the bornology generated by the sets
\be
\bigcup_{n\ge 0}[n/2]!\, S^{\otimes n} \subset \Bb(\Ac)\ ,
\ee
for any bounded set $S$ in the bornology of $\Ac$. The {\it entire bar construction} $\Bbe(\Ac)$ is the completion of $\Bb(\Ac)$ with respect to this bornology. The operators $\Delta$ and $b'$ are bounded, hence extend to $\Bbe(\Ac)$ which becomes a DG coalgebra. The associated bicomodule $\Om_1\Bbe(\Ac)$ is given by
\be
\Om_1\Bbe(\Ac)=\Bbe(\Ac)\hotimes\Ac\hotimes\Bbe(\Ac)\ ,
\ee
and the whole construction of the $X$-complex of the entire bar DG coalgebra goes through with the obvious modifications. In particular we get an isomorphism of total $\zz_2$-graded complexes
\be
\Xb(\Bbe(\Ac))\cong \Ome\Ac\ .
\ee
\begin{definition}
Let $\Ac$ be a complete bornological algebra. The \emph{entire cyclic homology} of $\Ac$ is the homology of the $\zz_2$-graded complex of entire chains $\Ome\Ac$ endowed with the total differential $b+B$:
\be
HE_*(\Ac)=H_*(\Ome\Ac,b+B)\ ,\quad *=0,1\ .
\ee
Equivalently, it is equal to the homology of the reduced $X$-complex of the entire bar DG coalgebra $\Bbe(\Ac)$.
\end{definition}

Thus the (entire) cyclic homology is computed by the $X$-complex of a DG coalgebra. Quite remarkably, it is also equal to the homology of the $X$-complex of an {\it algebra}. This is the content of the Cuntz-Quillen formalism \cite{CQ1}, adapted to the entire framework by Meyer \cite{Me}. First, decompose the $\zz_2$-graded algebra of differential forms $\Om\Ac=\Om^+\Ac\oplus\Om^-\Ac$ into its even and odd parts. The even part $\Om^+\Ac$ is a trivialy graded subalgebra. We endow $\Om^+\Ac$ with a deformed associative product, the {\it Fedosov product} \cite{CQ1}
\be
\om_1\odot\om_2 =\om_1\om_2 -d\om_1d\om_2\ ,\quad \om_{1,2}\in\Om^+\Ac\ .
\ee
Associativity is easy to check. In fact the algebra $(\Om^+\Ac,\odot)$ is isomorphic to the non-unital tensor algebra $T\Ac=\bigoplus_{n\ge 1}\Ac^{\hotimes n}$, under the correspondence
\be
\Om^+\Ac\ni a_0da_1\ldots da_{2n} \longleftrightarrow a_0\otimes\om(a_1,a_2)\otimes\ldots \otimes\om(a_{2n-1},a_{2n}) \in T\Ac\ ,\label{cor}
\ee
where $\om(a_i,a_j):=a_ia_j- a_i\otimes a_j \in \Ac\oplus \Ac^{\hotimes 2}$ is the {\it curvature} of $(a_i,a_j)$. Now endow $\Om\Ac$ with the {\it analytic bornology} generated by the sets
\be
\bigcup_{n\ge 0}\, \St(dS)^n\subset \Om\Ac\ ,
\ee
for any bounded $S\subset\Ac$. The completion of $\Om\Ac$ for this bornology is denoted by $\Oman\Ac$, and splits into the direct sum $\Oman^+\Ac\oplus\Oman^-\Ac$. It turns out that the Fedosov product $\odot$ is bounded for the analytic bornology restricted to $\Om^+\Ac$ \cite{Me}, and thus extends to the analytic completion $\Oman^+\Ac$. The complete bornological algebra $(\Oman^+\Ac,\odot)$ is also denoted by $\Tc\Ac$ and called the {\it analytic tensor algebra} of $\Ac$.\\
In what follows, we will consider $\Tc\Ac$ as a trivially graded DG algebra, with differential simply equal to zero. The $\Tc\Ac$-bimodule of universal one-forms is given by
\be
\Om^1\Tc\Ac\cong \Tct\Ac\hotimes\Ac\hotimes\Tct\Ac\ ,\quad x\dd a\,y\leftrightarrow x\otimes a\otimes y
\ee
for any $a\in\Ac$ and $x,y\in\Tct\Ac$, where $\Tct\Ac$ is the unitalization of $\Tc\Ac$. The bimodule structure is the obvious one. This implies that the bornological space $\Om^1\Tc\Ac_{\nat}$ is isomorphic to $\Tct\Ac\hotimes\Ac$, which can further be identified with the analytic completion of odd forms $\Oman^-\Ac$, through the correspondence $x\otimes a\leftrightarrow xda$, $\forall a\in \Ac$, $x\in \Tct\Ac$. Thus collecting $\Tc\Ac$ and $\Om^1\Tc\Ac_{\nat}$ together yields a bornological vector space isomorphism
\be
X(\Tc\Ac)\cong \Oman\Ac\ .
\ee
Because $\Tc\Ac$ has no differential, the only boundary maps of its $X$-complex are given by $\nat\dd:\Tc\Ac\to\Om^1\Tc\Ac_{\nat}$ and $\bb:\Om^1\Tc\Ac_{\nat}\to \Tc\Ac$. We still denote by $(\nat \dd,\bb)$ the boundaries induced on $\Oman\Ac$ through the above isomorphism; Cuntz and Quillen have explicitly computed them in terms of the usual operators on differential forms \cite{CQ1}:
\beq
\bb&=&b-(1+\kappa)d\quad \mbox{on}\ \Om^{2n+1}\Ac\ ,\label{bou}\\
\nat \dd&=&\sum_{i=0}^{2n}\kappa^id -\sum_{i=0}^{n-1}\kappa^{2i}b\quad \mbox{on}\ \Om^{2n}\Ac\ .\non
\eeq
The crucial result \cite{CQ1,Me} is that the complex $(\Oman\Ac,\nat \dd\oplus\bb)=X(\Tc\Ac)$ is homotopy equivalent to the complex of entire chains $\Ome\Ac$ endowed with the differential $(b+B)$. Let $P_1:\Om\Ac\to\Om\Ac$ be the spectral projection for the eigenvalue 1 of the operator $\kappa^2$, and $P_1^{\bot}$ be the orthogonal projection. $P_1$ and $P_1^{\bot}$ commute with all operators commuting with $\kappa$, in particular $b$ and $B$. Then one shows that $P_1$ extends to the entire and analytic completions $\Ome\Ac$ and $\Oman\Ac$. The subcomplex $P_1^{\bot}\Ome\Ac$ endowed with the differential $b+B$ is contractible. The same result holds for $P_1^{\bot}\Oman\cong P_1^{\bot}X(\Tc\Ac)$ endowed with the differential $(\nat\dd\oplus\bb)$. Let $c:\Oman\Ac\to\Ome\Ac$ be the bornological vector space isomorphism induced by
\be
c(a_0da_1\ldots da_n)=(-)^{[n/2]}[n/2]!\, a_0da_1\ldots da_n\quad \forall n\in\nn\ .\label{res}
\ee
Then $c$ maps isomorphically $P_1\Oman\Ac$ onto $P_1\Ome\Ac$, and under this correspondence, the boundaries $\nat \dd\oplus\bb$ and $b+B$ coincide: $c^{-1}(b+B)c=\nat \dd\oplus\bb$ on $P_1\Oman\Ac$. It follows that the $X$-complex $X(\Tc\Ac)$ is homotopy equivalent to the $(b+B)$-complex of entire chains $\Ome\Ac$. We thus have two possibilities for computing the entire cyclic homology of an algebra. This leads to many equivalent definitions of the bivariant theory:
\begin{definition}
Let $\Ac$ and $\Bc$ be complete bornological algebras. Let $\hom(\Ome\Ac,\Ome\Bc)$ denote the space of \emph{bounded} linear maps between the entire cyclic bicomplexes $\Ome\Ac$ and $\Ome\Bc$. It is naturally a $\zz_2$-graded bornological complex, the differential of a map $f$ corresponding to the commutator $(b+B)\circ f-(-)^{|f|}f\circ(b+B)$. The bivariant entire cyclic cohomology of $\Ac$ and $\Bc$ is the cohomology of this complex:
\be
HE_*(\Ac,\Bc)=H_*(\hom(\Ome\Ac,\Ome\Bc))\ ,\qquad *=0,1\ .
\ee
Since $\Ome\Ac$ (resp. $\Ome\Bc$) is homotopy equivalent to $X(\Tc\Ac)$ (resp. $X(\Tc\Bc)$), the bivariant entire cyclic cohomology is also computed by any of the following homotopy equivalent $\zz_2$-graded complexes: $\hom(X(\Tc\Ac),X(\Tc\Bc))$, $\hom(X(\Tc\Ac),\Ome\Bc)$ and $\hom(\Ome\Ac,X(\Tc\Bc))$.
\end{definition}
The complex $\hom(\Ome\Ac,X(\Tc\Bc))$ is of special interest for us. Using the isomorphism $\Xb(\Bbe(\Ac))\cong\Ome\Ac$, we deduce that the bivariant entire cyclic cohomology is computed by
\be
HE_*(\Ac,\Bc)=H_*(\hom(\Xb(\Bbe(\Ac)),X(\Tc\Bc)))\ .
\ee
In the above formula, $\Bbe(\Ac)$ is a DG coalgebra and $\Tc\Ac$ a trivially graded algebra (without differential), thus we are in a position to obtain bivariant cocycles via the generalized Chern character of section \ref{sgen}. The following definition is intended to introduce an algebraic notion of unbounded Kasparov bimodules \cite{Bl}.
\begin{definition}
Let $\Ac$ and $\Bc$ be trivially graded complete bornological algebras. We have to consider the unitalization $\Bct$ because we work in a non-unital setting. Denote by $\Psi(\Ac,\Bct)$ the set of quintuples $\Ec=(\Lc,\ell^1,\tau,\rho,D)$ such that:
\begin{itemize}
\item $\Lc$ is a $\zz_2$-graded unital complete bornological algebra, viewed as an algebra of ``abstract pseudodifferential operators'';
\item $\ell^1\subset\Lc$ is a $\zz_2$-graded two-sided ideal endowed with its own complete bornology (``smoothing operators''), such that the inclusion $\ell^1\to \Lc$ is bounded, as well as the left and right multiplications by $\Lc$;
\item $\tau:\ell^1\to\cc$ is a bounded graded trace of homogeneous degree on $\ell^1$ viewed as a bimodule on $\Lc$, i.e. $\tau([\ell^1,\Lc])=0$;
\item $\rho:\Ac\to \Lc\hotimes\Bct$ is a bounded homomorphism carrying $\Ac$ to the even degree component of the algebra $\Lc\hotimes\Bct$;
\item $D\in \Lc\hotimes\Bct$ is an odd element (Dirac operator) provided with a heat kernel $H(t)=\exp(-tD^2)$, i.e. an element $H\in\Lc\hotimes\Bct\hotimes\cinf[0,1]$ verifying the heat equation
\be
\frac{d}{dt}H(t)=-D^2 H(t)\ ,\quad H(0)=1\in\Lc\hotimes\Bct\ ,
\ee
and $H(t)\in\ell^1\hotimes\Bct$ for any $t>0$ (trace-class condition).
\end{itemize}
$\Psi(\Ac,\Bct)$ splits into the set $\Psi_0(\Ac,\Bct)$ of quintuples for which the homogeneous trace $\tau$ has even degree, and the set $\Psi_1(\Ac,\Bct)$ of quintuples with odd trace. The direct sum of two elements $\Ec_1$, $\Ec_2$ of the same degree is obviously defined by $\Lc=\Lc_1\oplus \Lc_2$, $\tau=\tau_1\oplus\tau_2$, $\rho=\rho_1\oplus \rho_2$ and $D=D_1\oplus D_2$. This turns both $\Psi_i(\Ac,\Bct)$ into a semigroup. \\
If $\Ec_0=(\Lc,\ell^1,\tau,\rho_0,D_0)$ and $\Ec_1=(\Lc,\ell^1,\tau,\rho_1,D_1)$ differ only by $\rho_i$ and $D_i$, we say that they are smoothly homotopic iff there exists an interpolating element $\Ec=(\Lc[0,1],\ell^1[0,1],\tau\otimes\id,\rho,D)$, with $\Lc[0,1]=\Lc\hotimes\cinf[0,1]$, such that the evaluations $\ev_t\circ\rho$ and $\ev_tD$ coincide with $\rho_t$, $D_t$ for $t=0,1$.
\end{definition}
\begin{example}\textup{The above definition is motivated by the classical example of fibered manifolds. Let $X\stackrel{Z}{\longrightarrow}Y$ be a fibration of smooth compact manifolds. Let $\Ac=\cinf(X)$ and $\Bc=\cinf(Y)$ be the Fr\'echet algebras of smooth complex-valued functions over the total space and the base manifold respectively (since $Y$ is compact, $\Bc$ is already unital and we don't consider its unitalization). An element $(\Lc,\ell^1,\tau,\rho,D)\in\Psi(\Ac,\Bc)$ will roughly correspond to a family of pseudodifferential operators ($\Psi$DO) over $Y$ acting on sections of some vector bundle over the fiber $Z$. For notational simplicity, we shall assume that $Y$ is a $n$-torus $\TT_n$, and that pseudodifferential operators act on the scalar functions $\cinf(\TT_n)$. Let $d\in\rr$ and $\psi^d(\TT_n)$ denote the space of $\Psi$DOs of order $\le d$, that is, operators $P:\cinf(\TT_n)\to\cinf(\TT_n)$ defined by their symbols $p(x,\xi)$ in the variables $x\in\TT_n$ and $\xi\in\zz^n$ (dual lattice of the torus)
\be
(Pf)(x)=\sum_{\xi}e^{ix\cdot\xi}p(x,\xi)\widehat{f}(\xi)\quad \forall f\in\cinf(\TT_n)\ ,
\ee
where $\widehat{f}$ is the Fourier transform of $f$, and $p(x,\xi)$ verifies the estimates
\be
||p||^d_{\al}=\sup_{x,\xi}|(1+|\xi|)^{-d}\d_x^{\al}p(x,\xi)|<\infty\ ,
\ee
for any $\al\in\nn$. Then $\psi^d(\TT_n)$ endowed with the seminorms $||\cdot ||^d_{\al}$ is a complete locally convex space. We consider the bornology given by the collection of all bounded subsets. For $d'\ge d$, the inclusion $\psi^d(\TT_n)\subset \psi^{d'}(\TT_n)$ is bounded, and the inductive limit $\psi(\TT_n)=\cup_d \psi^d(\TT_n)$ endowed with the obvious bornology is a complete bornological space. Moreover, the product of $\Psi$DOs is bounded, and we let $\Lc=\psi(\TT_n)$ be this complete bornological algebra. The two-sided ideal of smoothing operators $\psi^{-\infty}(\TT_n)=\cap_d \psi^d(\TT_n)$ is a complete algebra for the locally convex topology given by the family of seminorms
\be
||p||^{-\infty}_{\al,\beta}=\sup_{x,\xi}|(1+|\xi|)^{\beta}\d_x^{\al}p(x,\xi)|
\ee
for any $\al\in\nn$ and $\beta\ge 0$. The corresponding complete bornological algebra is the ideal $\ell^1$. The natural inclusion $\ell^1\subset \Lc$ is bounded, as well as the left and right multiplications by $\Lc$ and the operator trace $\tau:\ell^1\to\cc$. Choosing a finite cover of $Y$ and a partition of unity, we may reduce to the case where the fibration $X\to Y$ is trivial (replace $\Lc$ by a matrix algebra over $\Lc$). Then the algebra $\Ac=\cinf(X)$ is represented by an homomorphism $\rho:\Ac\to\Lc\hotimes\Bc$, and a smooth family of elliptic $\Psi$DOs of order $>0$ corresponds to an element $D\in \Lc\hotimes\Bc$ with smoothing heat kernel $\exp(-tD^2)\in \ell^1\hotimes\Bc$. In realistic situations the scalar functions on the fiber $Z$ are replaced by some $\zz_2$-graded vector bundle and the algebra $\Lc$ of $\Psi$DOs is also graded.}
\end{example}

We now return to the general case and fix an element $\Ec=(\Lc,\ell^1,\tau,\rho,D)\in\Psi(\Ac,\Bct)$. Thus $\rho:\Ac\to\Lc\hotimes\Bct$ is a bounded homomorphism and $D\in\Lc\hotimes\Bct$ is an odd element. Let $\Cc=\Bbe(\Ac)$ be the entire bar DG coalgebra of $\Ac$, with its differential $-b'$. Let $\Rc=\Tc\Bc$ be the analytic tensor algebra of $\Bc$. The latter identifies with the analytic completion $\Oman^+\Bc$ of even forms endowed with the Fedosov product $\odot$. We regard it as a trivially graded algebra with differential equal to zero. Consider the $\zz_2$-graded convolution algebra $\hom(\Cc,\Lc\hotimes\Rct)$ endowed with the differential $\delta_1$ transposed of $-b'$:
\be
\delta_1 f=-(-)^{|f|}f\circ b'\ ,\quad \forall f\in\hom(\Cc,\Lc\hotimes\Rct)\ .
\ee
We shall denote also by $\odot$ the convolution product. Now remark that the projection $\Cc=\Bbe(\Ac)\to \Bb_1(\Ac)=\Ac$ is bounded, as well as the {\it linear} inclusion of bornological spaces $\Lc\hotimes\Bct\hookrightarrow \Lc\hotimes\Tct\Bc$. Consequently, the homomorphism $\rho$ may be considered as a linear map $\rho\in\hom(\Cc,\Lc\hotimes\Rct)$ via the composition
\be
\Cc\to \Ac\stackrel{\rho}{\longrightarrow} \Lc\hotimes\Bct\hookrightarrow \Lc\hotimes\Rct\ .
\ee
Similarly, $D\in\Lc\hotimes\Bct$ defines an element $D\in\hom(\Cc,\Lc\hotimes\Rct)$ by
\be
\Cc\stackrel{\eta}{\longrightarrow} \cc\stackrel{\otimes D}{\longrightarrow}\Lc\hotimes\Bct\hookrightarrow\Lc\hotimes\Rct\ .
\ee
Introduce the following Quillen superconnection
\be
A=\rho+D\ \in\hom(\Cc,\Lc\hotimes\Rct)_-\ ,
\ee
with curvature
\be
F=\delta_1A+A\odot A=\delta_1\rho +\rho^{\odot 2}+[D,\rho]_{\odot}+D^{\odot 2}\ \in \hom(\Cc,\Lc\hotimes\Rct)_+\ .
\ee
Recall that the products and commutators are taken in the convolution algebra $\hom(\Cc,\Lc\hotimes\Rct)$. As a map from $\Bbe(\Ac)$ to $\Lc\hotimes\Tct\Bc$, it is easy to see that $F$ vanishes on $\Bb_n(\Ac)$ for $n\ge 3$. Using the isomorphism $\Tc\Bc\cong(\Oman^+\Bc,\odot)$, we first evaluate $F$ on $\Bb_0(\Ac)=\cc$:
\be
F(1_{\cc})= D\odot D= D^2+dDdD\ \in \Lc\hotimes\Bct \oplus \Lc\hotimes\Om^2\Bc\ ,
\ee
where the product $D^2$ is now taken in $\Lc\hotimes\Bct$, and $dD$ is the image of $D$ under the derivation $d:\Lc\hotimes\Bct\to\Lc\hotimes\Om^1\Bc$. The sign $+$ in front of $dDdD$ comes from the fact that $D$ is odd and the Fedosov product equals $D^2-(-)^{|D|}dDdD$ in this case. In the same manner, we evaluate $F$ on $\Bb_1(\Ac)=\Ac$:
\beq
F(a)&=&([D,\rho]_{\odot})(a)\ =\ D\odot\rho(a)-\rho(a)\odot D\\
&=& [D,\rho(a)]+dDd\rho(a)+d\rho(a)dD\ \in\Lc\hotimes\Bct \oplus \Lc\hotimes\Om^2\Bc\ ,\non
\eeq
and on $\Bb_2(\Ac)=\Ac\hotimes\Ac$:
\beq
&&F(a_1,a_2) = (\delta_1\rho+\rho^{\odot 2})(a_1,a_2)= \rho\, b'(a_1,a_2)-\rho(a_1)\odot\rho(a_2)\\
&&= \rho(a_1a_2)-\rho(a_1)\rho(a_2)+d\rho(a_1)d\rho(a_2) = d\rho(a_1)d\rho(a_2)\ \in \Lc\hotimes\Om^2\Bc\ .\non
\eeq
Therefore, we may rewrite the curvature as
\be
F=D^2+[D,\rho]+d(\rho+D)d(\rho+D)\ .\label{curv}
\ee
In order to write down the bivariant Chern character of $\Ec$, we must take the exponential of $F$ in the convolution algebra $\hom(\Cc,\Lc\hotimes\Rct)$. It is shown in \cite{P3} how this Fedosov exponential $\exp_{\odot}(-F)$ viewed as a map $\Bbe(\Ac)\to \Omtan^+\Bc$ can be computed as a power series involving the heat kernel $\exp(-tD^2)\in \ell^1\hotimes\Bct$. In fact one has 
\be
\exp_{\odot}(-F)=\sum_{n\ge 0}\int_{\Delta_n}ds_1\ldots ds_n \, e^{-s_0F}dFde^{-s_1F}\ldots dFde^{-s_nF}\ ,
\ee
which may be rewritten as
\beq
\exp_{\odot}(-F)&=&\sum_{n\ge 0}(-)^n\int_{\Delta_{2n}}ds_1\ldots ds_n\, e^{-s_0F}dF e^{-s_1F} dF\ldots e^{-s_{2n-1}F}dF e^{-s_{2n}F}\non\\
&=& \exp(-F+\sqrt{-1}dF)|_+\ ,
\eeq
that is, the exponential for the ordinary (not Fedosov) product of differential forms $\Lc\hotimes\Omtan\Bc$, where we retain only the component in $\Lc\hotimes\Omtan^+\Bc$. We write $-F+\sqrt{-1}dF=-D^2+H$ with
\be
H=-[D,\rho]-d(\rho+D)d(\rho+D)+\sqrt{-1}d(D^2+[D,\rho])\ ,
\ee
and develop the latter exponential in power series of $H$:
\be
\exp_{\odot}(-F)=\sum_{n\ge 0} \int_{\Delta_n}ds_1\ldots ds_n\, e^{-s_0D^2}He^{-s_1D^2}H\ldots e^{-s_{n-1}D^2}He^{-s_nD^2}|_+\ .\label{exp}
\ee
Thus each term is a well-defined element of $\hom(\Bbe(\Ac),\ell^1\hotimes \Omtan^+\Bc)$, because the heat kernel is in $\ell^1\hotimes\Bct$. The requirement of convergence for the series leads to the following definition.
\begin{definition}\label{dte}
Let $\Ec=(\Lc,\ell^1,\tau,\rho,D)$ be an element of $\Psi(\Ac,\Bct)$, $A=\rho+D$ the superconnection in the convolution algebra $\hom(\Bbe(\Ac),\Lc\hotimes\Tct\Bc)$, and $F$ its curvature. Then $\Ec$ is called \emph{$\te$-summable} iff the exponential of $F$, taken in the convolution algebra, converges towards a bounded map
\be
\exp(-F)\ :\ \Bbe(\Ac)\to \ell^1\hotimes\Tct\Bc\ .
\ee
In other words, $\exp(-F)$ must be a ``trace-class'' bounded map. Two $\te$-summable quintuples $\Ec_0=(\Lc,\ell^1,\tau,\rho_0,D_0)$ and $\Ec_1=(\Lc,\ell^1,\tau,\rho_1,D_1)$ are (smoothly) homotopic iff there exists a $\te$-summable interpolating element $\Ec=(\Lc[0,1],\ell^1[0,1],\tau\otimes\id,\rho,D)$.
\end{definition}
If the element $\Ec\in\Psi(\Ac,\Bct)$ is $\te$-summable, we obtain a bivariant cocycle as explained in section \ref{sgen}. Recall that the $X$-complex of the DG coalgebra $\Cc=\Bbe(\Ac)$ is isomorphic to the cyclic bicomplex
\be
X(\Cc): \Cc\ \xymatrix@1{\ar@<0.5ex>[r]^{\beta=1-\la} &  \ar@<0.5ex>[l]^{\d\nat=N}}\ \Om_1\Cc^{\nat}\ ,
\ee
where $\Cc$ and $\Om_1\Cc^{\nat}$ are themselves complexes for the differentials $-b'$ and $b$. On the other hand, the (augmented) $X$-complex of the trivially graded algebra $\Rc=\Tc\Bc$ is given by
\be
\Xt(\Rc):\ \Rct\ \xymatrix@1{\ar@<0.5ex>[r]^{\nat\dd} &  \ar@<0.5ex>[l]^{\bb}}\ \Om^1\Rc_{\nat}\ .
\ee
Consider the derivations on the convolution algebra
\beq
\dd_1 &:& \hom(\Cc,\Lc\hotimes\Rct)\to \hom(\Om_1\Cc,\Lc\hotimes\Rct) \ ,\quad \dd_1 f=(-)^{|f|}f\circ \d\ ,\non\\
\dd_2 &:& \hom(\Cc,\Lc\hotimes\Rct)\to \hom(\Cc,\Lc\hotimes\Om^1\Rc) \ ,\quad \dd_2 f=\dd\circ f\ ,
\eeq
and the partial traces
\beq
\nat_1 &:& \hom(\Om_1\Cc,\Lc\hotimes\Rct)\to \hom(\Om_1\Cc^{\nat},\Lc\hotimes\Rct)\ ,\non\\
\nat_2 &:& \hom(\Cc,\Lc\hotimes\Om^1\Rc)\to \hom(\Cc,\Lc\hotimes\Om^1\Rc_{\nat})\ ,\\
\nat_{12} &:& \hom(\Om_1\Cc,\Lc\hotimes\Om^1\Rc)\to \hom(\Om_1\Cc^{\nat},\Lc\hotimes\Om^1\Rc_{\nat})\ ,\non
\eeq
Then we have
\beq
\dd_1A\ ,\ \dd_1F &\in& \hom(\Om_1\Cc,\Lc\hotimes\Rct)\ ,\non\\
\dd_2A\ ,\ \dd_2F &\in& \hom(\Cc,\Lc\hotimes\Om^1\Rc)\ ,\\
e^{-F} &\in& \hom(\Cc,\ell^1\hotimes\Rct)\ ,\non
\eeq
because $\Ec$ is $\te$-summable. As in section \ref{sgen}, we introduce the components of $\ch(A)$:
\beq
\hom(\Cc,\ell^1\hotimes\Rct) &\ni& \ch_0^0(A) =e^{-F}\ ,\non\\ 
\hom(\Om_1\Cc^{\nat},\ell^1\hotimes\Rct) &\ni& \ch_1^0(A) =-\nat_1 \dd_1Ae^{-F}\ ,\non\\
\hom(\Cc,\ell^1\hotimes\Om^1\Rc_{\nat}) &\ni& \ch_0^1(A) = \nat_2 e^{-F}\dd_2A\ ,\\
\hom(\Om_1\Cc^{\nat},\ell^1\hotimes\Om^1\Rc_{\nat}) &\ni& \ch_1^1(A)= -\nat_{12} \dd_1A e^{-F}\dd_2A - \non\\
&-&\int_{\Delta_2}ds_1ds_2\,\nat_{12} e^{-s_0F}\dd_2Fe^{-s_1F}\dd_1Fe^{-s_2F}\ .\non
\eeq
We used the fact that $\ell^1$ is a two-sided ideal in $\Lc$. Hence applying the trace $\tau:\ell^1\to\cc$, we get a cocycle (remark \ref{raux})
\be
\tau\ch(A)\in \hom(X(\Cc),\Xt(\Rc))\ .
\ee
Let $|\tau|=0$ or $1$ be the degree of the homogeneous graded trace $\tau$. This allows to write the cocycle condition explicitly in terms of the components of $\ch(A)$, together with the boundary maps $-b'$, $b$, $N$, $(1-\la)$ on the cyclic bicomplex $X(\Cc)$ and $(\nat\dd\oplus\bb)$ on $X(\Rc)$:
\beq
\tau\ch_0^0(A)\circ (-b')+\tau \ch_1^0(A)\circ(1-\la) &=& (-)^{|\tau|}\bb \circ \tau\ch_0^1(A)\ ,\non\\
\tau\ch_1^0(A)\circ b+\tau \ch_0^0(A)\circ N &=& (-)^{|\tau|}\bb \circ \tau\ch_1^1(A)\ ,\\
\tau\ch_0^1(A)\circ (-b')+\tau \ch_1^1(A)\circ(1-\la) &=& (-)^{|\tau|}\nat\dd \circ \tau\ch_0^0(A)\ ,\non\\
\tau\ch_1^1(A)\circ b+\tau \ch_0^1(A)\circ N &=& (-)^{|\tau|}\nat\dd \circ \tau\ch_1^0(A)\ .\non\\
\eeq
This shows that the cocycle $\tau\ch(A)$ has parity $|\tau|$. Equivalently, we may use the identification of the reduced $X$-complex $\Xb(\Cc)$ with the $(b+B)$-complex of entire forms $\Ome\Ac$:
\beq
&&\tau\ch(A) \in \hom(\Ome\Ac,\Xt(\Tc\Bc))\ ,\\
&&\tau\ch(A)\circ (b+B)-(-)^{|\tau|}(\nat\dd\oplus\bb)\circ \tau\ch(A)=0\ .\non
\eeq
Let us evaluate $\tau\ch(A)$ on a $n$-form over $\Ac$. $\ch_0^0(A)$ and $\ch_0^1(A)$ are defined on $\Bb_n(\Ac)$, which corresponds to the space of boundaries $d\Om^{n-1}\Ac$. On the other hand $\ch_1^0(A)$ and $\ch_1^1(A)$ are defined on $(\Om_1\Bb(\Ac)^{\nat})_n$, or equivalently $\Ac d\Om^{n-1}\Ac$. Thus for any differential form $da_1\ldots da_n\in\Om^n\Ac$, one has
\be
\tau\ch_0^0(da_1\ldots da_n)=\tau e^{-F}(a_1\otimes\ldots\otimes a_n)\ \in\Tct\Bc\ ,
\ee
where the string $a_1\otimes\ldots\otimes a_n$ is an element of $\Bb_n(\Ac)\subset\Cc$, whose evaluation on the cochain $\exp(-F)\in \hom(\Cc,\ell^1\hotimes\Rct)$ is given by formula (\ref{exp}). Similarly
\beq
\lefteqn{\tau\ch_0^1(A)(da_1\ldots da_n) = \tau\nat (e^{-F}\dd_2A) (a_1\otimes\ldots\otimes a_n)}\non\\
&=& \tau\nat e^{-F}(a_1\otimes\ldots\otimes a_{n-1})\dd\rho(a_n)+ \tau\nat e^{-F}(a_1\otimes\ldots\otimes a_n)\dd D
\eeq
in $\Om^1\Tc\Bc_{\nat}$. Furthermore, for any $n$-form $a_0da_1\ldots da_n\in\Om^n\Ac$, one has
\be
\tau\ch_1^0(A)(a_0da_1\ldots da_n)= -\tau\dd_1A e^{-F}\nat(a_0\otimes a_1\otimes\ldots\otimes a_n)\ ,
\ee
where $a_0\otimes a_1\otimes\ldots\otimes a_n$ is an element of $\Om_1\Cc^{\nat}$. Hence
\beq
\tau\ch_1^0(A)(a_0da_1\ldots da_n)&=& -\tau\, \big(\dd_1A(a_0)\big)\big(e^{-F}(a_1\otimes\ldots \otimes a_n)\big)\\
&=& \tau \rho(a_0)\, e^{-F}(a_1\otimes\ldots \otimes a_n)\ \in \Tct\Bc\ .\non
\eeq
The last component $\tau\ch_1^1(A)$ is a sum of two terms. The first one evaluated on $a_0da_1\ldots da_n\in\Om^n\Ac$ yields
\beq
\lefteqn{-\tau\nat\dd_1A\, e^{-F}\dd_2A\nat (a_0\otimes a_1\otimes\ldots\otimes a_n)=}\non\\
&=& -\tau\nat \big(\dd_1A(a_0)\big)\big((e^{-F}\dd_2\rho) \,(a_1\otimes\ldots \otimes a_n)\big)\\
&=& \tau\nat \rho(a_0)e^{-F}(a_1\otimes\ldots\otimes a_{n-1})\dd\rho(a_n)+ \tau\nat \rho(a_0)e^{-F}(a_1\otimes\ldots\otimes a_n)\dd D\ ,\non
\eeq
in $\Om^1\Tc\Bc_{\nat}$. The evaluation of the last term 
\be
-\int_{\Delta_2}ds_1ds_2\,\nat_{12} e^{-s_0F}\dd_2Fe^{-s_1F}\dd_1Fe^{-s_2F}
\ee
is slightly more lengthy, but this can be done without difficulty as above. We let it as an exercise.\\

For the applications of the bivariant Chern character, in particular when dealing with the composition product, it is better to work within the $\zz_2$-graded complex $\hom(\Ome\Ac,\Ome\Bc)$. Thus we first project $\Xt(\Tc\Bc)$ onto $X(\Tc\Bc)$ by dropping the direct factor $X(\cc):\cc\rlarrows 0$, and then use the Cuntz-Quillen homotopy equivalence \cite{CQ1,Cu2,Me}
\be
P_1\circ c : X(\Tc\Bc)\cong\Oman\Bc \to \Ome\Bc\ ,
\ee
where $P_1$ is the spectral projection for the eigenvalue 1 of the squared Karoubi operator $\kappa^2$, and $c$ is the rescaling (\ref{res}). Collecting all the preceding results together, one has the following
\begin{theorem}\label{t1}
Let $\Ac$ and $\Bc$ be complete bornological algebras, and let $\Ec=(\Lc,\ell^1,\tau,\rho,D)$ be a $\te$-summable element of $\Psi_i(\Ac,\Bct)$, $i=0,1$. Let $A=\rho+D$ be the Quillen superconnection associated to $\Ec$, viewed as an element of the DG convolution algebra $\hom(\Bbe(\Ac),\Lc\hotimes\Tct\Bc)$. Let $P_1 c: X(\Tc\Bc)\to \Ome\Bc$ be the homotopy equivalence. Then the map
\be
\ch(\Ec)=P_1c\, \tau\ch(A)\in \hom(\Ome\Ac,\Ome\Bc)
\ee
is an entire bivariant cyclic cocycle of parity $i$:
\be
(b+B)\circ \ch(\Ec)-(-)^i\ch(\Ec)\circ (b+B)=0\ .
\ee
Its cohomology class in $HE_i(\Ac,\Bc)$ is the bivariant Chern character of $\Ec$. The latter is additive for the semigroup law of $\Psi_i(\Ac,\Bct)$ and invariant with respect to smooth homotopies of $\rho$ and $D$ among $\te$-summable elements.
\end{theorem}
{\it Proof:} We only have to prove the homotopy invariance. If $\Ec=(\Lc[0,1],\ell^1[0,1],\tau\otimes\id,\rho,D)$ interpolates between two elements $\Ec_0$ and $\Ec_1$, we consider the auxiliary DG algebra tensor product $\Lc\hotimes\Om[0,1]$, endowed with the de Rham boundary of differential forms over $[0,1]$. Then as in the proof of proposition \ref{phomo} (see also remark \ref{raux}), the Chern character of $\Ec$ yields a cocycle in the complex $\hom(\Ome\Ac,\Ome\Bc\hotimes\Om[0,1])$, where the differential on $\Om[0,1]$ is taken into account. By projecting the cocycle condition over $\Om^1[0,1]$, one gets a transgression formula involving the Chern-Simons terms $cs$ as in proposition \ref{phomo}, and the homotopy invariance follows. \cqfd\\
\begin{remark}\label{rper}\textup{
The projection $\Ome\Bc\to\Om^n\Bc$ being bounded, there is a natural map from $HE_*(\Bc)$ to the periodic cyclic homology $HP_*(\Bc)$, i.e. the homology of the direct product $\Omh\Bc=\prod_{n\ge 0}\Om^n\Bc$ endowed with the boundary $b+B$. Hence we may forget the complex $\hom(\Ome\Ac,\Ome\Bc)$ and view the bivariant Chern character of $\Ec$ simply as a map
\be
\ch(\Ec): HE_*(\Ac)\to HP_{*+|\tau|}(\Bc)\ .
\ee
This allows to weaken somewhat the $\te$-summability condition \ref{dte}. This will also be the correct setting for the characteristic maps of Hopf algebras.}
\end{remark}
\begin{remark}\textup{
Following section \ref{sgen}, the process generalizes to an arbitrary number of algebras appearing in covariant or contravariant position. For example if $\Ac$ is covariant and $\Bc_1,\Bc_2$ are two contravariant algebras, we may work in the DG convolution algebra $\hom(\Bbe(\Ac),\Tct\Bc_1\hotimes\Tct\Bc_2)$. Then any odd connection $A$ in the convolution algebra such that $\exp(-F)$ converges yields a map
\be
\ch(A)\ :\  HE_i(\Ac)\to \bigoplus_{j+k=i}HE_j(\Bc_1)\otimes HE_k(\Bc_2)\ ,\quad i\in\zz_2\ .\label{cup}
\ee
In particular if $\Ac=\Bc_1\hotimes\Bc_2$ and if the connection $A$ is the composition
\be
\begin{CD}
\Bbe(\Ac) @>{\textup{\small proj}}>> \Ac = \Bc_1\hotimes\Bc_2\hookrightarrow \Tct\Bc_1\hotimes\Tct\Bc_2\ ,
\end{CD}
\ee 
then (\ref{cup}) is a coproduct in entire cyclic homology \cite{L}. This works also in periodic cyclic homology. As another related example, if we take two algebras $\Ac_1$, $\Ac_2$ in contravariant position, and a third algebra $\Bc$ in a covariant position, we may work in the DG convolution algebra $\hom(\Bbe(\Ac_1)\hotimes\Bbe(\Ac_2),\Tct\Bc)$. Then any odd connection $A$ in the convolution algebra such that $\exp(-F)$ converges gives rise to a bilinear pairing
\be
\ch(A)\ :\ HE_i(\Ac_1)\times HE_j(\Ac_2)\to HE_{i+j}(\Bc)\ ,\quad i,j\in\zz_2\ .\label{shuf}
\ee
Again, if $\Ac_1$ and $\Ac_2$ are \emph{unital} and $\Bc=\Ac_1\hotimes\Ac_2$, we can choose $A=\iota_1+\iota_2$ where $\iota_1$ is the linear map 
\be
\begin{CD}
\Bbe(\Ac_1)\hotimes\Bbe(\Ac_2) @>{\textup{\small proj}}>> \Ac_1 @>{\id\otimes 1}>> \Bc\hookrightarrow \Tct\Bc\ ,
\end{CD}
\ee
and similarly for $\iota_2$. Then (\ref{shuf}) is a shuffle product \cite{L}, working also in periodic theory. The large flexibility of our construction allows, in principle, to extend these kind of pairing to very general situations just by choosing an appropriate connection $A$.}
\end{remark}
\begin{remark}\textup{
Everything extends also to the cyclic homology of DG algebras (see \cite{G, L}). If $(\Ac,\delta)$ is a differential graded algebra, its bar coalgebra $\Bb(\Ac)$ is naturally endowed with an induced differential $\delta$ anticommuting with the usual acyclic operator $b'$. Then the entire cyclic homology of $(\Ac,\delta)$ is computed by the (reduced) $X$-complex of the DG coalgebra $(\Bbe(\Ac),-b'+\delta)$. Equivalently, the analytic tensor algebra $\Tc\Ac$ becomes a DG algebra with the induced action of $\delta$, and the total complex $X(\Tc\Ac)$ computes the entire cyclic homology of $(\Ac,\delta)$. Again, the connection and curvature method goes over to this situation just by taking into account this extra differential, and provides characteristic maps between the cyclic homologies of DG algebras.}
\end{remark}

\section{Hopf algebras}\label{shopf}

We shall present an illustration of the generalized Chern character in the context of Hopf algebras. Our aim is to show that the universal cocycles constructed in section \ref{sgen} yield characteristic maps with values in the cyclic homology of Hopf algebras in the sense of Connes-Moscovici \cite{CM98,CM99,CM00}. Let $\Hc$ be a Hopf algebra provided with a character $\delta:\Hc\to\cc$ such that the twisted antipode $\Sd=\delta*S$ is involutive. We adopt Crainic's interpretation \cite{Cr} of the Hopf algebra cyclic bicomplex as a quotient of the cyclic bicomplex of the {\it coalgebra} $\Hc$ by the subspace spanned by $\delta$-coinvariant cochains (see below). By duality, the cyclic bicomplex of multilinear maps from $\Hc$ to $\cc$ restricted to the subspace of $\delta$-invariants is still a cyclic bicomplex, computing the periodic cyclic homology of $\Hc$ denoted $HP^{\delta}_*(\Hc)$. \\
Given a complete bornological algebra $\Ac$, we introduce the set of $\Hc$-equivariant ``$K$-cycles'' $\Psi^{\Hc}(\Ac,\cc)$. It corresponds to quintuples $\Ec=(\Lc,\ell^1,\tau,\rho,D)$, where $\Lc$ is an auxiliary $\zz_2$-graded $\Hc$-algebra, $\rho:\Ac\to\Lc$ an homomorphism and $D\in\Lc$ a Dirac operator with heat kernel. The ideal $\ell^1\subset\Lc$ is stable by the action of $\Hc$ and $\tau:\ell^1\to\cc$ is a $\delta$-invariant graded trace of homogeneous degree, that is, $\tau(h(x))=\delta(h)\tau(x)$ for any $h\in\Hc$, $x\in\ell^1$. Using a suitable Quillen superconnection, we show that a formula analogous to the bivariant Chern character of section \ref{sbiv} gives rise to an homotopy invariant characteristic map
\be
\ch(\Ec): HE_*(\Ac)\to HP^{\delta}_{*+|\tau|}(\Hc)
\ee
under some $\te$-summability hypotheses on $\Ec$. This construction contains the Connes-Moscovici characteristic map \cite{CM98} as a very special case (it corresponds to the purely algebraic situation $D=0$). \\

Let $\Hc$ be a (trivially graded) Hopf algebra. We shall consider that $\Hc$ is endowed with a complete bornology. In some situations however there is no relevant topology or bornology and all constructions involving $\Hc$ are performed purely algebraically. This is the case for example when $\Hc$ is a discrete algebra acting by unbounded operators on a Hilbert space (section \ref{ssec}). In this case we endow $\Hc$ with the {\it fine bornology} \cite{Me}, that is, the bornology generated by the elements of $\Hc$. Thus a set is small iff it is contained in the convex hull of a finite number of points in $\Hc$. This bornology is a always complete, any linear map on $\Hc$ is automatically bounded, and the completed tensor products coincide with the algebraic tensor products. Hence there is no restriction by considering $\Hc$ endowed with a complete bornology.\\
Denote by $1\in\Hc$ the unit of $\Hc$ and $\eps:\Hc\to\cc$ its counit, related by the condition $\eps(1)=1$. The coproduct $\Delta:\Hc\to \Hc\hotimes\Hc$ (with completed tensor product) is written according to Sweedler's notation \cite{Sw}
\be
\Delta h=\sum h_{(0)}\otimes h_{(1)}\ ,\quad\forall h\in\Hc\ ,
\ee
so that coassociativity $(\Delta\otimes \id)\circ\Delta = (\id\otimes\Delta)\circ \Delta$ reads
\beq
\sum h_{(0)(0)}\otimes h_{(0)(1)} \otimes h_{(1)} &=& \sum h_{(0)}\otimes h_{(1)(0)} \otimes h_{(1)(1)}\non\\
&:=& \sum h_{(0)}\otimes h_{(1)} \otimes h_{(2)}\ .
\eeq
More generaly, applying the coproduct $n$ times we write
\be
\Delta^n h = \sum h_{(0)}\otimes h_{(1)}\otimes\ldots \otimes h_{(n)}\ ,\quad \forall h\in\Hc\ .
\ee
Recall the compatibility relations with the counit
\be
\sum \eps(h_{(0)})h_{(1)}=\sum h_{(0)}\eps(h_{(1)})=h\ ,
\ee
and with the product
\be
\Delta(h^1h^2)=\Delta(h^1)\Delta(h^2)\quad\forall h^1,h^2\in\Hc\ ,\quad \Delta 1= 1\otimes 1\ .
\ee
Finally, let $S:\Hc\to\Hc$ be the antipode. By definition it is a bounded linear map verifying the relation
\be
\sum S(h_{(0)})h_{(1)}= \sum h_{(0)} S(h_{(1)})=\eps(h)1\ ,\quad\forall h\in\Hc\ .
\ee
We also recall the basic properties of $S$:
\beq
&& S(h^1h^2)=S(h^2)S(h^1)\ ,\quad \Delta S(h)=\sum S(h_{(1)})\otimes S(h_{(0)})\ ,\non\\
&& \eps(S(h))=\eps(h)\ ,\quad S(1)=1\ .
\eeq
Now let $\Hcb=\ker \eps$ be the augmentation ideal. Since $\eps(1)=1$, the Hopf algebra splits linearly: $\Hc=\cc\oplus \Hcb$, with $\cc$ the line generated by the unit. As in section \ref{sgen} for coalgebras, it follows that the dual space $\hom(\Hc,\cc)$ is canonically an augmented associative algebra: let $\Bc$ be the space of bounded linear maps $\Hc\to\cc$ vanishing on $1\in\Hc$. Then it is an associative algebra for the product
\be
(b_1b_2)(h):= \sum b_1(h_{(0)})b_2(h_{(1)})\ ,\quad\forall b_i\in\Bc\ ,\ h\in\Hc\ .
\ee
The linear splitting of $\Hc$ allows to identify $\Bc$ with $\hom(\Hcb,\cc)$ and the augmented algebra $\Bct=\cc\oplus\Bc$ with $\hom(\Hc,\cc)$. We first define the cyclic cohomology of $\Hc$ {\it viewed as a coalgebra only} (with linear splitting), by dualizing formally the construction of the cyclic homology of the algebra $\Bc$. By {\it formally}, we mean for example that an algebraic tensor product like $\Bc^{\otimes n}$ is replaced by the space of linear maps $\hom(\Hcb^{\hotimes n},\cc)$, and all operators acting on $\Bc^{\otimes n}$ are transposed to $\Hcb^{\hotimes n}$ according to formal manipulations. For any $n\ge 0$, we introduce the space of $n$-cochains $C^n(\Hc)$ and its {\it normalized subspace} $\Cb^n(\Hc)$
\be
C^n(\Hc)= \Hc^{\hotimes (n+1)}\ ,\quad \Cb^n(\Hc)= \Hc\hotimes \Hcb^{\hotimes n}\ .
\ee
This leads to the following dictionnary:
\be
\left. \begin{array}{c}
\Bc \\
\Bct \\
\Bct^{\otimes n} \\
\Om^n\Bc \end{array} \right\} \leftrightarrow
\left\{ \begin{array}{l}
\hom(\Hcb,\cc) \\
\hom(\Hc,\cc) \\
\hom(C^n(\Hc),\cc)\\
\hom(\Cb^n(\Hc),\cc) \end{array} \right.
\ee
where $\Om^n\Bc= \Bct\otimes\Bc^{\otimes n}$ is the space of $n$-forms over $\Bc$. Thus, the operators $d$, $b$, $\kappa$, $B$ defined on $\Om^n\Bc$ can be transposed into operators acting on the space of normalized cochains $\Cb^n(\Hc)$. Let us start with the Hochschild operator $b$. Its action on the subspace $\Om^n\Bc\subset \Bct^{\otimes (n+1)}$ is the restriction of the following map $b: \Bct^{\otimes (n+1)}\to \Bct^{\otimes n}$, $b^2=0$:
\beq
\lefteqn{b(\xt_0\otimes \ldots\otimes \xt_n) = \xt_0\xt_1\otimes \xt_2\otimes \ldots \otimes\xt_n - \xt_0\otimes \xt_1\xt_2\otimes\ldots\otimes\xt_n+\ldots} \non\\
&\ldots&+(-)^{n-1}\xt_0\otimes\ldots\otimes \xt_{n-1}\xt_n +(-)^n\xt_n\xt_0\otimes\ldots\otimes \xt_{n-1}\ ,
\eeq
for any $\xt_i\in\Bct$. Hence at the dual level, we may use the coalgebra structure of $\Hc$ to get the transposed operator $b: C^{n}(\Hc)\to C^{n+1}(\Hc)$, $b^2=0$:
\beq
\lefteqn{b(h^0\otimes \ldots\otimes h^n)=\Delta h^0\otimes h^1\otimes\ldots \otimes h^n- h^0\otimes\Delta h^1\otimes \ldots\otimes h^n+\ldots}\\
&\ldots&+(-)^nh^0\otimes\ldots\otimes \Delta h^n+(-)^{n+1}\sum h^0_{(1)}\otimes h^1\ldots h^n\otimes h^0_{(0)}\ ,\non
\eeq
for any $h^i\in\Hc$. The decomposition $\Hc=\cc\oplus\Hcb$ implies that $C^n(\Hc)$ splits into the direct sum of $\Cb^n(\Hc)$ and its orthogonal $\Cb^n_{\bot}(\Hc)$ spanned by elements $h^0\otimes h^1\ldots\otimes h^n$ such that at least one of the $h^i$'s, $i\ge 1$, is proportional to $1\in\Hc$. Hence $\Cb^n(\Hc)$ is canonically isomorphic, as a vector space, to the quotient $C^n/\Cb^n_{\bot}$. Since $\Delta 1=1\otimes 1$, it follows that $\Cb^n_{\bot}(\Hc)$ is stable by $b$, whence an induced boundary map $b:\Cb^n(\Hc)\to\Cb^{n+1}(\Hc)$ with $b^2=0$.\\
Next, the differential on non-commutative forms $d:\Om^n\Bc\to\Om^{n+1}\Bc$ transposes to $d:\Cb^n(\Hc)\to\Cb^{n-1}(\Hc)$ simply by
\be
d(h^0\otimes \hb^1\otimes\ldots\otimes\hb^n)= \eps(h^0)\hb^1\otimes\ldots\otimes\hb^n\ ,\quad d^2=0\ ,
\ee
for any $h^0\in\Hc$ and $\hb^i\in\Hcb$. As usual, we introduce the Karoubi cyclic operator $\kappa:\Cb^n(\Hc)\to \Cb^n(\Hc)$ and Connes boundary map $B:\Cb^n(\Hc)\to\Cb^{n-1}(\Hc)$, which are the transposed of the usual operators on $\Om\Bc$:
\be
\kappa=1-(bd+db)\ ,\quad B=d\circ(1+\kappa+\ldots+\kappa^{n-1}) \ \mbox{on}\ \Cb^n(\Hc)\ ,
\ee
with the relations $b^2=B^2=Bb+bB=0$, $\kappa B=B=B\kappa$. 
\begin{definition}
Let $\Hc$ be a Hopf algebra. The periodic cyclic cohomology of $\Hc$ viewed only as a coalgebra is computed by the \emph{direct sum} complex
\be
\Cb^*(\Hc)=\bigoplus_{n\ge 0} \Cb^n(\Hc)
\ee
endowed with the total differential $(b+B)$. At the dual level, the periodic cyclic homology of the coalgebra $\Hc$ is computed by the \emph{direct product} complex
\be
\Cb_*(\Hc)=\prod_{n\ge 0} \hom(\Cb^n(\Hc),\cc)
\ee
with differential the transposed of $(b+B)$.
\end{definition}
So far we have only exploited the coalgebra structure of $\Hc$ and its unit. We now take into account the product and antipode in order to get the cyclic cohomology of $\Hc$ viewed as a Hopf algebra. First, it is necessary to fix another datum, namely a character (bounded algebra homomorphism) $\delta:\Hc\to\cc$,
\be
\delta(h^1h^2)=\delta(h^1)\delta(h^2)\ ,\quad \forall h^1,h^2\in\Hc\ .
\ee
Then define the twisted antipode $\Sd:\Hc\to\Hc$ by convoluting the ordinary antipode with the character:
\be
\Sd(h)=\sum \delta(h_{(0)})h_{(1)}\ ,\quad \forall h\in\Hc\ .
\ee
Direct computations show that the twisted antipode fulfill the following relations:
\beq
&& \Sd(h^1h^2) = \Sd(h^2)\Sd(h^1)\ ,\quad \Delta \Sd(h)=\sum S(h_{(1)})\otimes \Sd(h_{(0)})\ ,\label{rela}\\
&& \sum\Sd(h_{(0)})h_{(1)} = \delta(h)1\ ,\quad \eps\circ\Sd=\delta\ ,\quad \delta\circ\Sd=\eps\ ,\quad \Sd(1)=1\ .\non
\eeq
The following important lemma is due to Crainic \cite{Cr}:
\begin{lemma}\label{lequi}
For the twisted antipode $\Sd:\Hc\to\Hc$, the following two conditions are equivalent:
$$\Sd^2=\id_{\Hc}\ ;$$
$$\sum \Sd(h_{(1)})h_{(0)}=\delta(h)1\quad\forall h\in\Hc\ .$$
\end{lemma}
{\it Proof:} See \cite{Cr}.\cqfd\\

Let us now introduce the space of (co)invariants. The space of normalized cochains $\Cb^n(\Hc)$ is naturally a left $\Hc$-module for the diagonal action:
\be
h\cdot(h^0\otimes \hb^1\otimes\ldots\otimes\hb^n)= \sum h_{(0)}h^0\otimes h_{(1)}\hb^1\otimes\ldots\otimes h_{(n)}\hb^n\ ,
\ee
for any $h\in\Hc$ and $h^0\otimes \hb^1\otimes\ldots\otimes\hb^n\in\Cb^n(\Hc)$. This action is well-defined because $\Hcb=\ker\eps$ is an ideal in $\Hc$. We denote by $\Cb^n_{\delta}(\Hc)$ the quotient of $\Cb^n(\Hc)$ by the subspace of $\delta$-coinvariants (i.e. the subspace spanned by the elements $h\cdot x-\delta(h)x$, for any $x\in\Cb^n(\Hc)$, $h\in\Hc$): 
\be
\Cb^n_{\delta}(\Hc)=\Cb^n(\Hc)/\delta\textup{-coinv}\ .
\ee
By duality, the space of linear maps $\Cb_n(\Hc)=\hom(\Cb^n(\Hc),\cc)$ is a right $\Hc$-module:
\be
(f\cdot h)(h^0\otimes \hb^1\otimes\ldots\otimes\hb^n)= f\big(h\cdot(h^0\otimes \hb^1\otimes\ldots\otimes\hb^n)\big)\ ,
\ee
for any $f\in\Cb_n(\Hc)$ and $h\in\Hc$. We denote by $\Cb_n^{\delta}(\Hc)$ the subspace of $\delta$-invariant chains:
\be
f\in \Cb_n^{\delta}(\Hc)\ \Leftrightarrow\ f\cdot h=\delta(h)f\quad \forall h\in\Hc\ .
\ee
The following proposition is the fundamental result leading to the cyclic cohomology of Hopf algebras \cite{CM99,Cr}. Under the present form it is adapted to the normalized framework:
\begin{proposition}
Let $\Hc$ be a Hopf algebra, $\delta:\Hc\to\cc$ a character and $\Sd$ the twisted antipode. If $\Sd^2=\id$, then the operators $d,b,\kappa,B$ defined on the complex of cochains $\Cb^*(\Hc)$ pass to the quotient by coinvariants $\Cb^*_{\delta}(\Hc)$. Thus $\Cb^*_{\delta}(\Hc)$ endowed with the boundaries $b$ and $B$ is a mixed complex. By duality, the subspace of invariant chains $\Cb_*^{\delta}(\Hc)$ is stable by the transposed operators $d,b,\kappa,B$, whence a mixed complex. 
\end{proposition}
{\it Proof:} The Hopf algebra $\Hc$ acts on the non-normalized space of cochains $C^n(\Hc)$ by the diagonal action. We first show that the subspace of coinvariants 
$$
\textup{coinv}(C^n)=\{h\cdot x-\delta(h)x\ |\ h\in\Hc, x\in C^n(\Hc)\}
$$
is stable by the Hochschild operator $b$:
\beq
\lefteqn{b(h^0\otimes \ldots\otimes h^n):=\Delta h^0\otimes h^1\otimes\ldots \otimes h^n- h^0\otimes\Delta h^1\otimes \ldots\otimes h^n+\ldots}\non\\
&\ldots&+(-)^nh^0\otimes\ldots\otimes \Delta h^n+(-)^{n+1}\sum h^0_{(1)}\otimes h^1\ldots h^n\otimes h^0_{(0)}\ .\non
\eeq
Remark that in the above expression, $b$ involves operators of the form $\id\otimes\ldots\otimes\Delta\otimes\ldots\otimes\id$ and, only for the last term, a cyclic permutation of $n+1$ elements
$$
h^0\otimes h^1\ldots\otimes h^{n+1}\mapsto h^1\otimes\ldots h^{n+1}\otimes h^0\ .
$$
Hence it is sufficient to show that the coinvariants of $C^n(\Hc)$ are stable under these operators. From the properties of the coproduct, it is easy to see that the action of $\Hc$ commutes with the former operators:
\beq
\lefteqn{h\cdot(h^0\otimes\ldots\Delta h^i\otimes\ldots\otimes h^n)=}\non\\
&=& \sum h_{(0)}h^0\otimes\ldots \otimes h_{(i)}h^i_{(0)}\otimes h_{(i+1)}h^i_{(1)}\otimes\ldots\otimes h_{(n+1)}h^n\non\\
&=& (\id\otimes\ldots \otimes\Delta\otimes\ldots\otimes\id)\big( h\cdot(h^0\otimes\ldots\otimes h^n)\big)\ ,\non
\eeq
for any $h\in\Hc$ and $h^0\otimes\ldots\otimes h^n\in C^n(\Hc)$. The only difficulty thus comes from the cyclic permutation. This is the critical step involving the involution condition $\Sd^2=\id$. We have to show that, modulo $\delta$-coinvariants,
$$
\sum h_{(1)}h^1\otimes h_{(2)}h^2\otimes\ldots h_{(n)}h^n\otimes h_{(0)}h^0 \equiv \delta(h) h^1\otimes h^2\otimes\ldots h^n\otimes h^0\ \textup{mod coinv}\ .
$$
Using the properties of the counit and the relation $\delta\circ\Sd=\eps$, we have
\beq
\lefteqn{\sum h_{(1)}h^1\otimes h_{(2)}h^2\otimes\ldots h_{(n)}h^n\otimes h_{(0)}h^0=}\non\\
&=& \sum \eps(h_{(1)})\, h_{(2)}h^1\otimes h_{(3)}h^2\otimes\ldots h_{(n+1)}h^n\otimes h_{(0)}h^0\non\\
&=& \sum \delta(\Sd(h_{(1)}))\,h_{(2)}h^1\otimes h_{(3)}h^2\otimes\ldots h_{(n+1)}h^n\otimes h_{(0)}h^0\non\\
&\equiv& \sum \Sd(h_{(1)})_{(0)}h_{(2)}h^1\otimes \Sd(h_{(1)})_{(1)}h_{(3)}h^2\otimes\ldots \non\\
&&\quad \ldots\otimes\Sd(h_{(1)})_{(n-1)}h_{(n+1)}h^n\otimes \Sd(h_{(1)})_{(n)}h_{(0)}h^0\quad \textup{mod coinv}\ .\non
\eeq
Next, the second relation of (\ref{rela}) implies for any $h\in\Hc$
\beq
\lefteqn{\sum \Sd(h)_{(0)}\otimes\ldots\otimes \Sd(h)_{(n-1)}\otimes \Sd(h)_{(n)} :=\Delta^n \Sd(h)}\non\\
&&= \sum S(h_{(n)})\otimes\ldots\otimes S(h_{(1)})\otimes \Sd(h_{(0)})\non
\eeq
hence we can write
\beq
\lefteqn{\sum \Sd(h_{(1)})_{(0)}h_{(2)}h^1\otimes \Sd(h_{(1)})_{(1)}h_{(3)}h^2\otimes\ldots} \non\\
&&\qquad\qquad \ldots\otimes\Sd(h_{(1)})_{(n-1)}h_{(n+1)}h^n\otimes \Sd(h_{(1)})_{(n)}h_{(0)}h^0\non\\
&=& \sum S(h_{(1)(n)})h_{(2)}h^1\otimes S(h_{(1)(n-1)})h_{(3)}h^2\otimes\ldots \non\\
&&\qquad\qquad \ldots\otimes S(h_{(1)(1)})h_{(n+1)}h^n\otimes \Sd(h_{(1)(0)})h_{(0)}h^0\non\\
&=& \sum S(h_{(n+1)})h_{(n+2)}h^1\otimes S(h_{(n)})h_{(n+3)}h^2\otimes\ldots \non\\
&&\qquad\qquad \ldots\otimes S(h_{(2)})h_{(2n+1)}h^n\otimes \Sd(h_{(1)})h_{(0)}h^0\ ,
\eeq
by reindexing the subscripts of $h$. Using repeatedly the relation $S(h_{(0)})h_{(1)}=\eps(h)$ starting from the left, we have
\beq
\lefteqn{ \sum S(h_{(n+1)})h_{(n+2)}h^1\otimes S(h_{(n)})h_{(n+3)}h^2\otimes\ldots} \non\\
&&\qquad\qquad \ldots\otimes S(h_{(2)})h_{(2n+1)}h^n\otimes \Sd(h_{(1)})h_{(0)}h^0\non\\
&=& \sum h^1\otimes S(h_{(n)})h_{(n+1)}h^2\otimes\ldots \non\\
&&\qquad\qquad \ldots\otimes S(h_{(2)})h_{(2n-1)}h^n\otimes \Sd(h_{(1)})h_{(0)}h^0\non\\
&\vdots&\non\\
&=& \sum h^1\otimes h^2\otimes \ldots\otimes S(h_{(2)})h_{(3)}h^n\otimes \Sd(h_{(1)})h_{(0)}h^0\non\\
&=& \sum h^1\otimes h^2\otimes \ldots\otimes h^n\otimes \Sd(h_{(1)})h_{(0)}h^0\non\\
&=& \sum \delta(h)\, h^1\otimes h^2\otimes \ldots\otimes h^n\otimes h^0\ ,
\eeq
where the last equality comes from the condition $\Sd(h_{(1)})h_{(0)}=\delta(h)1$, equivalent to $\Sd^2=\id$ by lemma \ref{lequi}. Thus we have proved that $\textup{coinv}(C^n)$ is stable by $b$. Using the fact that the direct factor $\Cb^n_{\bot}(\Hc)$ in $C^n(\Hc)$ is stable by $b$, the canonical isomorphism
$$
\Cb^n_{\delta}(\Hc)\cong C^n(\Hc)/(\Cb^n_{\bot}+\textup{coinv}(C^n))\ ,
$$
implies that $b$ is well-defined on $\Cb^n_{\delta}(\Hc)$. \\
Consider now the differential $d:\Cb^n(\Hc)\to\Cb^{n-1}(\Hc)$. One has
\beq
d(h\cdot(h^0\otimes \hb^1\otimes\ldots\otimes\hb^n)) &=& \sum \eps(h_{(0)}h^0)\, h_{(1)}\hb^1\otimes\ldots \otimes h_{(n)}\hb^n\non\\
&=& \sum \eps(h^0)\, h_{(0)}\hb^1\otimes\ldots \otimes h_{(n-1)}\hb^n\non\\
&=& h\cdot(d(h^0\otimes \hb^1\otimes\ldots\otimes\hb^n))\ ,\non
\eeq
hence $d$ preserves the coinvariant subspace of $\Cb^n(\Hc)$ and passes to $\Cb^n_{\delta}(\Hc)$. Since the other operators $\kappa, B$ are made out of $b$ and $d$, we have proved the proposition for $\Cb^*_{\delta}(\Hc)$. The stability of the invariant subspace $\Cb_*^{\delta}(\Hc)$ under the transposed operators is just the dual statement. \cqfd\\
\begin{definition}
Let $\Hc$ be a complete bornological Hopf algebra, $\delta:\Hc\to\cc$ a character and $\Sd$ the twisted antipode such that $\Sd^2=\id$. The \emph{periodic cyclic cohomology} of $\Hc$ is the cohomology of the $\zz_2$-graded complex $\Cb^*_{\delta}(\Hc)=\bigoplus_{n\ge 0}\Cb^n_{\delta}(\Hc)$ of cochains modulo coinvariants, endowed with the total boundary $(b+B)$:
\be
HP^*_{\delta}(\Hc)=H^*(\Cb^*_{\delta}(\Hc))\ ,\quad *=0,1\ .
\ee
The \emph{periodic cyclic homology} of $\Hc$ is the homology of the $\zz_2$-graded complex of $\delta$-invariant chains $\Cb_*^{\delta}(\Hc)=\prod_{n\ge 0}\Cb_n^{\delta}(\Hc)$ endowed with the transposed of the total boundary $(b+B)$:
\be
HP_*^{\delta}(\Hc)=H_*(\Cb_*^{\delta}(\Hc))\ ,\quad *=0,1\ .
\ee
\end{definition}
\begin{remark}\label{rcm}\textup{
In fact the quotient $\Cb^n_{\delta}(\Hc)$ is isomorphic to the vector space $\Hcb^{\hotimes n}$. The correspondence is given by
\beq
h^0\otimes \hb^1\otimes\ldots \otimes \hb^n\ \textup{mod coinv} &\mapsto& \Sd(h^0)\cdot (\hb^1\otimes\ldots \otimes \hb^n)\\
&& = \sum \Sd(h^0)_{(0)}\hb^1\otimes\ldots \otimes \Sd(h^0)_{(n-1)}\hb^n\ ,\non
\eeq
with explicit inverse 
\be
\hb^1\otimes\ldots \otimes \hb^n\mapsto 1\otimes\hb^1\otimes\ldots \otimes \hb^n \ \textup{mod coinv}\ .
\ee
One can show that under this correspondence the mixed complex $(\Cb^*_{\delta}(\Hc), b, B)$ is isomorphic to the normalized complex of the Hopf algebra $\Hc$ introduced by Connes and Moscovici in \cite{CM00}. In particular in low degrees, the Hochschild operator $b:\Cb^0_{\delta}(\Hc)\cong\cc\to\Cb^1_{\delta}(\Hc)\cong\Hcb$ vanishes, whereas Connes' boundary $B:\Cb^1_{\delta}(\Hc)\to \Cb^0_{\delta}(\Hc)$ is given by $B(\hb)=\delta(\hb)$.}
\end{remark}

As in the case of associative algebras, the cyclic cohomology of Hopf algebras has an $X$-complex interpretation. First, we view $\Hc$ only as a coalgebra and formally dualize all the constructions for the algebra $\Bc$: the Fedosov product on even differential forms $\Om^+\Bc$ is transposed into a Fedosov {\it coproduct} on the subspace of even degree cochains $\Cb^+(\Hc)=\bigoplus_{n\ge 0}\Cb^{2n}(\Hc)$
\be
\odot: \Cb^+(\Hc)\to \Cb^+(\Hc)\hotimes\Cb^+(\Hc)\ .
\ee
We don't need to write down explicitly this coproduct and leave it as an exercise to the interested reader. In fact, the trivially graded coalgebra $(\Cb^+(\Hc),\odot)$ is isomorphic to a sub-coalgebra of the {\it completed} (counital) tensor coalgebra $\widehat{T}(\Hcb)=\prod_{n\ge 0}\Hcb^{\hotimes n}$ over the vector space $\Hcb$, endowed with the coproduct
\be
(\hb^1\otimes\ldots\otimes\hb^n)\mapsto \sum_{i=0}^n(\hb^1\otimes\ldots\otimes\hb^i)\otimes(\hb^{i+1}\otimes\ldots\otimes\hb^n)\ .
\ee
The counit of $(\Cb^+(\Hc),\odot)$ is given by the projection $\Cb^+(\Hc)\to \Cb^0(\Hc)=\Hc$ followed by the counit $\eps:\Hc\to\cc$. In particular, observe that this coalgebra splits linearly in the sense of section \ref{sgen}. As a vector space, the $X$-complex of the trivially graded coalgebra (without differential) $\Cb^+(\Hc)$ is isomorphic to the direct sum of even and odd cochains $\Cb^+(\Hc)\oplus \Cb^-(\Hc)=\Cb^*(\Hc)$:
\be
\Cb^*(\Hc)\cong X(\Cb^+(\Hc))\ :\ \Cb^+(\Hc)\ \xymatrix@1{\ar@<0.5ex>[r]^{\beta} &  \ar@<0.5ex>[l]^{\d\nat}}\ \Om_1\Cb^+(\Hc)^{\nat}\ .
\ee
The boundary maps $\beta$ and $\d\nat$ induced by this isomorphism are given by the transposed of (\ref{bou}), up to a sign:
\beq
-\beta &=& b- d(1+\kappa)\quad \textup{ on }\ \Cb^{2n}(\Hc)\\
\d\nat &=& d\sum_{i=0}^{2n}\kappa^i-b \sum_{i=0}^n\kappa^{2i}\quad \textup{ on }\ \Cb^{2n+1}(\Hc)\ .\non
\eeq
The sign in front of $\beta$ comes from the conventions used in the definition of the $X$-complex for coalgebras (section \ref{sgen}): morally, the transposition of the structure maps from the $X$-complex of algebras to the $X$-complex of coalgebras is done in the graded category. As in the case of algebras, the $(b+B)$-complex of cochains $\Cb^*(\Hc)$ is homotopy equivalent to the $X$-complex $(X(\Cb^+(\Hc)), -\beta\oplus\d\nat)$ via the map $cP_1:\Cb^*(\Hc)\to X(\Cb^+(\Hc))$, where $P_1$ is the spectral projection for the eigenvalue 1 of the Karoubi operator $\kappa^2$ (a polynomial in $\kappa$ for each degree $\Cb^n(\Hc)$), and $c:\Cb^*(\Hc)\to X(\Cb^+(\Hc))$ is the rescaling
\be
c(h^0\otimes \hb^1\otimes\ldots\otimes\hb^n)=(-)^{[n/2]}[n/2]!\, h^0\otimes \hb^1\otimes\ldots\otimes\hb^n\ .
\ee
We now take into account the full Hopf algebra structure and project $\Cb^*(\Hc)$ onto the quotient by coinvariants $\Cb^*_{\delta}(\Hc)$. In the same way, $\Hc$ acts on the vector space $X(\Cb^+(\Hc))$ by the diagonal action and we define the quotient
\be
X_{\delta}(\Cb^+(\Hc))=X(\Cb^+(\Hc))/\delta\textup{-coinv}\ .
\ee
Then if $\Sd^2=\id$, all the operators $d,b,\kappa,B$ and $P_1$ of $\Cb^*(\Hc)$ pass to the quotient $\Cb^*_{\delta}(\Hc)$. Similarly for the $X$-complex, all maps descend to $X_{\delta}(\Cb^+(\Hc))$, and we get as before an homotopy equivalence $cP_1:\Cb^*_{\delta}(\Hc)\to X_{\delta}(\Cb^+(\Hc))$ in the coinvariant setting. Thus the periodic cyclic cohomology $HP^*_{\delta}(\Hc)$ is computed by $X_{\delta}(\Cb^+(\Hc))$. Similarly, the periodic cyclic homology of $\Hc$ is computed by the complex $\hom(X_{\delta}(\Cb^+(\Hc)),\cc)$, with differentials induced by
\beq
\nat\dd f &:=& f\circ \d\nat \ ,\quad \forall f\in\hom(\Cb^+(\Hc),\cc)\ ,\\
\bb \gamma &:=& - \gamma\circ\beta\ ,\quad \forall \gamma\in\hom(\Om_1\Cb^+(\Hc),\cc)\ .\non
\eeq
The transposed of $cP_1$ induces a homotopy equivalence between the complex $\hom(X_{\delta}(\Cb^+(\Hc)),\cc)$ endowed with the differential $(\nat\dd\oplus\bb)$, and the complex of Hopf cyclic chains $\hom(\Cb^*_{\delta}(\Hc),\cc)=\Cb_*^{\delta}(\Hc)$ endowed with the boundary $(b+B)$.\\

We shall now introduce the algebraic notion of ``$\Hc$-equivariant $K$-cycle'' for a bornological algebra $\Ac$, and associate to it a characteristic map $HE_*(\Ac)\to HP_*^{\delta}(\Hc)$. The construction is formally identical to the bivariant Chern character of section \ref{sbiv} on $\Psi(\Ac,\Bct)$ (see remark \ref{rper}), with the bornological algebra $\Bct$ replaced, at a dual level, by the Hopf algebra $\Hc$.\\
We first have to introduce Hopf actions. If $\Lc$ is a complete bornological algebra and $\Hc$ a complete bornological Hopf algebra, a left action of $\Hc$ is given by a bounded map $\Hc\hotimes\Lc\to\Lc$, $(h\otimes x)\mapsto h(x)$, such that
\beq
(h^1h^2)(x) &=& h^1(h^2(x))\quad \forall h^i\in\Hc\ ,\ x\in\Lc\ ,\\
h(xy) &=& \sum h_{(0)}(x)h_{(1)}(y)\quad \forall h\in\Hc\ ,\ x,y\in\Lc\ .\non
\eeq
If $\Lc$ is $\zz_2$-graded, we assume that the action of $\Hc$ preserves the grading. If $\Lc$ has a unit, we assume $h(1)=\eps(h)1$, $\forall h\in\Hc$.
\begin{definition}\label{dest}
Let $\Ac$ be a trivially graded complete bornological algebra. Let $\Hc$ be a complete bornological Hopf algebra, provided with a character $\delta:\Hc\to\cc$. The set of $\Hc$-equivariant $K$-cycles $\Psi^{\Hc}_i(\Ac,\cc)$ of degree $i=0,1$ over $\Ac$ consists in quintuples $\Ec=(\Lc,\ell^1,\tau,\rho,D)$ such that:
\begin{itemize}
\item $\Lc$ is a unital $\zz_2$-graded complete bornological algebra provided with an $\Hc$-action;
\item $\ell^1\subset\Lc$ is a two-sided ideal endowed with its own complete bornology, stable by the action of $\Hc$ on $\Lc$;
\item $\tau:\ell^1\to\cc$ is a bounded $\delta$-invariant trace of degree $i$ on the $\Lc$-bimodule $\ell^1$, that is, $\tau([\Lc,\ell^1])=0$ and $\tau(h(x))=\delta(h)\tau(x)$ for any $h\in\Hc$, $x\in \ell^1$;
\item $\rho:\Ac\to \Lc$ is a bounded homomorphism of even degree;
\item
$D\in\Lc$ is an odd element (Dirac operator) provided with a heat kernel $H\in\Lc[0,1]$, with $H(0)=1$ and $H(t)=\exp(-tD^2)\in\ell^1$ for any $t>0$. 
\end{itemize}
One defines in the obvious fashion the sum of two elements of $\Psi^{\Hc}_i(\Ac,\cc)$. A smooth homotopy between two $\Hc$-equivariant elements $\Ec_0=(\Lc,\ell^1,\tau,\rho_0,D_0)$ and $\Ec_1=(\Lc,\ell^1,\tau,\rho_1,D_1)$, where $\Lc$ is endowed with the same action of $\Hc$, is given by an $\Hc$-equivariant interpolation $\Ec=(\Lc[0,1],\ell^1[0,1],\tau\otimes\id,\rho,D)$, where the action of $\Hc$ on the suspension $\Lc[0,1]$ is constant, and $\ev_t\circ\rho=\rho_t$, $\ev_tD=D_t$ for $t=0,1$.
\end{definition}
Let $\Ec=(\Lc,\ell^1,\tau,\rho,D)$ be an element of $\Psi^{\Hc}(\Ac,\cc)$. Consider the entire DG bar coalgebra $\Cc_1=\Bbe(\Ac)$ endowed with its differential $-b'$, together with the trivially graded coalgebra $\Cc_2=\Cb^+(\Hc)$ of even degree cochains provided with the Fedosov coproduct $\odot$. Here $\Hc$ is considered only as a coalgebra. We work in the DG convolution algebra $\hom(\Cc_1\hotimes\Cc_2,\Lc)$ with differential $\delta_1$ induced by $-b'$:
\be
\delta_1f=-(-)^{|f|}f\circ b'\ ,\quad \forall f\in \hom(\Cc_1\hotimes\Cc_2,\Lc)\ .
\ee
The homomorphism $\rho:\Ac\to\Lc$ induces another homomorphism
\be
\rhoh:\Ac\to \hom(\Hc,\Lc)\ ,\quad a\in\Ac\mapsto \Big(h\in\Hc\mapsto h(\rho(a))\in\Lc\Big)
\ee
with target the convolution algebra $\hom(\Hc,\Lc)$, where $\Hc$ is viewed only as a coalgebra. The Hopf action on $\Lc$ shows that $\rhoh$ is indeed an homomorphism:
\beq
\big(\rhoh(a_0a_1)\big)(h)&=& h(\rho(a_0a_1))\ =\ h(\rho(a_0)\rho(a_1))\ =\ \sum h_{(0)}(\rho(a_0))h_{(1)}(\rho(a_1))\non\\
&=&\sum (\rhoh(a_0))(h_{(0)})\,(\rhoh(a_1))(h_{(1)})\ =\ \big(\rhoh(a_0)\rhoh(a_1)\big)(h)\ ,
\eeq
where the last equality comes from the definition of the convolution product on $\hom(\Hc,\Lc)$. Next, we may view $\rhoh$ as a {\it linear} map $\Ac\hotimes\Hc\to\Lc$ (here $\Hc$ enters only as a coalgebra), and use the natural projections $p_1:\Cc_1=\Bbe(\Ac)\to\Bb_1(\Ac)=\Ac$ and $p^0:\Cc_2=\Cb^+(\Hc)\to\Cb^0(\Hc)=\Hc$ to get a bounded linear map
\be
\begin{CD}
\Cc_1\hotimes\Cc_2 @>{p_1\otimes p^0}>> \Ac\hotimes\Hc\stackrel{\rhoh}{\longrightarrow} \Lc\ .
\end{CD}
\ee
Hence we consider $\rhoh$ as an odd degree element of the DG convolution algebra $\hom(\Cc_1\hotimes\Cc_2,\Lc)$. In the same way, define the element $\Dh\in\hom(\Hc,\Lc)$ by
\be
\Dh(h)=h(D)\quad \forall h\in\Hc\ ,
\ee
and regard it as an odd element of $\hom(\Cc_1\hotimes\Cc_2,\Lc)$ via the composition
\be
\begin{CD}
\Cc_1\hotimes\Cc_2 @>{\eta\otimes p^0}>> \Hc \stackrel{\Dh}{\longrightarrow} \Lc\ ,
\end{CD}
\ee
where $\eta:\Cc_1\to\cc$ is the counit of $\Bbe(\Ac)$. We introduce now the Quillen superconnection
\be
A=\rhoh+\Dh\ \in\ \hom(\Cc_1\otimes\Cc_2,\Lc)_-\ ,
\ee
and its curvature
\be
F=\delta_1A+A^2\ \in\ \hom(\Cc_1\otimes\Cc_2,\Lc)_+\ .
\ee
To become familiar with the notations, it is instructive to evaluate explicitly the curvature on the elements of $\Cc_1\hotimes\Cc_2=\Bbe(\Ac)\hotimes \Cb^+(\Hc)$. Since the coalgebra $\Cb^+(\Hc)$ replaces formally the algebra $\Bct$ of section \ref{sbiv} in a dual fashion, we can use formula (\ref{curv}) provided the differential forms over $\Bc$ are reinterpreted as chains over $\Hc$:
\be
F \ =\ \left\{ \begin{array}{cll}
\Dh^2 &\in& \hom(\Cb^0(\Hc),\Lc)\\
+ [\Dh,\rhoh] &\in& \hom(\Ac\hotimes\Cb^0(\Hc),\Lc)\\
+ d\Dh d\Dh &\in& \hom(\Cb^2(\Hc),\Lc)\label{curva}\\
+ d\Dh d\rhoh+d\rhoh d\Dh &\in& \hom(\Ac\hotimes\Cb^2(\Hc),\Lc)\\
+ d\rhoh d\rhoh &\in& \hom(\Bb_2(\Ac)\hotimes\Cb^2(\Hc),\Lc)\ . \end{array}\right.
\ee
For any elements $a_1\otimes\ldots \otimes a_n\in\Bb_n(\Ac)$ and $h^0\otimes \hb^1\otimes \ldots \hb^n \in \Cb^n(\Hc)$, we evaluate explicitly
\beq
F(h^0) &=& \Dh^2(h^0)\ =\ h^0(D^2)\ , \non\\
F(a_1\otimes h^0) &=&  [\Dh,\rhoh(a_1)](h^0)\ =\ h^0([D,\rho(a_1)])\ ,\non\\
F(h^0\otimes\hb^1\otimes\hb^2)&=& d\Dh d\Dh (h^0\otimes\hb^1\otimes\hb^2)\ =\ \eps(h^0)\hb^1(D)\hb^2(D)\ ,\non\\
F(a_1\otimes (h^0\otimes\hb^1\otimes\hb^2)) &=& (d\Dh d\rhoh(a_1)+d\rhoh(a_1)d\Dh)(h^0\otimes\hb^1\otimes\hb^2)\non \\
&=& \eps(h^0)(\hb^1(D)\hb^2(\rho(a_1))+\hb^1(\rho(a_1))\hb^2(D))\ ,\non\\
F((a_1\otimes a_2)\otimes(h^0\otimes\hb^1\otimes\hb^2)) &=& (d\rhoh(a_1)d\rhoh(a_2))(h^0\otimes\hb^1\otimes\hb^2)\non\\ 
&=& \eps(h^0)\hb^1(\rho(a_1))\hb^2(\rho(a_2))\ .
\eeq
Since we want to take the exponential of $F$ in the convolution algebra, we are led to the following definition of $\te$-summability:
\begin{definition}\label{dte2}
Let $\Ac$ and $\Hc$ be respectively a complete bornological algebra and Hopf algebra, $\Ec=(\Lc,\ell^1,\tau,\rho,D)$ an element of $\Psi^{\Hc}(\Ac,\cc)$, $A=\rhoh+\Dh$ the associated superconnection in the DG convolution algebra $\hom(\Bbe(\Ac)\hotimes \Cb^+(\Hc),\Lc)$, and $F$ its curvature. Then $\Ec$ is called $\te$-summable iff the exponential of $F$ converges in the convolution algebra towards a bounded map
\be
\exp(-F): \Bbe(\Ac)\hotimes \Cb^+(\Hc)\to\ell^1\ .
\ee
Two $\te$-summable elements $\Ec_0$ and $\Ec_1$ are smoothly homotopic iff there exists a $\te$-summable interpolating element.
\end{definition}

Fix a $\te$-summable element $\Ec\in\Psi^{\Hc}(\Ac,\cc)$. We construct a bivariant Chern character for the superconnection $A$ in exactly the same way as in section \ref{sbiv}, the only difference being the formal replacement of $\Bct$ with $\hom(\Hc,\cdot)$, etc... For convenience, let us translate explicitly the construction into this new setting. Introduce the differentials and partial traces
\beq
\dd_1 &:& \hom(\Cc_1\hotimes\Cc_2,\Lc)\to \hom(\Om_1\Cc_1\hotimes\Cc_2,\Lc) \ ,\non\\
\dd_2 &:& \hom(\Cc_1\hotimes\Cc_2,\Lc)\to \hom(\Cc_1\hotimes\Om_1\Cc_2,\Lc) \ ,\non\\
\nat_1 &:& \hom(\Om_1\Cc\hotimes\Cc_2,\Lc)\to \hom(\Om_1\Cc^{\nat}\hotimes\Cc_2,\Lc)\ ,\\
\nat_2 &:& \hom(\Cc_1\hotimes\Om_1\Cc_2,\Lc)\to \hom(\Cc_1\hotimes\Om_1\Cc_2^{\nat},\Lc)\ ,\non\\
\nat_{12} &:& \hom(\Om_1\Cc_1\hotimes\Om_1\Cc_2,\Lc)\to \hom(\Om_1\Cc_1^{\nat}\hotimes\Om_1\Cc_2^{\nat},\Lc)\ .\non
\eeq
Then we have
\beq
\dd_1A\ ,\ \dd_1F &\in& \hom(\Om_1\Cc_1\hotimes\Cc_2,\Lc)\ ,\non\\
\dd_2A\ ,\ \dd_2F &\in& \hom(\Cc_1\hotimes\Om_1\Cc_2,\Lc)\ ,\\
e^{-F} &\in& \hom(\Cc_1\hotimes\Cc_2,\ell^1)\ ,\non
\eeq
because $\Ec$ is $\te$-summable. Again, formula (\ref{exp}) computing explicitly the exponential of $F$ in the convolution algebra $\hom(\Bbe(\Ac)\hotimes \Cb^+(\Hc),\ell^1)$ still holds,
\be
e^{-F}= \sum_{n\ge 0} \int_{\Delta_n}ds_1\ldots ds_n\, e^{-s_0\Dh^2}He^{-s_1\Dh^2}H\ldots e^{-s_{n-1}\Dh^2}He^{-s_n\Dh^2}|_+\ ,\label{exph}
\ee
with $H=-[\Dh,\rhoh]-d(\rhoh+\Dh)d(\rhoh+\Dh)+\sqrt{-1}d(\Dh^2+[\Dh,\rhoh])$ interpreted as in (\ref{curva}). Then, the components of $\ch(A)$ are given by
\beq
\hom(\Cc_1\otimes\Cc_2,\ell^1) &\ni& \ch_0^0(A) =e^{-F}\ ,\non\\ 
\hom(\Om_1\Cc_1^{\nat}\otimes\Cc_2,\ell^1) &\ni& \ch_1^0(A) =-\nat_1 \dd_1Ae^{-F}\ ,\non\\
\hom(\Cc_1\otimes\Om_1\Cc_2^{\nat},\ell^1) &\ni& \ch_0^1(A) = \nat_2 e^{-F}\dd_2A\ ,\\
\hom(\Om_1\Cc_1^{\nat}\otimes\Om_1\Cc_2^{\nat},\ell^1) &\ni& \ch_1^1(A)= -\nat_{12} \dd_1A e^{-F}\dd_2A - \non\\
&-&\int_{\Delta_2}ds_1ds_2\,\nat_{12} e^{-s_0F}\dd_2Fe^{-s_1F}\dd_1Fe^{-s_2F}\ ,\non
\eeq
and applying the trace $\tau:\ell^1\to\cc$, we get a cocycle
\be
\tau\ch(A)\in \hom(X(\Cc_1)\hotimes X(\Cc_2),\cc)\ .
\ee
We may compose $\tau\ch(A)$ with the homotopy equivalence $cP_1:\Cb^*(\Hc)\to X(\Cc_2)$ defined on the $(b+B)$-complex of cochains on the coalgebra $\Hc$, and use the identification of the reduced $X$-complex $\Xb(\Cc_1)$ with the $(b+B)$-complex of entire forms $\Ome\Ac$. This yields a chain map
\be
\tau\ch(A)\circ cP_1: \Ome\Ac\to \hom(\Cb^*(\Hc),\cc)=\Cb_*(\Hc)
\ee
from $(\Ome\Ac,b+B)$ to $(\Cb_*(\Hc),b+B)$, of parity $|\tau|$. The point is that the image of this map actually lies in the $\delta$-invariant subspace of $\Cb_*(\Hc)$, which is a subcomplex computing the cyclic homology of $\Hc$ considered as a Hopf algebra whenever the twisted antipode $\Sd$ verifies the involution condition.
\begin{theorem}\label{t2}
Let $\Ac$ be a complete bornological algebra, $\Hc$ a complete bornological Hopf algebra and $\delta:\Hc\to\cc$ a bounded character. If $\Sd^2=\id$, then any $\te$-summable element $\Ec=(\Lc,\ell^1,\tau,\rho,D)$ of $\Psi^{\Hc}_i(\Ac,\cc)$ gives rise to a characteristic map
\be
\ch(\Ec): HE_j(\Ac)\to HP_{j+i}^{\delta}(\Hc)\ ,\quad j\in\zz_2\ ,\label{char}
\ee
where $\ch(\Ec)$ is represented by the chain map $\tau\ch(A)\circ cP_1$, with $A=\rhoh+\Dh$ the superconnection associated to $\Ec$. This characteristic map is additive and invariant for smooth homotopies of $\rho$ and $D$.
\end{theorem}
{\it Proof:} The $\delta$-invariance of the cocycle $\tau\ch(A)\in\hom(X(\Cc_1)\hotimes X(\Cc_2),\cc)$ is a direct consequence of the construction. Let $h^0\otimes\hb^1\otimes\ldots\otimes\hb^n\in\Cb^n(\Hc)$ be a coalgebra $n$-cochain. Using the isomorphisms $\Xb(\Cc_1)\cong\Ome\Ac$ and $X(\Cc_2)\cong\Cb^n(\Hc)$, the evaluation of $\ch(A)$ on this cochain is a bounded linear map $\Ome\Ac\to\ell^1$:
$$
\ch(A)(h^0\otimes\hb^1\otimes\ldots\otimes\hb^n)\ \in \ \hom(\Ome\Ac,\ell^1)\ ,
$$ 
whose further evaluation on a $p$-form $a_0da_1\ldots da_p\in\Om^p\Ac$ yields a polynomial 
$$
Q= h^0(Q_0)\prod_{i=1}^n \hb^i(Q_i)\ \in\ \ell^1\ ,
$$
where the $Q_i$'s are polynomials in the Dirac operator $D$, the heat kernel $e^{-tD^2}$ and the $\rho(a_j)$'s. If $h\in\Hc$ acts on $h^0\otimes\hb^1\otimes\ldots\otimes\hb^n$ by the diagonal action, $Q$ changes into
$$
\sum h_{(0)}h^0(Q_0)\prod_{i=1}^n h_{(i)}\hb^i(Q_i)=h\cdot Q\ ,
$$
hence applying the $\delta$-invariant trace $\tau:\ell^1\to\cc$, one sees that 
$$
\tau\ch(A)\big(h\cdot(h^0\otimes\hb^1\otimes\ldots\otimes\hb^n)\big)=\delta(h)\tau\ch(A)(h^0\otimes\hb^1\otimes\ldots\otimes\hb^n)\ .
$$
This shows that the target of the chain map $\tau\ch(A)\circ cP_1: \Ome\Ac\to\Cb_*(\Hc)$ lies in the $\delta$-invariant subcomplex of Hopf cyclic chains. The homotopy invariance with respect to $\rho$ and $D$ is a consequence of the existence of a transgression formula in the total complex $\hom(\Ome\Ac\hotimes\Cb^*(\Hc),\Om[0,1])$ analogous to the proof of theorem \ref{t1}, compatible with the $\delta$-invariance as before. \cqfd
\begin{example}\textup{
The characteristic map (\ref{char}) is a far-reaching generalization of the Connes-Moscovici map introduced in \cite{CM98}. The latter corresponds to the following particular case. Let $\Ac$ and $\Hc$ be as in the theorem. Suppose further that $\Hc$ acts on $\Ac$ and $\tau:\Ac\to\cc$ is a $\delta$-invariant trace. Set $\Lc=\ell^1=\Ac$, and let $\rho:\Ac\to\Ac$ be the identity homomorphism and $D=0$. Then $\Ec=(\Lc,\ell^1,\tau,\rho,D)$ is an element of $\Psi^{\Hc}_0(\Ac,\cc)$, and the characteristic map (\ref{char}) actually factors through the periodic cyclic homology of $\Ac$,
\be
\ch(\Ec): HP_*(\Ac)\to HP_*^{\delta}(\Hc)\ ,
\ee
because this situation is purely algebraic. In fact it is defined for any $\Hc$-algebra $\Ac$ not necessarily bornological. Then $\ch(\Ec)$ is the characteristic map of Connes-Moscovici associated to $\Ac$, $\Hc$ and the $\delta$-invariant trace $\tau$.}
\end{example}
\begin{remark}\textup{
The whole construction can be dualized with $\Hc$ a complete bornological Hopf algebra {\it coacting} on $\Lc$. In this case, $\Hc$ replaces exactly the algebra $\Bct$ of section \ref{sbiv}, while adding some extra structure (invariance with respect to coactions). The previous character $\delta:\Hc\to\cc$ must be dualized into a group-like element $\si\in\Hc$, and the twisted antipode corresponds to the product $S_{\si}(h)=\si S(h)$. In particular, the (augmented) $(b,B)$-complex of entire chains $\Omte\Hcb$ is endowed with a left coation $\Delta_l: \Omte\Hcb\to\Hc\hotimes\Omte\Hcb$ coming from the coproduct,
\be
\Delta_l(h^0dh^1\ldots dh^n)= \sum (h^0_{(0)}h^1_{(0)}\ldots h^n_{(0)})\otimes h^0_{(1)}dh^1_{(1)}\ldots dh^n_{(1)}\ ,
\ee
and the subspace of $\si$-invariant chains
\be
\Delta_l(h^0dh^1\ldots dh^n)=\si\otimes h^0dh^1\ldots dh^n
\ee
is a $(b,B)$-subcomplex iff $S_{\si}^2=\id$. We denote by $HP_*^{\si}(\Hc)$ the corresponding periodic cyclic homology. Starting from a quintuple $\Ec=(\Lc,\ell^1,\tau,\rho,D)$ in $\Psi_i(\Ac,\cc)$ such that $\Lc$ and $\ell^1$ are endowed with a coation of $\Hc$ and $\tau$ is a $\si$-invariant trace (i.e. $\tau(x)\to \si\tau(x)$ under the coaction), an obvious modification of the construction above yields a characteristic map
\be
\ch(\Ec): HE_*(\Ac)\to HP_{*+i}^{\si}(\Hc)\ .
\ee
This dual formulation is well-adapted to Hopf algebras viewed as algebras of functions over a quantum group, instead of algebras of operators (enveloping algebras). It should be possible to generalize the construction with a modular pair $(\si,\delta)$ as introduced by Connes-Moscovici \cite{CM99,CM00}, which allows to replace the invariant trace $\tau$ by a Haar measure on $\Hc$. This may presumably have some applications in the theory of locally compact quantum groups in the sense of Woronowicz \cite{W}.}
\end{remark}

The entire cyclic homology of $\Ac$ is the natural target for the Chern character on {\it topological} $K$-theory \cite{Me}. The definition of topological $K$-theory for complete bornological algebras follows the definition of Phillips for Fr\'echet algebras \cite{Ph}. \\
Let $\lg$ be the algebra of infinite matrices $(x_{ij})_{i,j\in\nn}$ over $\cc$ with rapidly decreasing entries. It is a Fr\'echet algebra for the family of norms $p_n$
\be
p_n((x_{ij}))=\sum_{i,j}(1+i+j)^n|x_{ij}|\ .
\ee
Given a complete bornological algebra $\Ac$, we first stabilize by taking the completed tensor product $\Ac\hotimes\lg$, add a unit to the resulting algebra, and consider the unital algebra of $2\times 2$ matrices $\Asf=M_2(\cc)\hotimes (\widetilde{\Ac\hotimes\lg})$. Then the topological $K$-theory of even degree $\Kt_0(\Ac)$ is the set of smooth homotopy classes of idempotents $e\in\Asf$ such that $e-\left(\begin{array}{cc}
				1 & 0 \\
				0 & 0 \end{array}\right) \in M_2(\cc)\hotimes \Ac\hotimes\lg$. The direct sum of idempotents turns $\Kt_0(\Ac)$ into an abelian group. Meyer shows in \cite{Me} that the Chern character on even $K$-theory is a well-defined additive map 
\be
\cht_0: \Kt_0(\Ac)\to HE_0(\Ac)\ ,
\ee
It is given in terms of the idempotent $e$ by the usual formula (see \cite{C1}).\\
The odd $K$-theory $\Kt_1(\Ac)$ is the abelian group of smooth homotopy classes of elements $x\in\Ac\hotimes\lg$ such that $1+x\in \widetilde{\Ac\hotimes\lg}$ is invertible. Again, there is a well-defined additive map \cite{Me}
\be
\cht_1: \Kt_1(\Ac)\to HE_1(\Ac)\ .
\ee
As a consequence of the above theorem, we get a pairing between the set of homotopy classes of equivariant $K$-cycles over the complete bornological algebra $\Ac$ and its topological $K$-theory:
\begin{corollary}
Let $\Ac$ be a complete bornological algebra, and $\Hc$ be a complete bornological Hopf algebra provided with a bounded character $\delta:\Hc\to\cc$ such that the twisted antipode $\Sd$ is involutive. Then there is a bilinear pairing between the topological $K$-theory of $\Ac$ and the set of smooth homotopy classes of $\te$-summable $\Hc$-equivariant $K$-cycles $\Psi^{\Hc,\te}(\Ac,\cc)$,
\be
\Kt_i(\Ac)\times \Psi^{\Hc,\te}_j(\Ac,\cc)\to HP_{i+j}^{\delta}(\Hc)\ ,\quad i,j\in\zz_2 \ ,
\ee
with values in the periodic cyclic homology of $\Hc$. Given a representative $\Ec\in\Psi^{\Hc,\te}_j(\Ac,\cc)$, its pairing with $K$-theory is obtained by composing the characteristic map $\ch(\Ec): HE_*(\Ac)\to HP_{*+j}^{\delta}(\Hc)$ with the Chern character on $\Kt_*(\Ac)$. \cqfd
\end{corollary}

\section{Secondary characteristic classes}\label{ssec}

We shall now consider the same situation as in the previous section, but restrict ourselves to Hopf algebras acting by the {\it ajoint representation}. Let $\Ac$ be a complete bornological algebra and $\Hc$ a Hopf algebra with involutive antipode. We choose the character $\delta:\Hc\to\cc$ simply equal to the counit $\eps$. If $\Ec=(\Lc,\ell^1,\tau,\rho,D)\in\Psi^{\Hc}(\Ac,\cc)$ is an $\Hc$-equivariant $K$-cycle such that $\Hc$ acts on $\Lc$ by the adjoint representation, the image of the characteristic map $\ch(\Ec):HE_*(\Ac)\to HP^{\eps}_*(\Hc)$ is contained in the trivial factor $\cc\subset HP^{\eps}_*(\Hc)$. This implies that the action of $\Hc$ is unfortunately not detected by $\ch(\Ec)$. However, in some circumstances this leads us to exhibit {\it secondary characteristic classes}. Since this goal requires some analysis, we must consider that all algebras are represented as operators acting on a Hilbert space. For any integer $p$, we define the set of $p_+$-summable $\Hc$-equivariant spectral triples $\Esf^{\Hc}_p(\Ac,\cc)$. Such elements $\Ec$ are given by triples $(\Ac,\Hg,D)$ in the sense of Connes' non-commutative geometry \cite{C1}: $\Hg$ is a separable Hilbert space; $D$ is an unbounded selfadjoint operator with compact resolvent (Dirac operator), such that $(1+D^2)^{-(p/2+\eps)}$ is trace-class for any $\eps>0$ (we call this property $p_+$-summability); and $\Ac$ is represented as bounded operators on $\Hg$ such that the commutator $[D,a]$ is bounded for any $a\in\Ac$. In addition, the Hopf algebra $\Hc$ acts on $\Hg$ by (possibly unbounded) operators, but we assume that the adjoint action of $\Hc$ on the Dirac operator induces only bounded perturbations of $D$, and that $[h,a]=0=[h,[D,a]]$ for any $h\in \Hc$, $a\in\Ac$. Under some extra analytical conditions, this gives rise to secondary classes
\be
\psi(\Ec):HC_n(\Ac)\to HC^{\eps,+}_{n-p}(\Hc)\ ,\quad  \forall n\ge 0\ ,
\ee
defined on the {\it non-periodic} cyclic homology $HC_n(\Ac)$, with target a cyclic homology of $\Hc$ obtained from the vertical filtration of its cyclic bicomplex. Remarkably, these classes are still invariant with respect to (at least) bounded homotopies of the Dirac operator. They are {\it not} homotopy invariant with respect to the representation homomorphism $\rho:\Ac\to\End(\Hg)$ (the functors $HC_n$ are not homotopy invariant).  As a consequence, we get induced characteristic maps on the {\it higher algebraic} $K$-theory of $\Ac$ \cite{R} rather than the topological $K$-theory.\\

Let $\Hc$ be a Hopf algebra endowed with a complete bornology, such that its antipode is involutive: $S^2=\id$. This allows to choose the character $\delta:\Hc\to\cc$ simply equal to the counit $\eps$, so that the twisted antipode $S_{\delta}$ coincides with $S$. The corresponding periodic cyclic cohomology (resp. homology) is thus denoted $HP^*_{\eps}(\Hc)$ (resp. $HP_*^{\eps}(\Hc)$). Recall that from remark \ref{rcm}, for any character $\delta$ the Hochschild coboundary operator on the Hopf cochains of degree zero vanishes:
\be
b:\Cb^0_{\delta}(\Hc)\cong\cc\to \Cb^1_{\delta}(\Hc)\cong\Hcb\ ,\quad \la\mapsto 0\ ,
\ee
($\Hcb=\ker\eps$) whereas Connes' boundary map $B$ on one-cochains is given by
\be
B:\Cb^1_{\delta}(\Hc)\to \Cb^0_{\delta}(\Hc)\ ,\quad \hb\mapsto \delta(\hb)\ ,
\ee
for any $\hb\in\Hcb$. Hence in the present case $\delta=\eps$, one has $B(\hb)=0$. It follows that the $(b,B)$-bicomplex of Hopf cochains $\Cb^*_{\eps}(\Hc)=\bigoplus_{n\ge 0}\Cb^n_{\eps}(\Hc)$ splits into the direct sum of complexes
\be
\Cb^*_{\eps}(\Hc)=\cc\oplus \bigoplus_{n\ge 1}\Cb^n_{\eps}(\Hc)\ ,
\ee
where $\cc$ is the trivial complex, and similarly for the bicomplex of Hopf chains $\Cb_*^{\eps}(\Hc)=\cc\oplus\prod_{n\ge 1}\hom(\Cb^n_{\eps}(\Hc),\cc)$. As a consequence, the periodic cyclic homology of $\Hc$ splits in even degree,
\be
HP_0^{\eps}(\Hc)=\cc\oplus \HPb_0^{\eps}(\Hc)\ ,\quad HP_1^{\eps}(\Hc)=\HPb_1^{\eps}(\Hc)\ ,
\ee
and similarly for cohomology. We call $\HPb_*^{\eps}(\Hc)$ the {\it reduced} (periodic) {\it cyclic homology} of $\Hc$. \\

Let $\Ac$ be a complete bornological algebra, and $\Ec=(\Lc,\ell^1,\tau,\rho,D)$ an $\Hc$-equivariant $K$-cycle in $\Psi^{\Hc}_i(\Ac,\cc)$, $i=0,1$ (definition \ref{dest}). We recall that $\Hc$ acts from the left on the auxiliary algebras $\Lc,\ell^1$ and $\tau:\ell^1\to\cc$ is an $\eps$-invariant trace of degree $i$, that is, $\tau(h(x))=\eps(h)\tau(x)$ for any $x\in\ell^1, h\in\Hc$. If $\Ec$ is $\te$-summable, then theorem \ref{t2} yields a characteristic map $\ch(\Ec): HE_j(\Ac)\to HP_{j+i}^{\eps}(\Hc)$ for any $j\in\zz_2$. Using the splitting $HP_*^{\eps}(\Hc)=\cc\oplus \HPb_*^{\eps}(\Hc)$, the characteristic map $\ch(\Ec)$ splits into a $\cc$-valued component
\be
\ch^{\cc}(\Ec): HE_i(\Ac)\to \cc\ ,
\ee
and a {\it reduced characteristic map}
\be
\chb(\Ec): HE_j(\Ac)\to \HPb_{j+i}^{\eps}(\Hc)\ ,\quad j\in\zz_2\ .
\ee
It is easy to see that the $\cc$-valued component exactly corresponds to the JLO cocycle \cite{JLO}, whose evaluation on a $n$-form $a_0da_1\ldots da_n \in \Om^n\Ac$ reads
\beq
\lefteqn{\ch^{\cc}(\Ec)(a_0da_1\ldots da_n)=}\label{jlo}\\
&&(-)^n\int_{\Delta_n}ds_1\ldots ds_n\, \tau\big(\rho(a_0)e^{-s_0D^2}[D,\rho(a_1)]e^{-s_1D^2}\ldots [D,\rho(a_n)]e^{-s_nD^2}\big)\ ,\non
\eeq
where $\Delta_n=\{(s_0,\ldots s_n)\in[0,1]^{n+1}\ |\ \sum_is_i=1\}$ is the standard $n$-simplex. Remark that due to the odd parity of $D$, this expression vanishes whenever the degree $i$ of the trace $\tau$ is not congruent to $n$ mod 2, thus $\ch^{\cc}(\Ec)$ is indeed an entire cyclic cocycle of degree $i$ over $\Ac$. It is clear that the JLO formula does not involve the Hopf algebra action. All the relevant information about the cyclic homology of $\Hc$ is therefore transfered into the reduced characteristic map.\\
We now restrict to the case where $\Hc$ viewed as an associative algebra is represented in $\Lc$ via a unital algebra homomorphism $g:\Hc\to\Lc$, and as a Hopf algebra, acts by the adjoint representation on $\Lc$ and $\ell^1$. In the following we shall denote the image of an element $g(h)\in\Lc$ simply by $h$ when no confusion arises. The left adjoint action $\Hc\hotimes\Lc\to\Lc$ reads
\be
h\cdot x :=\sum h_{(0)}x S(h_{(1)})\ ,\quad \forall h\in\Hc\ ,\ x\in\Lc\ .
\ee
One checks easily the compatibility with the product on $\Lc$, hence $\Lc$ endowed with the adjoint representation of $\Hc$ is indeed a $\Hc$-algebra. Furthermore, the two-sided ideal $\ell^1\subset \Lc$ is necessarily stable under this action, and {\it any} bounded trace $\tau:\ell^1\to\cc$ is $\eps$-invariant:
\be
\tau(h\cdot x)=\sum\tau(h_{(0)}x S(h_{(1)}))=\sum \tau(S(h_{(1)})h_{(0)}x)=\eps(h)\tau(x)\ ,\quad \forall h\in\Hc\ ,\ x\in\ell^1\ ,
\ee
because the hypothesis $S^2=\id$ is equivalent to $\sum S(h_{(1)})h_{(0)}=\eps(h)1$, $\forall h\in\Hc$. The proposition below states that if $\Hc$ acts by the adjoint representation, then the characteristic map $\ch(\Ec)$ retracts onto the JLO cocycle.
\begin{proposition}\label{ptriv}
Let $\Ac$ be a complete bornological algebra, $\Hc$ a complete bornological Hopf algebra such that $S^2=\id$. Let $\Ec=(\Lc,\ell^1,\tau,\rho,D)\in \Psi^{\Hc}_i(\Ac,\cc)$ be an $\Hc$-equivariant $K$-cycle such that $\Hc$ is represented in $\Lc$ and acts by the adjoint representation. Then the reduced characteristic map $\chb(\Ec): HE_j(\Ac)\to \HPb_{j+i}^{\eps}(\Hc)$ vanishes.
\end{proposition}
{\it Proof:} Let $g:\Hc\to\Lc$ be the representation (unital algebra homomorphism) of $\Hc$ into $\Lc$. If we regard $\Hc$ as a coalgebra, the linear map $g$ is an element of the convolution algebra $\hom(\Hc,\Lc)$. The latter has a unit given by $1_{\Lc}\eps$. Then $g$ is invertible in the convolution algebra, with inverse corresponding to the composition $g^{-1}=g\circ S\in\hom(\Hc,\Lc)$. \\
We shall build an interpolation between $\Ec$ and a quintuple whose Chern character just reduces to the JLO cocycle. Consider the unital algebra of $2\times 2$ matrices $M_2(\Lc[0,1])$ over the suspension $\Lc[0,1]=\Lc\hotimes\cinf[0,1]$, and the associated convolution algebra $\hom(\Hc,M_2(\Lc[0,1]))$. Let $f_1, f_2:[0,1]\to\rr$ be smooth functions over the interval, such that $f_1^2+f_2^2=1$, $f_1(0)=0$, $f_1(1)=1$. We may choose $f_1(t)=t$ and $f_2(t)=\sqrt{1-t^2}$. Then from the homomorphism $\rho:\Ac\to\Lc$, we construct an algebra homomorphism
$$
\rhoh:\Ac\to \hom(\Hc,M_2(\Lc[0,1]))\ ,\quad \rhoh(a)=\left( \begin{array}{cc}
f_1^2\, g\rho(a)g^{-1} & f_1f_2 \, g\rho(a) \\
f_2f_1\, \rho(a)g^{-1} & f_2^2 \, \rho(a) \end{array}\right)\ .
$$
We view $\rhoh$ as a bounded linear map $\Ac\hotimes\Hc\to M_2(\Lc[0,1])$, $a\otimes h\mapsto \rhoh(a)(h)$. Next, from $D$ we introduce an odd element:
$$
\Dh= \left( \begin{array}{cc}
 gDg^{-1} & 0 \\
 0  &  D \end{array}\right)\ \in\ \hom(\Hc,M_2(\Lc[0,1]))\ .
$$
For $t=1$, we have
$$
\ev_1\rhoh= \left( \begin{array}{cc}
 g\rho g^{-1} & 0 \\
 0  & 0 \end{array}\right)\ ,\quad \ev_1 \Dh= \left( \begin{array}{cc}
 gDg^{-1} & 0 \\
 0  &  D \end{array}\right)\ .
$$
The evaluation of $g\rho g^{-1}\in \hom(\Ac\hotimes\Hc,\Lc)$ on a product $a\otimes h$ yields
$$
g\rho g^{-1}(a\otimes h)=\sum h_{(0)}\rho(a) S(h_{(1)})=h\cdot(\rho(a))\ ,
$$
and similarly $gDg^{-1}(h)=h\cdot D$, which shows that for $t=1$, the upper left corner of $\rhoh$ and $\Dh$ coincide with the construction of section \ref{shopf}, leading to the characteristic map associated to $\Ec$. For $t=0$, one has
$$
\ev_0\rhoh= \left( \begin{array}{cc}
 0 & 0 \\
 0  & \rho \end{array}\right)\ ,\quad \ev_0 \Dh= \left( \begin{array}{cc}
 gDg^{-1} & 0 \\
 0  &  D \end{array}\right)\ ,
$$
and $\rho(a\otimes h)=\eps(h)\rho(a)$, $D(h)=\eps(h)D$. This shows that for $t=0$, the bottom right corner of the matrices $\rhoh$ and $\Dh$ vanish on the augmentation ideal $\Hcb=\ker\eps$.\\
Let $(\Lc\hotimes\Om[0,1], d)$ be the DG algebra of differential forms over $[0,1]$ with values in $\Lc$. As in section \ref{shopf}, we will construct a bivariant Chern character by working in the DG convolution algebra $\Rc=\hom(\Bbe(\Ac)\hotimes \Cb^+(\Hc),\Lc\hotimes\Om[0,1])$, where the differential of $\Lc\hotimes\Om[0,1]$ is taken into account. Recall that $\Bbe(\Ac)$ is the entire bar DG coalgebra of $\Ac$, and $\Cb^+(\Hc)$ is the Fedosov coalgebra (without differential) of cochains over $\Hc$. First, we consider $\rhoh$ and $\Dh$ as odd elements of $\Rc$ by the same trick as in section \ref{shopf}, and form the superconnection $A=\rhoh+\Dh\in \Rc_-$. The trace $\tau:\ell^1\to\cc$ extends to a trace $\tau\otimes\tr_2:M_2(\ell^1\hotimes\Om[0,1])\to \Om[0,1]$, so that the usual Chern character
$$
(\tau\otimes\tr_2)\,\ch(A) \in \hom(X(\Bbe(\Ac))\hotimes X(\Cb^+(\Hc)),\Om[0,1])
$$ 
is a cocycle in the corresponding complex, where the de Rham differential $d$ over $\Om[0,1]$ is taken into account. As in the proof of proposition \ref{phomo}, we let $\chi$ be the projection of this cocycle onto the subspace of numerical functions $\cinf[0,1]\subset\Om[0,1]$ and $cs$ the projection onto the one-forms $\Om^1[0,1]$. The cocycle property implies that
$$
d\chi= \pm cs\circ \d\ ,
$$
where $\d$ denotes the total differential on $X(\Bbe(\Ac))\hotimes X(\Cb^+(\Hc))$. Moreover, this construction is compatible with the $\eps$-invariance (same proof as theorem \ref{t2}), hence holds at the level of Hopf cochains. Using the homotopy equivalence $cP_1$, we may identify the complexes mod $\eps$-coinvariants $X_{\eps}(\Cb^+(\Hc))$ and $\Cb^*_{\eps}(\Hc)$. For $t=0$, we know that the bottom right corner of the matrix $A$ vanishes on $\Hcb$, and a careful inspection shows that after taking the trace $\tau\otimes\tr_2$, this implies the vanishing of $\chi(t=0)$ on a chain $h^0\otimes\hb^1\otimes\ldots\otimes\hb^n \in \Cb_{\eps}^n(\Hc)$ whenever $n>0$. Hence we deduce that $\chi_{t=0}=\ch^{\cc}(\Ec)$ on $\Ome\Ac=\Xb(\Bbe(\Ac))$. On the other hand, for $t=1$ the upper left corner of the matrix $A$ corresponds to the connection entering in the construction of the Chern character of $\Ec$, so that after taking the trace $\tau\otimes\tr_2$, one sees that $\chi_{t=1}=\ch(\Ec)$ on $\Ome\Ac\hotimes X_{\eps}(\Cb^+(\Hc))$. Therefore
$$
\chb(\Ec)=\ch(\Ec)-\ch^{\cc}(\Ec)=\int_{[0,1]}d\chi=\pm\int_{[0,1]}cs\circ\d\ ,
$$
is a coboundary in the complex $\hom(\Ome\Ac\hotimes X_{\eps}(\Cb^+(\Hc)),\cc)$.\cqfd\\

Thus when $\Hc$ acts by the adjoint representation, it is not detected by the reduced characteristic map $\chb(\Ec)$ which becomes trivial. This may be regarded as a drawback, however in some circumstances it leads to the appearance of {\it secondary characteristic classes}. The latter are much finer invariants associated to the Hopf algebra action, and in fact correspond to characteristic classes for the higher {\it algebraic} $K$-theory of $\Ac$ instead of the topological one. Since the treatment involves a fine control of the components of the cyclic cochains, we shall assume from now on that $\Lc$, $\ell^1$ are realized as algebras of (not necessarily bounded) operators on a Hilbert space. Moreover, the Hopf algebra $\Hc$ is now considered as a discrete algebra (fine bornology).
\begin{definition}
Let $\Ac$ be a complete bornological algebra, $\Hc$ a Hopf algebra with fine bornology. For $p\in\rr_+$, a $p$-summable $\Hc$-equivariant spectral triple $\Ec=(\Ac,\Hg,D)$ over $\Ac$ corresponds to the following data:
\begin{itemize}
\item A separable $\zz_2$-graded Hilbert space $\Hg=\Hg_+\oplus\Hg_-$ together with a bounded representation $\rho:\Ac\to \End(\Hg)$ of $\Ac$ into the algebra of bounded endomorphisms of even degree over $\Hg$;
\item A representation $g$ of the algebra $\Hc$ as \emph{not necessarily bounded} operators of even degree over $\Hg$ (recall $\Hc$ has fine bornology);
\item A selfadjoint unbounded operator $D$ of odd degree with compact resolvent, such that $(1+D^2)^{-p/2}$ is trace-class (Dirac operator);
\item The commutator $[D,\rho(a)]$ is a bounded endomorphism of $\Hg$ for any $a\in\Ac$;
\item The adjoint action of $\Hc$ on the Dirac operator is a bounded perturbation of $D$: 
\be
\sum g(h_{(0)})Dg(S(h_{(1)}))- \eps(h)D\ \mbox{bounded for any}\  h\in\Hc\ .
\ee
\end{itemize}
We say that an equivariant spectral triple $(\Ac,\Hg,D)$ is $p_+$-summable iff it is $(p+\eps)$-summable for any $\eps>0$.
\end{definition}
The $p$-summability assumption $(1+D^2)^{-p/2}$ trace-class implies that the heat kernel $\exp(-tD^2)$ is a matrix with rapid decay in a basis of eigenvectors of $D$ for any $t>0$. Morally, the previous auxiliary algebra $\Lc$ is generated by the operators $g(h)$, $D$, $\rho(a)$ and $\exp(-tD^2)$ for any $a\in\Ac$, $h\in\Hc$, and $\ell^1$ is the ideal generated by $\exp(-tD^2)$, consisting in matrices with rapid decay. The trace $\tau:\ell^1\to\cc$ corresponds to the ordinary operator trace, whose degree depends on the parity of the spectral triple:\\
\noindent i) Even case: the $\zz_2$-grading $\Hg=\Hg_+\oplus\Hg_-$ implies that $\Lc$ and $\ell^1$ split into even and odd parts. In matricial form one has
\be
\rho=\left( \begin{array}{cc}
 \rho_+ & 0 \\
 0  & \rho_- \end{array}\right)\ ,\quad D=\left( \begin{array}{cc}
 0 & D_- \\
 D_+  & 0 \end{array}\right)\ .
\ee
Then $\tau$ is the ordinary supertrace of operators $\Tr_s$. It is a map of even degree from $\ell^1$ to $\cc$.\\
\noindent ii) Odd case: one has $\Hg_+\cong\Hg_-$, and in matricial form the homomorphism $\rho$ and the Dirac operator decompose into
\be
\rho=\left( \begin{array}{cc}
 \al & 0 \\
 0  & \al \end{array}\right)\ ,\quad D=\left( \begin{array}{cc}
 0 & Q \\
 Q  & 0 \end{array}\right)\ .
\ee
Then $\tau$ evaluated on a trace-class operator of the form $\left( \begin{array}{cc}
 0 & x \\
 x  & 0 \end{array}\right)$ is equal to $\sqrt{2i}\Tr(x)$, and it vanishes on diagonal matrices. Then $\tau$ is a trace of odd degree from $\ell^1$ to $\cc$. The factor $\sqrt{2i}$ assures the compatibility of the Chern character with the cup-product on $K$-homology. It is also compatible with Bott periodicity in the bivariant context \cite{P3}.\\

We may assume the representation of $\Ac$ endowed with the norm $||\rho(a)||_{\infty}+||[D,\rho(a)]||_{\infty}$ (with $||\cdot||_{\infty}$ the operator norm), and the map $\rho:\Ac\to \rho(\Ac)$ bounded. Also, we suppose that $\Hc$ has fine bornology. The $p$-summability assumption actually implies the $\te$-summability property stated in definition \ref{dte2} (use for example the same kind of estimates as in the proof of proposition \ref{pvan}). Consequently, if $\Hc$ is a Hopf algebra with involutive antipode, any $p$-summable $\Hc$-equivariant spectral triple $\Ec=(\Ac,\Hg,D)$ of parity $i$ yields a characteristic map $\ch(\Ec): HE_*(\Ac)\to HP_{*+i}^{\eps}(\Hc)$ which retracts onto the JLO cocycle (\ref{jlo}). However, the $p$-summability assumption gives further estimates on the behaviour of the {\it cocycle} representing $\ch(\Ec)$.\\
Thus we fix an $\Hc$-equivariant spectral triple $\Ec=(\Ac,\Hg,D)$ of parity $i$. Recall that in the construction of $\ch(\Ec)$ (section \ref{shopf}), we consider the linear map $\rhoh\in\hom(\Ac\hotimes\Hc,\Lc)$ and the element $\Dh\in\hom(\Hc,\Lc)$ defined by
\be
\rhoh (a\otimes h)=h\cdot \rho(a)\ ,\quad \Dh(h)=h\cdot D\ ,
\ee
where $h\in\Hc$ acts on the operators of $\Hg$ by the adjoint representation. Then we combine $\rhoh$ and $\Dh$ into the superconnection $A=\rhoh+\Dh$ viewed as an odd element of the DG convolution algebra $\hom(\Bbe(\Ac)\hotimes \Cb^+(\Hc),\Lc)$. Here $\Bbe(\Ac)$ is the bar DG coalgebra of $\Ac$ and $\Cb^+(\Hc)$ is the trivially graded Fedosov coalgebra of cochains over $\Hc$ (itself viewed only as a coalgebra). The corresponding cocycle $\tau\ch(A)\in \hom(X(\Bbe(\Ac))\hotimes X(\Cb^+(\Hc)),\cc)$ is $\eps$-invariant, and after composition with the homotopy equivalence $cP_1:\Cb^*_{\eps}(\Hc)\to X_{\eps}(\Cb^+(\Hc))$ we get a chain map
\be
\chi(\rho,D):=\tau\ch(A)\circ cP_1\ :\ \Xb(\Bbe(\Ac))\cong \Ome\Ac \to \hom(\Cb^*_{\eps}(\Hc),\cc)\cong \Cb_*^{\eps}(\Hc)\ ,
\ee
from the $(b+B)$-complex of entire chains over $\Ac$ to the $(b+B)$-complex of periodic chains over the Hopf algebra ($S^2=\id$). Decompose $\chi$ into its components $(\chi^n_k)_{n,k\in\nn}$:
\be
\chi^n_k(\rho,D): \Om^n\Ac\to \Cb_k^{\eps}(\Hc)\ .
\ee
We now assume that the representation $g:\Hc\to\Lc$ {\it commutes} with the operators $\rho(a)$ and $[D,\rho(a)]$ for any $a\in\Ac$, that is $g(h)\rho(a)=\rho(a)g(h)$ and $g(h)[D,\rho(a)]=[D,\rho(a)]g(h)$ for any $h\in\Hc$. Then if we rescale $D$ by $tD$ for $t>0$, we have the following vanishing result at the limit $t\to 0$:
\begin{proposition}\label{pvan}
Let $\Ac$ be a complete bornological algebra, $\Hc$ a Hopf algebra with fine bornology such that $S^2=\id$. Let $\Ec=(\Ac,\Hg,D)$ be a $p$-summable $\Hc$-equivariant spectral triple such that the representation of $\Hc$ as operators on $\Hg$ commute with $\rho(a)$ and $[D,\rho(a)]$ for any $a\in\Ac$. Then the map $\chi^n_k(\rho,tD):\Om^n\Ac\to \Cb_k^{\eps}(\Hc)$ vanishes at the limit $t\to 0$ whenever $n+k>p$.
\end{proposition}
{\it Proof:} Let $A=\rhoh+t\Dh$ be the superconnection associated to $\Ec$ and $F\in \hom(\Bbe(\Ac)\hotimes \Cb^+(\Hc),\Lc)$ its curvature. Recall that the $X$-complex $X(\Cb^+(\Hc))$ is isomorphic to $\Cb^*(\Hc)$ as a vector space, and the homotopy equivalence $cP_1:\Cb^*_{\eps}(\Hc)\to X_{\eps}(\Cb^+(\Hc))$ (quotients by $\eps$-coinvariants) preserves the degree of Hopf cochains. Therefore, it is sufficient to show that the component of $\tau\ch(A)$
$$
\tau\ch(A)^n_k:\Om^n\Ac\hotimes X(\Cb^+(\Hc))^k\to\cc
$$
vanishes for $n+k>p$, where $X(\Cb^+(\Hc))^k$ is the space of $k$-cochains $\Cb^k(\Hc)=\Hc\hotimes\Hcb^{\hotimes n}$ (we forget the $\eps$-invariance). First, we use the computation (\ref{curva}) for the curvature $F$,
$$
F=t^2\Dh^2+t[\Dh,\rhoh]+ d(\rhoh+t\Dh)d(\rhoh+t\Dh)\ ,
$$
where we recall that the symbol $d$ of non-commutative differential forms has to be interpreted in a dual context for the Fedosov coalgebra $\Cb^+(\Hc)$, for example 
$$
d\Dh d\Dh\in \hom(\Cb^2(\Hc),\Lc)\ ,\quad d\Dh d\Dh(h^0\otimes\hb^1\otimes\hb^2)=\eps(h^0)\hb^1(D)\hb^2(D)\ ,
$$
and
$$
d\rhoh d\Dh\in\hom(\Ac\hotimes\Cb^2(\Hc),\Lc)\ ,\quad d\rhoh d\Dh(a\otimes h^0\otimes\hb^1\otimes\hb^2)=\eps(h^0)\hb^1(\rho(a))\hb^2(D)\ .
$$
From the definition of the equivariant spectral triple, $\Dh(\hb)=\hb(D)$ is a bounded operator for any $\hb$ in the augmentation ideal $\Hcb=\ker \eps$, therefore the symbol $d\Dh$ is always bounded. Moreover, by hypothesis $\rho$ commutes with the representation of $\Hc$, hence one has $\rhoh(a\otimes h)=h(\rho(a))=\eps(h)\rho(a)$ and subsequently $d\rhoh=0$. Since $\Hc$ commutes also with $[D,\rho(a)]$, one has $d[\Dh,\rhoh]=0$.\\
We have to evaluate explicitly the component $\tau\ch(A)^n_k$ on a $n$-form over $\Ac$. Again, the computation is identical to section \ref{sbiv}. For simplicity, we will focus only on the part $\tau e^{-F}$ of $\tau\ch(A)$. This amounts to restrict to the particular case $k=2m$, and evaluate on the $n$-form $da_1\ldots da_n\in\Om^n\Ac$. Therefore, we have
$$
\tau\ch(A)^n_{2m}(da_1\ldots da_n)=\tau e^{-F}(a_1\otimes\ldots \otimes a_n)\ \in \hom(\Cb^{2m}(\Hc),\cc)\ ,
$$
so that we just want to evaluate $\tau e^{-F}$ on $\Bb_n(\Ac)\hotimes \Cb^{2m}(\Hc)$. To keep the notations simple, we forget the hat over $\rhoh$ and $\Dh$ and write just $\rho$ and $D$. Hence in the manipulations below, we must keep in mind that $\rho$, $D$ and $dD$ are in fact operator-valued functions over the fine (=discrete) bornological space $\Hc$, evaluated on some elements $h^0,\hb^1,\ldots,\hb^{2m}$ or their coproducts. Note that the condition $(1+D^2)^{-p/2}$ is trace-class holds when $D$ is evaluated on any element $h\in \Hc$, since $h(D)$ is a bounded perturbation of the Dirac operator. The Fedosov exponential is given by formula (\ref{exph})
$$
e^{-F}=\sum_{n\ge 0} \pm \int_{\Delta_n}e^{-s_0t^2D^2}H_1e^{-s_1t^2D^2}H_2\ldots e^{-s_{n-1}t^2D^2}H_ne^{-s_nt^2D^2}\ ,
$$
where the $H_i$'s are taken among the three possibilities $t[D,\rho]$, $t^2dDdD$ and $t^2d(D^2)$ (recall $d\rho=0$ and $d[D,\rho]=0$). Since $dD$ is a bounded operator on $\Hg$, as well as the commutator $[D,\rho]$, one sees that only the last possibility $d(D^2)=[dD,D]$ may be unbounded. Applying the trace $\tau$, proportional to the operator (super)trace $\Tr$ in Hilbert space, we want to control the behaviour of
$$
\int_{\Delta_n}\Tr(e^{-s_0t^2D^2}H_1e^{-s_1t^2D^2}H_2\ldots e^{-s_{n-1}t^2D^2}H_ne^{-s_nt^2D^2})
$$
when $t\to 0$. Consider the integrand at a point of the $n$-simplex $(s_0,\ldots, s_n)$, $\sum_is_i=1$, with all $s_i\ne 0$. Using H\"older's inequality for the norms $|| x ||_{s^{-1}}=(\Tr|x|^{1/s})^s$, we have
\be
\Tr|e^{-s_0t^2D^2}H_1e^{-s_1t^2D^2}H_2\ldots e^{-s_{n-1}t^2D^2}H_ne^{-s_nt^2D^2}| \le \prod_{i=0}^n || H_i e^{-s_it^2D^2}||_{s_i^{-1}}\ .\label{meuh}
\ee
We focus on a given $i$. If $H_i$ is one of the bounded operators $t[D,\rho]$ or $t^2dDdD$, the $p$-summability assumption that $(1+D^2)^{-p/2}$ is trace-class implies
\beq
|| H_i e^{-s_it^2D^2}||_{s_i^{-1}} &\le& || H_i ||_{\infty}||e^{-s_it^2D^2}||_{s_i^{-1}} \non\\
&\le& || H_i ||_{\infty} (\Tr (e^{-t^2D^2}))^{s_i}\non\\
&\le& || H_i ||_{\infty} \big(\Tr ((t^2+t^2D^2)^{-p/2}(t^2+t^2D^2)^{p/2}e^{-t^2D^2})\big)^{s_i}\non\\
&\le& || H_i ||_{\infty} (t^{-p}\Tr (1+D^2)^{-p/2} ||(t^2+t^2D^2)^{p/2}e^{-t^2D^2}||_{\infty})^{s_i}\ ,\non
\eeq
where $||\cdot||_{\infty}$ is the operator norm. Since $||(t^2+t^2D^2)^{p/2}e^{-t^2D^2})||_{\infty}$ is bounded by the maximum of the numerical function $x\mapsto (t^2+x^2)^{p/2}e^{-x^2}$, we deduce there is a bounded function $C_t$ as $t\to 0$ such that
$$
|| H_i e^{-s_it^2D^2}||_{s_i^{-1}} \le (C_tt^{-p})^{s_i} || H_i ||_{\infty}  \ .
$$
Let us now consider the unbounded term $t^2[dD,D]=t^2(dDD-DdD)$. For example if $H_i$ is equal to $t^2 dD D$, we may write
$$
|| H_i e^{-s_it^2D^2}||_{s_i^{-1}} \le ||tdD||_{\infty}||tD e^{-s_it^2D^2}||_{s_i^{-1}}\ ,
$$
and using $p$-summability as before,
\beq
||tD e^{-s_it^2D^2}||_{s_i^{-1}} &=& \big(\Tr((t^2+t^2D^2)^{-p/2}(t^2+t^2D^2)^{p/2}|tD|^{1/s_i}e^{-t^2D^2})\big)^{s_i}\non\\
&\le& \big( t^{-p}\Tr(1+D^2)^{-p/2}|| (t^2+t^2D^2)^{p/2}|tD|^{1/s_i}e^{-t^2D^2}||_{\infty}\big)^{s_i}\ . \non
\eeq
For $t$ sufficiently small and $s_i\le 1$, the maximum of the function $x\mapsto \big((t^2+x^2)^{p/2}x^{1/s_i}e^{-x^2}\big)^{s_i}$ is bounded by a term proportional to $1/\sqrt{s_i}$, so that we can write
$$
||tD e^{-s_it^2D^2}||_{s_i^{-1}} \le {C'}_t\frac{t^{-ps_i}}{\sqrt{s_i}}\ ,
$$
where ${C'}_t$ is a bounded function of $t$. This expression is integrable over $s_i$ near zero, hence we can integrate (\ref{meuh}) over the simplex $\Delta_n$ and estimate for small $t$
$$
\int_{\Delta_n}\Tr|e^{-s_0t^2D^2}H_1e^{-s_1t^2D^2}H_2\ldots e^{-s_{n-1}t^2D^2}H_ne^{-s_nt^2D^2}| \le C\,t^{-p}\prod_{i}||A_i||_{\infty}\ ,
$$
where $A_i$ are bounded operators chosen among $t[D,\rho]$, $t^2dDdD$ and $tdD$. Thus the product of $||A_i||_{\infty}$'s carries one power of $t$ for each $dD$ and $\rho$, so that $\tau e^{-F}$ evaluated on an element of $\Bb_n(\Ac)\hotimes \Cb^k(\Hc)$ behaves like $t^{n+k-p}$, hence is vanishes at the limit $t\to 0$ when $n+k>p$. The same discussion holds for the other components of $\tau\ch(A)$, which also involve the exponential of $F$, with the same kind of estimates.\cqfd\\

Under reasonable analytical hypotheses on the spectral triple $(\Ac,\Hg,D)$ (see \cite{CM95}), one shows that the cochain $\chi^n_k(\rho,tD)$ evaluated on an element of $\Om^n\Ac\hotimes \Cb^k_{\eps}(\Hc)$ has an asymptotic expansion when $t\to 0$ of the form
\be
\chi^n_k(\rho,tD)=\sum_{l,m\in\nn}t^{p_m}(\log t)^l (\al^n_k)_{l,m}\ ,\label{asy}
\ee
where the $p_m$'s are complex numbers with real part bounded below, and $(\al^n_k)_{l,m}$ are bounded maps $\Om^n\Ac\to \Cb^{\eps}_k(\Hc)$. The homotopy invariance of $\ch(\Ec)$ implies that the cohomology class of the cocycle $\chi(\rho,tD)$ is independent of $t$, hence corresponds to the cohomology class of the constant term $l=p_m=0$ in the above expansion, or equivalently to the finite part $\Pf\chi(\rho,tD)$. The preceding proposition shows that the $(n,k)$-component vanishes whenever $n+k>p$, therefore we actually get a {\it periodic} cyclic cocycle over $\Ac$ (i.e. finite-dimensional). For $k=0$, we deduce that the periodic cocycle
\be
(\Pf\chi^n_0(\rho,tD))_{0\le n\le p}\ \in\ \hom(\Om\Ac, \cc)
\ee
is a representative of the $\cc$-valued characteristic map $\ch^{\cc}(\Ec)$. This yields the local formula of Connes-Moscovici \cite{CM95}, corresponding to the retraction of the JLO cocycle for $p$-summable spectral triples. For $k>0$ and $n+k\le p$, the collection 
\be
\psi(\rho,D):=(\Pf\chi^n_k(\rho,tD))_{k>0,n+k\le p}\ \in\ \hom(\Om\Ac\hotimes \Cb^*_{\eps}(\Hc),\cc)
\ee
is a local representative of the reduced characteristic map $\chb(\Ec)$. Its periodic cyclic cohomology class $HP_*(\Ac)\to \HPb_{*+i}^{\eps}(\Hc)$ vanishes, but by considering the natural filtrations on the cyclic bicomplexes, we are going to exhibit non-trivial characteristic maps in the {\it non-periodic} cyclic cohomology of $\Ac$.\\

We first recall the non-periodic version of the cyclic homology of $\Ac$ and its relationship with the periodic and negative theories \cite{L}. We know that the cyclic bicomplex (\ref{bcp}) is isomorphic to the $(b,B)$-complex of non-commutative differential forms over the {\it non-unital} algebra $\Ac$
\be
\xymatrix{
 & \ar[d]_{b}  & \ar[d]_{b} \ar@{--}@<7ex>[dddd] & \ar[d]_{b}   & \ar[d]_{b}   & \ar[d]_{b}   \\
 &  \Om^4\Ac \ar[l]_{B} \ar[d]_{b} & \Om^3\Ac \ar[l]_{B} \ar[d]_{b} & \Om^2\Ac \ar[l]_{B} \ar[d]_{b} & \Om^1\Ac \ar[l]_{B} \ar[d]_{b} & \Ac \ar[l]_{B} \\
 & \Om^3\Ac \ar[l]_{B} \ar[d]_{b} & \Om^2\Ac \ar[l]_{B} \ar[d]_{b} & \Om^1\Ac \ar[l]_{B} \ar[d]_{b} & \Ac \ar[l]_{B} &  \\
 & \Om^2\Ac \ar[l]_{B} \ar[d]_{b} & \Om^1\Ac \ar[l]_{B} \ar[d]_{b} & \Ac \ar[l]_{B} & & \\
 & \Om^1\Ac \ar[l]_{B} \ar[d]_{b} & \Ac \ar[l]_{B} &  &  &  \\
 & \Ac \ar[l]_{B} &  &   &   &  }
\ee
with $\Om^n\Ac=\Act\hotimes\Ac^{\hotimes n}$. We draw a vertical dashed line in order to split the bicomplex into a subcomplex on the left and a quotient complex on the right. By definition, the {\it non-periodic} cyclic homology of $\Ac$ is computed by the quotient bicomplex $CC_*(\Ac)$ situated on the right of the dashed line. The component of degree $n$, for $n\in\zz$, of this bicomplex corresponds to the diagonal
\be
CC_n(\Ac)=\bigoplus_{i\ge 0} \Om^{n-2i}\Ac\ ,
\ee
with the convention that $\Om^k\Ac=0$ for $k<0$. The total complex $CC_*(\Ac)$ is endowed with the boundary $b+B: CC_n(\Ac)\to CC_{n-1}(\Ac)$, after suppressing the map $B$ acting on $\Om^n\Ac\subset CC_n(\Ac)$. The corresponding cyclic homology is denoted
\be
HC_n(\Ac):=H_n(CC_*(\Ac),b+B)\ ,\quad \forall n\in\zz\ .
\ee
By construction one has $HC_n(\Ac)=0$ for $n<0$. We can also compute $HC_n(\Ac)$ by introducing the {\it Hodge filtration} of the bicomplex $\Om\Ac$ \cite{CQ1}. It is given by the decreasing family of subcomplexes $F_n(\Ac)\subset \Om\Ac$, $F_{n+1}(\Ac)\subset F_n(\Ac)$:
\be
F_n(\Ac):= b\Om^{n+1}\Ac \oplus \bigoplus_{k> n}\Om^k\Ac\ ,\quad \forall n\in\zz\ .
\ee
$F_n(\Ac)$ is clearly stable under $b$ and $B$, so that the quotient $\Om\Ac/F_n(\Ac)$ is a $\zz_2$-graded complex for the total boundary $b+B$, and one has
\be
HC_n(\Ac)=H_i(\Om\Ac/F_n(\Ac), b+B)\ ,\quad i\equiv n\ \mathrm{mod}\ 2\ ,\quad \forall n\in\zz\ .
\ee
By deleting the dashed line, one recovers the periodic complex $CP_*(\Ac)$, whose degree $n\in\zz$ is the diagonal corresponding to the direct product
\be
CP_n(\Ac)=\prod_{i\in\zz}\Om^{n+2i}\Ac\ .
\ee
Periodicity means $CP_n(\Ac)\cong CP_{n+2}(\Ac)$ for any $n\in\zz$. The corresponding periodic cyclic homology is $HP_n(\Ac)$. The subcomplex situated on the left of the dashed line is the {\it negative cyclic bicomplex} $CC^-_*(\Ac)$. Its component of degree $n\in\zz$ is given by the direct product (there is a conventional shift by two in the degree)
\be
CC^-_n(\Ac)=\prod_{i\ge 0}\Om^{n+2i}\Ac\ .
\ee
In particular one has $CC^-_n(\Ac)\cong CP_n(\Ac)$ for $n\le 0$. The total boundary $b+B$ carries $CC^-_n(\Ac)$ into $CC^-_{n-1}(\Ac)$, whence the definition of negative cyclic homology:
\be
HC^-_n(\Ac)=H_n(CC^-_*(\Ac),b+B)\ ,\quad \forall n\in\zz\ .
\ee
One has $HC^-_n(\Ac)\cong HP_n(\Ac)$ for $n\le 0$. By definition, the three complexes fit into a short exact sequence
\be
0 \longrightarrow CC^-_*(\Ac) \stackrel{I}{\longrightarrow} CP_*(\Ac) \stackrel{S}{\longrightarrow} CC_*(\Ac)[2] \longrightarrow 0 
\ee
where $S$ is the periodicity shift, and $(CC_*(\Ac)[2])_n=CC_{n-2}(\Ac)$. The associated homology long exact sequence relates the non-periodic and negative theories to the periodic one:
\be
\ldots \stackrel{S}{\longrightarrow} HC_{n-1}(\Ac) \stackrel{B}{\longrightarrow} HC^-_n(\Ac) \stackrel{I}{\longrightarrow} HP_n(\Ac) \stackrel{S}{\longrightarrow} HC_{n-2}(\Ac) \stackrel{B}{\longrightarrow} HC^-_{n-1}(\Ac) \stackrel{I}{\longrightarrow} \ldots \label{exa}
\ee
Indeed, it is easy to see that the connecting map of the exact sequence is given by the operator $B$ crossing the dashed line in the cyclic bicomplex.\\

Now consider the Hopf algebra $\Hc$. When it is viewed only as a coalgebra, the normalized bicomplex of chains $\Cb_*(\Hc)=\hom(\Cb^*(\Hc),\cc)$, with $\Cb^n(\Hc)=\Hc\hotimes\Hcb^{\hotimes n}$, plays (at a formal level) a role analogous to the augmented bicomplex $\Omt\Bc$ for the dual algebra $\Bc=\hom(\Hcb,\cc)$. When the full Hopf algebra structure is taken into account, with involutive antipode, we restrict to the subcomplex of $\eps$-invariant chains $\Cb^{\eps}_*(\Hc)$. By remark \ref{rcm}, its component $\Cb^{\eps}_n(\Hc)$ of degree $n$ is isomorphic to $\hom(\Hcb^{\hotimes n},\cc)$. We know that this bicomplex splits in even degree into the direct sum of $\cc=\Cb_0^{\eps}(\Hc)$ and the ``interesting'' part which yields is the {\it reduced} cyclic bicomplex. Therefore, the reduced cyclic bicomplex of Hopf chains is given by
\be
\xymatrix{
 & \ar[d]_{b}  & \ar[d]_{b}  & \ar[d]_{b}   & \ar[d]_{b}   & \ar[d]_{b}   \\
 &  \Cb_4^{\eps}(\Hc) \ar[l]_{B} \ar[d]_{b} & \Cb_3^{\eps}(\Hc) \ar[l]_{B} \ar[d]_{b} & \Cb_2^{\eps}(\Hc) \ar[l]_{B} \ar[d]_{b} & \Cb_1^{\eps}(\Hc) \ar[l]_{B} \ar[d] & 0 \ar[l] \\
\ar@{--}@<-4ex>[rrrr] & \Cb_3^{\eps}(\Hc) \ar[l]_{B} \ar[d]_{b} & \Cb_2^{\eps}(\Hc) \ar[l]_{B} \ar[d]_{b} & \Cb_1^{\eps}(\Hc) \ar[l]_{B} \ar[d] & 0 \ar[l] &  \\
 & \Cb_2^{\eps}(\Hc) \ar[l]_{B} \ar[d]_{b} & \Cb_1^{\eps}(\Hc) \ar[l]_{B} \ar[d] & 0 \ar[l] & & \\
 & \Cb_1^{\eps}(\Hc) \ar[l]_{B} \ar[d] & 0 \ar[l] &  &  &  \\
 & 0 \ar[l] &  &   &   &  }\label{hbcp}
\ee
Here we draw an horizontal dashed line in order to separate the cyclic bicomplex into new (non-periodic) complexes. We shall call the quotient complex above the line the {\it positive} cyclic bicomplex $CC^{\eps,+}_*(\Hc)$. Its component of degree $n\in\zz$ is the direct product
\be
CC^{\eps,+}_n(\Hc)=\prod_{i\ge 0}\Cb_{2i-n}^{\eps}(\Hc)\ ,\quad \forall n\in\zz\ ,
\ee
with the convention that $\Cb_k^{\eps}(\Hc)=0$ for $k\le 0$. The total boundary $b+B: CC^{\eps,+}_n(\Hc)\to CC^{\eps,+}_{n-1}(\Hc)$ is obtained by truncating the Hochschild operator $b$ acting on $\Cb_{-n}^{\eps}(\Hc)$. The corresponding positive cyclic homology is therefore
\be
HC^{\eps,+}_n(\Hc)=H_n(CC^{\eps,+}_*(\Hc),b+B)\ ,\quad \forall n\in\zz\ .
\ee
The latter can also be computed by means of an analogue of the Hodge filtration. We introduce the following increasing filtration $G_n(\Hc)\subset \Cb_*^{\eps}(\Hc)$:
\be
G_n(\Hc)=B\big(\Cb^{\eps}_{n-1}(\Hc)\big)\oplus \bigoplus_{k<n}\Cb^{\eps}_k(\Hc)\ ,\quad \forall n\in\zz\ .
\ee
Then $G_n(\Hc)$ is a $\zz_2$-graded subcomplex of $\Cb^{\eps}_*(\Hc)$ and one has
\be
HC^{\eps,+}_n(\Hc)=H_i(\Cb^{\eps}_*(\Hc)/G_{-n}(\Hc),b+B)\ ,\quad i\equiv n\ \mathrm{mod}\ 2\ ,\quad \forall n\in\zz\ .
\ee
As usual, the reduced periodic complex $\bar{CP}^{\eps}_*(\Hc)$ is the total complex (\ref{hbcp}) (direct products) obtained by ignoring the dashed line. One has $\HPb^{\eps}_n(\Hc)\cong \HPb^{\eps}_{n+2}(\Hc)$ for any $n\in\zz$, and $\HPb^{\eps}_n(\Hc)\cong HC^{\eps,+}_n(\Hc)$ whenever $n\ge -1$. The subcomplex below the dashed line will be denoted $CN^{\eps}_*(\Hc)$, with a shift by two in degrees
\be
CN^{\eps}_n(\Hc)=\bigoplus_{i\ge 0} \Cb_{-2i-n}^{\eps}(\Hc)\ ,\quad \forall n\in\zz\ .
\ee
One has $CN^{\eps}_n(\Hc)=0$ for any $n\ge 0$. Therefore the corresponding homology
\be
HN^{\eps}_n(\Hc)=H_n(CN^{\eps}_*(\Hc),b+B)
\ee
vanishes in positive degrees. By construction these total complexes fit into a short exact sequence
\be
0 \longrightarrow CN^{\eps}_*(\Hc)[-2] \stackrel{S}{\longrightarrow} \bar{CP}^{\eps}_*(\Hc) \stackrel{I}{\longrightarrow} CC^{\eps,+}_*(\Hc) \longrightarrow 0 \ ,
\ee
whose connecting map is given by the Hochschild operator $b$ crossing the dashed line in (\ref{hbcp}), whence the long exact sequence relating the homologies
\be
\stackrel{I}{\longrightarrow} HC^{\eps,+}_{n+1}(\Hc) \stackrel{b}{\longrightarrow} HN^{\eps}_{n+2}(\Hc) \stackrel{S}{\longrightarrow} \HPb^{\eps}_n(\Hc) \stackrel{I}{\longrightarrow} HC^{\eps,+}_n(\Hc) \stackrel{b}{\longrightarrow} HN^{\eps}_{n+1}(\Hc) \stackrel{S}{\longrightarrow}
\ee

Let $\Ec=(\Ac,\Hg,D)$ be a $p_+$-summable $\Hc$-equivariant spectral triple such that $\Hc$ commutes with the representation $\rho:\Ac\to \End(\Hg)$ as well as with the commutators $[D,\rho(a)]$. We suppose that $p$ is an integer congruent to the parity of the triple modulo 2. If the characteristic map $\chi(\rho,tD)$ verifies the asymptotic expansion (\ref{asy}), its finite part for $t\to 0$ yields the chain map $\psi(\rho,D)$ given in components by
\be
\psi^n_k(\rho,D)=\Pf\chi^n_k(\rho,tD)\ :\  \Om^n\Ac\to \Cb_k^{\eps}(\Hc)\ .
\ee
The $p_+$-summability assumption in connection with proposition \ref{pvan} implies that for any $\eps>0$, $\psi^n_k(\rho,D)$ vanishes whenever $n+k>p+\eps$. Since $p$ is an integer, $\psi^n_k(\rho,D)=0$ whenever $n+k>p$. Consequently, the cocycle $\psi$ carries the degree $n$ of the Hodge filtration $F_n(\Ac)$ into the subcomplex $G_{p-n}(\Hc)$ for any $n\in\zz$, whence a well-defined characteristic map $\psi(\Ec): HC_n(\Ac)\to HC^{\eps,+}_{n-p}(\Hc)$ for any $n\ge 0$.
\begin{theorem}\label{t3}
Let $\Ac$ be a complete bornological algebra, and $\Hc$ a Hopf algebra with fine bornology whose antipode is involutive. Let $p$ be an integer and $\Ec=(\Ac,\Hg,D)$ a $p_+$-summable $\Hc$-equivariant spectral triple of degree $i\equiv p \mod 2$. We assume that $\Hc$ commutes with the representation $\rho:\Ac\to \End(\Hg)$ and the commutators $[D,\rho(a)]$, and that the character $\chi(\rho,tD)$ satisfies the asymptotic expansion (\ref{asy}). Then the finite part $\psi(\rho,D)=\Pf \chi(\rho,tD)$ defines a secondary characteristic class
\be
\psi(\Ec): HC_n(\Ac)\to HC^{\eps,+}_{n-p}(\Hc)\ ,\quad \forall n\ge 0\ .
\ee
It vanishes on the image of the periodicity map $S:HP_{n+2}(\Ac)\to HC_n(\Ac)$, and is invariant with respect to bounded perturbations of the Dirac operator $D$.
\end{theorem}
{\it Proof:} The fact that the periodic cyclic cohomology class of the cocycle $\psi(\rho,D)$ vanishes implies the triviality of $\psi(\Ec):HC_n(\Ac)\to HC^{\eps,+}_{n-p}(\Hc)$ on the image of $HP_{n+2}(\Ac)$. \\
The homotopy invariance with respect to bounded perturbations of the Dirac operator is proved using the Chern-simons transgressions. If $D_0$ and $D_1$ are Dirac operators with bounded difference, the interpolation $D_u=(1-u)D_0+uD_1$ is a family of Dirac operators with parameter $u\in[0,1]$. We work as usual in the complex $\hom(X(\Bbe(\Ac))\hotimes X(\Cb^+(\Hc)),\Lc\hotimes\Om[0,1])$, where the differential $d_u$ on $\Om[0,1]$ is taken into account. The curvature of the superconnection $A=\rhoh+t\Dh_u$ reads (cf. proof of proposition \ref{pvan})
$$
F=td_u\Dh_u+ t^2\Dh_u^2 + t[\Dh_u,\rhoh] + t^2d\Dh_ud\Dh_u\ \in \hom(\Bbe(\Ac)\hotimes\Cb^+(\Hc),\Lc\hotimes\Om[0,1])\ ,
$$
because $d\rhoh=0$. Let $cs(\rho,tD_u)$ be the component of $\tau\ch(A)$ proportional to $du$, and $\chi(\rho,tD_u)$ the other component. The cocycle condition gives as usual an identity
$$
d_u \chi(\rho,tD_u)=\pm cs(\rho,tD_u)\circ\d\ ,
$$
where $\d$ is the total boundary of the complex $\Om\Ac\hotimes \Cb_{\eps}^*(\Hc)$. Taking the finite part as $t\to 0$ gives a transgression of $\psi$. The boundedness of the variation $d_u\Dh_u=du(\Dh_1-\Dh_0)$ implies as in the proof of proposition \ref{pvan}, that the components $\Pf cs^n_k(\rho,tD_u):\Om^n\Ac\to \Cb^{\eps}_k(\Hc)$ vanish whenever $n+k>p$ (in fact we can achieve the better estimate $n+k\ge p$ because of the factor $t$ in front of $d_u\Dh_u$), hence one gets a transgression of $\psi$ as a chain map $CC_n(\Ac)\to CC^{\eps,+}_{n-p}(\Hc)$. Details are left to the reader. \cqfd\\
\begin{remark}\textup{
It is not true that the secondary characteristic class is homotopy invariant with respect to the homomorphism $\rho:\Ac\to \End(\Hg)$. This would require to stabilize by the the image of the periodicity map $S$, which annihilates $\psi(\Ec)$. On the other hand, $\psi(\Ec)$ is certainly invariant for more general homotopies of $D$ (not necessarily bounded), but the determination of the correct class of homotopies allowed may be more difficult.}
\end{remark}

In fact cyclic homology is the right target for the Chern character defined on the higher algebraic $K$-theory of Quillen \cite{Kar}. We are going to combine the preceding theorem with previous constructions of Karoubi \cite{Kar} and Connes-Karoubi \cite{CK} giving rise to secondary characteristic classes in algebraic $K$-theory. We follow essentially the exposition of Loday \cite{L}.\\
Let $\Ac$ be a unital Banach algebra, endowed with the bounded bornology. For any $q\in\nn$, the group of invertible matrices $GL_q(\Ac)$ is a topological group. The natural inclusions $GL_q(\Ac)\hookrightarrow GL_{q+1}(\Ac)$ are continuous, and $GL(\Ac)$ denotes the direct limit
\be
GL(\Ac)=\limind GL_q(\Ac)\ .
\ee
We can associate to this topological group a classifying space $BGL(\Ac)$ with abelian $\pi_1$. By definition, the topological $K$-theory of $\Ac$ corresponds to the homotopy groups
\be
\Kt_n(\Ac)=\pi_n(BGL(\Ac))\ ,\quad \forall n\ge 1\ ,
\ee
whereas for $n=0$, the abelian group $\Kt_0(\Ac)$ was introduced in section \ref{shopf}. The Bott periodicity theorem of topological $K$-theory \cite{Bl} states that $\Kt_n(\Ac)$ is isomorphic to $\Kt_{n+2}(\Ac)$ for any $n$. \\
We can also forget the topology of $GL(\Ac)$ and consider it as a discrete group $GL^{\delta}(\Ac)$, with classifying space $BGL^{\delta}(\Ac)$. Quillen's $+$-construction \cite{R} yields a topological space $BGL^{\delta}(\Ac)^+$ with abelian $\pi_1$. The higher algebraic $K$-theory of $\Ac$ corresponds to the homotopy groups:
\be
K_n(\Ac)=\pi_n(BGL^{\delta}(\Ac)^+)\ ,\quad \forall n\ge 1\ .
\ee
The identity homomorphism $GL^{\delta}(\Ac)\to GL(\Ac)$ is continuous and induces a classifying map $BGL^{\delta}(\Ac)\to BGL(\Ac)$ which factors through $BGL^{\delta}(\Ac)^+$, and gives rise to a homotopy fibration \cite{Kar}
\be
BGL^{\mathrm{rel}}(\Ac)\to BGL^{\delta}(\Ac)^+\to BGL(\Ac)\ .
\ee 
By definition, the relative $K$-theory groups are the homotopy groups of the corresponding homotopy fiber
\be
\Kr_n(\Ac)=\pi_n(BGL^{\mathrm{rel}}(\Ac))\ ,\quad \forall n\ge 1\ .
\ee
The long exact sequence of homotopy groups associated to the homotopy fibration $BGL^{\delta}(\Ac)^+\to BGL(\Ac)$ implies the exact sequence relating the topological and algebraic $K$-theories of $\Ac$:
\be
\ldots \longrightarrow \Kt_{n+1}(\Ac) \longrightarrow \Kr_n(\Ac) \longrightarrow K_n(\Ac) \longrightarrow \Kt_n(\Ac) \longrightarrow \Kr_{n-1}(\Ac) \longrightarrow \ldots
\ee
The Chern character on topological $K$-theory (section \ref{shopf}) yields an additive map $\cht_n:\Kt_n(\Ac)\to HP_n(\Ac)$ compatible with Bott periodicity of $K$-theory and the natural periodicity of $HP_n(\Ac)$. Connes and Karoubi construct in \cite{CK} a {\it relative Chern character} on $\Kr(\Ac)$ with values in non-periodic cyclic homology:
\be
\chr_n: \Kr_n(\Ac)\to HC_{n-1}(\Ac)\ ,\quad \forall n\ge 1\ .
\ee
There is also defined a {\it negative Chern character} \cite{J} on algebraic $K$-theory with values in negative cyclic homology:
\be
\ch^-_n: K_n(\Ac)\to HC^-_n(\Ac)\ ,\quad \forall n\ge 1\ .
\ee
By results of Connes and Karoubi \cite{CK,L}, the Chern characters relate in a commutative diagram the $K$-theory exact sequence (\ref{exa}) and the long exact sequence involving negative cyclic homology:
\be
\xymatrix{
 \Kt_{n+1}(\Ac) \ar[d]^{\cht_{n+1}} \ar[r] & \Kr_n(\Ac) \ar[d]^{\chr_{n}} \ar[r] & K_n(\Ac) \ar[d]^{\ch^-_{n}} \ar[r] & \Kt_n(\Ac) \ar[d]^{\cht_{n}} \ar[r] &  \Kr_{n-1}(\Ac) \ar[d]^{\chr_{n-1}}  \\
 HP_{n+1}(\Ac) \ar[r]^{S} & HC_{n-1}(\Ac) \ar[r]^{B} & HC^-_n(\Ac) \ar[r]^{I} & HP_n(\Ac) \ar[r]^{S} & HC_{n-2}(\Ac)  }
\ee
In fact the same discussion holds for unital Fr\'echet algebras, see \cite{CK,Kar}.\\

If $\Ec=(\Ac,\Hg,D)$ is a $p_+$-summable $\Hc$-equivariant spectral triple over a unital Banach algebra $\Ac$ for some integer $p$ as in theorem \ref{t3}, we may compose the characteristic map $\psi(\Ec)$ with the Chern character on relative $K$-theory to get a secondary characteristic class
\be
\begin{CD}
\Kr_{n+1}(\Ac) @>{\chr_{n+1}}>> HC_n(\Ac) @>{\psi(\Ec)}>> HC^{\eps,+}_{n-p}(\Hc)
\end{CD}
\ee
which vanish on the image of the homomorphism $\Kt_{n+2}(\Ac)\to \Kr_{n+1}(\Ac)$ in view of the above commutative diagram. The exactness of the $K$-theory sequence therefore implies that the secondary class is actually defined on the image of the homomorphism $\Kr_{n+1}\to K_{n+1}(\Ac)$. 
\begin{corollary}\label{cpair}
Let $\Ac$ be a unital Banach (or Fr\'echet) algebra, and $\Hc$ a Hopf algebra with fine bornology and involutive antipode. For any integer $p$, let $\Esf^{\Hc}_p(\Ac,\cc)$ be the set of $p_+$-summable $\Hc$-equivariant spectral triples verifying the conditions of theorem \ref{t3}. Then there is a bilinear pairing with values in the positive cyclic homology of $\Hc$
\be
\Ka_{n+1}(\Ac)\times \Esf^{\Hc}_p(\Ac,\cc)\to HC^{\eps,+}_{n-p}(\Hc)\ ,\quad \forall n\ge 0\ ,
\ee
where $\Ka_*(\Ac)$ denotes the kernel of the homomorphism $K_*(\Ac)\to\Kt_*(\Ac)$. This pairing is invariant with respect to bounded perturbations of the Dirac operator contained in the corresponding spectral triples. \cqfd
\end{corollary}
\begin{remark}\textup{
Connes and Karoubi construct in \cite{CK} a bilinear pairing between the algebraic $K$-theory of degree $p+1$ and the group generated by $p_+$-summable Fredholm modules (spectral triples) over $\Ac$,
\be
K_{p+1}(\Ac)\times \Esf_p(\Ac,\cc)\to \cc^{\times}\ ,\label{CK}
\ee
with values in the multiplicative abelian group of non-zero complex numbers. These ``higher regulators'' and our characteristic classes are not obviously related a priori. In particular the Connes-Karoubi pairing involves the algebraic $K$-theory of degree $p+1$, whereas our secondary characteristic classes are defined on the kernels $\Ka_n(\Ac)=\ker(K_n(\Ac)\to \Kt_n(\Ac))$ in all degrees $n\ge 1$. These kernels may be considered as retaining the ``purely algebraic part'' of algebraic $K$-theory, and the pairings detect invariants of $\Esf^{\Hc}_p(\Ac,\cc)$ which are not related to the topology of $\Ac$. The remarkable fact is that these pairings are nevertheless stable with respect to (at least bounded) homotopies of Dirac operators. This will be interpreted in section \ref{sBRS} in relation with BRS cohomology in Quantum Field Theory.}
\end{remark}
\begin{remark}\textup{
Taking the finite part $\Pf$ of the Chern-Simons terms used in the proof of proposition \ref{ptriv} yields a transgression $\varphi$ of the periodic chain map $\psi(\rho,D)=(b+B)\varphi\pm \varphi(b+B)$. The vanishing properties of $\psi$ in appropriate degrees implies that $\varphi$ induces a characteristic map $HC^-_{n+1}(\Ac)\to HC^{\eps,+}_{n-p}(\Hc)$ for any integer $n$ (including negative values of $n$), which fits into the commutative diagram
\be
\xymatrix{
\Kr_{n+1}(\Ac) \ar[r] \ar[d]_{\chr_{n+1}} & K_{n+1}(\Ac) \ar[d]^{\ch^-_{n+1}} \\
HC_n(\Ac) \ar[r]^{B} \ar[dr]_{\psi(\Ec)} & HC^-_{n+1}(\Ac) \ar[d]^{\varphi(\Ec)} \\
 & HC^{\eps,+}_{n-p}(\Hc) }
\ee
Hence $\varphi(\Ec)$ enables to extend (non canonically) the pairing of corollary \ref{cpair} to all the algebraic $K$-theory of $\Ac$. For negative values of $n$, one has $HC^-_{n+1}(\Ac)\cong HP_{n+1}(\Ac)$ and in fact the pairing involves the topological $K$-theory:
\be
\begin{CD}
\Kt_{n+1}(\Ac) @>{\cht_{n+1}}>> HP_{n+1}(\Ac) @>{\varphi(\Ec)}>> HC^{\eps,+}_{n-p}(\Hc)\ ,\quad \forall n<0\ .
\end{CD}
\ee
}
\end{remark}

\section{BRS cohomology}\label{sBRS}

The situation of $\Hc$-equivariant spectral triples fulfilling the requirements of theorem \ref{t3} arises naturally in connection with Yang-Mills theories in non-commutative geometry. There is a canonical way to build an equivariant spectral triple starting from an unbounded Fredholm module $(\Hg,D)$ whith the property of being a {\it bimodule} over a unital $*$-algebra $\Ac$, or equivalently a module over $\Ac\hotimes \Acop$. $\Acop$ is the opposite algebra of $\Ac$, i.e. the product of elements $a_1\cdot a_2$ in $\Acop$ corresponds to the opposite $a_2a_1$ with respect to the product in $\Ac$. Thus consider a $p_+$-summable spectral triple $(\Ac\hotimes\Acop,\Hg,D)$. This means that $D$ is an unbounded selfadjoint operator in the Hilbert space $\Hg$ such that $(1+D^2)^{-(p/2+\eps)}$ is trace-class for any $\eps>0$, the $*$-representations of $\Ac$ and $\Acop$ as bounded operators commute, and the commutators $[D,a]$ and $[D,b]$ are bounded and well-defined for any $a\in\Ac$, $b\in\Acop$. For compatibility with the technical hypotheses of theorem \ref{t3}, we also impose the {\it first order condition} 
\be
[a,[D,b]]=0\quad \forall\ a\in\Ac,\ b\in\Acop\ .
\ee
We may twist $(\Hg,D)$ by a finitely generated projective (right) $\Ac$-module $E=e(\Ac^N)$ represented by a projector $e=e^2=e^*$ in the matrix algebra $M_N(\Ac)$ for some $N\in\nn$. Hence $\Hg$ is replaced by its projection $e(\Hg^N)$ or $e\Hg$ on short, acted on by the compression of the $p_+$-summable Dirac operator $D_e:=eDe$. Since the representation of $\Acop$ commutes with $\Ac$, $e\Hg$ is still a module for $\Acop$ and the commutator $[D_e,b]$ is bounded for any $b\in\Acop$. The subalgebra $\Ac_e=eM_N(\Ac)e$ is also represented in $e\Hg$ and commutes with $\Acop$. The commutator $[D_e,a_e]$ is bounded for any $a_e\in\Ac_e$, and $[a_e,[D_e,b]]=0$ for any $b\in\Acop$. The {\it gauge group} $\GG$ is the unitary group of $\Ac_e$. Its complexified Lie algebra $\Lie \GG$ identifies with $(\Ac_e)_{\Lie}$. We let $\Hc=\Uc(\Lie\GG)$ be its enveloping algebra. It is a cocommutative algebra with involutive antipode (see below). Hence $\Ec=(\Acop,e\Hg,D_e)$ is a $p_+$-summable $\Hc$-equivariant spectral triple verifying the conditions of theorem \ref{t3}, so that it yields characteristic maps
\be
\psi(\Ec): HC_n(\Acop)\to HC^{\eps,+}_{n-p}(\Hc)\ ,\quad \forall n\ge 0\ .
\ee
When $\Ac=\Acop$ is the commutative algebra of smooth functions over a spin manifold, we may choose $(\Ac,e\Hg,D_e)$ as the ordinary Dirac module with coefficients in a vector bundle $E$. We will show that in this situation of ordinary (commutative) geometry, the secondary classes can be interpreted terms of BRS cohomology \cite{D}.\\

Let $M$ be a smooth compact riemannian spin manifold without boundary, of dimension $p$. We consider the unital Fr\'echet algebra $\Ac=\cinf(M)$ of smooth complex-valued functions over $M$. Let $G$ be a compact connected Lie group and $P\to M$ a $G$-principal bundle. We consider that $G$ is linearly represented on an hermitian finite-dimensional vector space $V$ by unitary operators. Let $E=P\times_G V$ be the hermitian complex vector bundle associated to $P$ over $M$. For any choice of $G$-connection $A$ over $P$ (Yang-Mills connection), we can form the associated connection $\nabla^E_A:\cinf(E)\to \cinf(E\otimes T^*M)$ acting on smooth sections of $E$. We denote by $\Sc\to M$ the spin bundle endowed with the spin connection $\nabla^{\Sc}$ associated to the metric on $M$. If the dimension $p$ is even, then $\Sc$ is $\zz_2$-graded. The total connection $\nabla_A=\nabla^{\Sc}\otimes\id + \id\otimes\nabla^E_A$ acts on smooth sections of the tensor product $\Sc\otimes E$. The associated Dirac operator
\be
D_A: \cinf(\Sc\otimes E)\to \cinf(\Sc\otimes E)
\ee
is obtained as usual from $\nabla_A$ via Clifford multiplication. It is of odd degree when $p$ is even. We let $\Hg=L^2(\Sc\otimes E)$ be the Hilbert space of square-integrable sections of $\Sc\otimes E$, provided with its norm inherited from the metrics on $M$ and $E$, and its natural grading. The pointwise multiplication by $\cinf(M)$ induces a faithful continuous $*$-representation $\rho:\Ac\to \End(\Hg)$. The Dirac operator acting on $\Hg$ is unbounded, selfadjoint and $p_+$-summable, i.e. $(1+D_A^2)^{-(p/2+\epsilon)}$ is trace-class for any $\epsilon>0$.\\ 
The gauge group $\GG$ is the subgroup of automorphisms of $P$ leaving the base manifold $M$ invariant. $\GG$ acts on $E$ and $\Hg$ by unitary endomorphisms. Its complexified Lie algebra $\Lie\GG$ identifies with the space of smooth endomorphisms of the vector bundle $E$ endowed with the commutator as Lie braket. We let $\Hc=\Uc(\Lie\GG)$ be its enveloping algebra. It is a cocommutative Hopf algebra, which implies in particular that its antipode is involutive. We recall that the coproduct and the antipode are given on primitive elements $X\in\Lie\GG\subset\Hc$ by
\be
\Delta X=1\otimes X+X\otimes 1\ ,\quad S(X)=-X\ ,
\ee
and the counit verifies $\eps(X)=0$, $\eps(1)=1$, so that $\Hcb=\ker\eps$ is the ideal generated by $\Lie\GG$. $\Hc$ acts on $\Hg$ by bounded endomorphisms commuting with the representation of $\Ac$ and the commutators $[D_A,\Ac]$. Moreover, the commutators $[D_A,\rho(a)]$ and $[D_A,h]$ are bounded for any $a\in\Ac$, $h\in\Hc$, so that $\Ec=(\Ac,\Hg,D_A)$ is a $p_+$-summable $\Hc$-equivariant spectral triple satisfying the requirements of theorem \ref{t3}. We thus get characteristic maps
\be
\psi(\Ec):HC_n(\Ac)\to HC^{\eps,+}_{n-p}(\Hc)\ ,\quad \forall n\ge 0\ .
\ee
The cyclic homologies corresponding to the classical space/symmetry $\Ac$ and $\Hc$ are known. First, the periodic cyclic homology of the enveloping algebra $\Hc=\Uc(\Lie\GG)$ is isomorphic to the Lie algebra cohomology of $\Lie\GG$ with coefficients in the trivial module $\cc$ \cite{CM98,CM00}:
\be
\HPb^{\eps}_0(\Hc)=\bigoplus_{i\ge 1}H^{2i}(\Lie\GG)\ ,\quad \HPb^{\eps}_1(\Hc)=\bigoplus_{i\ge 0}H^{2i+1}(\Lie\GG)\ ,
\ee
and periodicity implies $\HPb^{\eps}_n(\Hc)=\HPb^{\eps}_{n+2}(\Hc)$ for any $n$. One builds a chain map $\al$ between the $(b,B)$-complex $\Cb^{\eps}_*(\Hc)$ and the Chevalley-Eilenberg complex $C^*(\Lie\GG)$ as follows (we denote by $s$ the Lie algebra coboundary in $C^*(\Lie\GG)$ for consistency with the BRS construction). For any $f\in \Cb^{\eps}_n(\Hc)=\hom(\Hcb^{\hotimes n},\cc)$, its image $\al(f)\in C^n(\Lie\GG)$ evaluated on a $n$-chain $X_1\wedge \ldots \wedge X_n\in \La^n \Lie\GG$ is \cite{Cr}
\be
\al(f)(X_1\wedge \ldots \wedge X_n)= \frac{1}{n!}\sum_{\si \in S_n}\mathrm{sign}(\si)f(X_{\si(1)}\otimes \ldots \otimes X_{\si(n)})\ ,
\ee
where $S_n$ is the permutation group of $n$ elements. Then one has $s\circ\al =\al\circ B$ and $\al\circ b=0$, whence the induced map $\al: \HPb^{\eps}_*(\Hc)\to \bigoplus_{n>0}H^n(\Lie\GG)$ which turns out to be an isomorphism. In the same fashion, one sees that $\al$ induces a chain map
\be
\al: HC^{\eps,+}_{n}(\Hc)\to \bigoplus_{i\ge 0} H^{2i-n}(\Lie\GG)\ ,\quad \forall n\in\zz\ .\label{chain}
\ee

The cyclic (co)homology of the Fr\'echet algebra $\Ac=\cinf(M)$ has been computed by Connes \cite{C0}. We let $\Om^n(M)$ be the space of smooth differential forms of degree $n$ over $M$, $d$ the de Rham differential, and $B^n(M)=d\Om^{n-1}(M)$, $Z^n(M)$, $H^n(M)=Z^n(M)/B^n(M)$ respectively the spaces of de Rham coboundaries, cocycles and cohomology of degre $n$. Then the cyclic homology of $\Ac$ in degree $n$ is given by the sum
\be
HC_n(\Ac)=\Om^n(M)/B^n(M)\oplus H^{n-2}(M)\oplus H^{n-4}(M)\oplus\ldots\ .
\ee
The interesting part is the term of highest degree $\Om^n(M)/B^n(M)$; it is not homotopy invariant in the sense that an automorphism of $\Ac$ (i.e. a diffeomorphism of $M$) connected to the identity does not leave this space invariant. The situation is different with the periodic cyclic homology. The latter gives a purely topological information in terms of de Rham cohomology:
\be
HP_n(\Ac)=\bigoplus_{i\in\zz} H^{n+2i}(M)\ .
\ee
The negative cyclic homology is the difference between the two preceding homologies:
\be
HC^-_n(\Ac)=Z^n(M)\oplus H^{n+2}(M)\oplus H^{n+4}(M)\oplus\ldots\ .
\ee
It is easy to decompose explicitly the $SBI$ long exact sequence of cyclic homology (\cite{L}):
\be
\xymatrix{
HC_{n-1}(\Ac) \ar[r]^{B} \ar@{}[d] |{\stackrel{||}{\vdots}} & HC^-_n(\Ac) \ar[r]^{I} \ar@{}[d] |{\stackrel{||}{\vdots}} & HP_n(\Ac) \ar[r]^{S} \ar@{}[d] |{\stackrel{||}{\vdots}} & HC_{n-2}(\Ac) \ar[r]^{B} \ar@{}[d] |{\stackrel{||}{\vdots}} & HC^-_{n-1}(\Ac) \ar@{}[d] |{\stackrel{||}{\vdots}} \\
0 \ar[r] \ar@{}[d]|{\oplus} &  H^{n+2}(M) \ar[r] \ar@{}[d]|{\oplus} &  H^{n+2}(M) \ar[r] \ar@{}[d]|{\oplus} & 0 \ar[r] \ar@{}[d]|{\oplus} & H^{n+3}(M) \ar@{}[d]|{\oplus} \\
\Om^{n-1}/B^{n-1} \ar[r]^{d} \ar@{}[d]|{\oplus} &  Z^n(M) \ar[r] \ar@{}[d]|{\oplus} & H^n(M) \ar[r] \ar@{}[d]|{\oplus} & 0 \ar[r] \ar@{}[d]|{\oplus} & H^{n+1}(M) \ar@{}[d]|{\oplus} \\
H^{n-3}(M) \ar[r] \ar@{}[d]|{\oplus} & 0 \ar[r] \ar@{}[d]|{\oplus} & H^{n-2}(M) \ar[r] \ar@{}[d]|{\oplus} & \Om^{n-2}/B^{n-2} \ar[r]^{d} \ar@{}[d]|{\oplus} &  Z^{n-1}(M) \ar@{}[d]|{\oplus} \\
H^{n-5}(M) \ar[r] \ar@{}[d] |{\vdots} &  0 \ar[r] \ar@{}[d] |{\vdots} & H^{n-4}(M) \ar[r] \ar@{}[d] |{\vdots} & H^{n-4}(M) \ar[r] \ar@{}[d] |{\vdots} & 0 \ar@{}[d] |{\vdots}\\
 & & & & }
\ee

We know that the characteristic class $\psi(\Ec):HC_n(\Ac)\to HC^{\eps,+}_{n-p}(\Hc)$ vanishes on the range of the periodicity map $S:HP_{n+2}(\Ac)\to HC_n(\Ac)$, which corresponds to the sum of de Rham cohomology groups $H^n(M)\oplus H^{n-2}(M)\oplus\ldots$. Hence $\psi(\Ec)$ is actually defined on the quotient $\Om^n(M)/Z^n(M)$ for any $n<p$, and vanishes whenever $n\ge p$. Moreover, it is not difficult to see that the chain map $\psi(\rho,D): CC_n(\Ac)\to CC^{\eps,+}_{n-p}(\Hc)$ representing $\psi(\Ec)$ (section \ref{ssec}) carries $\Om^n\Ac$ to the space of $(p-n)$-chains over the Hopf algebra $\Cb^{\eps}_{p-n}(\Hc)$. Consequently, after composition with (\ref{chain}), the secondary characteristic classes reduce to a collection of maps
\be
\psi(\Ec): \Om^n(M)/Z^n(M)\to H^{p-n}(\Lie\GG)\ ,\quad 0\le n<p\ .
\ee
We shall interpret them in terms of BRS cohomology \cite{D, DTV, MSZ}.\\
Consider the $G$-principal bundle $P\stackrel{G}{\longrightarrow}M$, and let $\gg=\Lie G$ denote the finite-dimensional Lie algebra of the structure group, say of dimension $N$. The gauge group $\GG$ acts on $P$ by vertical diffeomorphisms commuting with the action of $G$. Denote by $\{e_{\al}\}_{1\le\al\le N}$ a basis of $\gg$. The infinitesimal action of $\gg$ (from the right) on $P$ yields a basis $\{e_{\al}\}$ of vector fields for each fiber. Any element $X\in\Lie\GG$ is a vertical vector field on $P$. We can decompose $X$ in the basis $\{e_{\al}\}$:
\be
X=X^{\al}e_{\al}\ ,
\ee
where $X^{\al}$ are smooth functions over $P$. By definition the vector field $X$ preserves the structure of $G$-principal bundle, which implies that its Lie braket with the vector field corresponding to any $\la\in\gg$ vanishes: one has $[X,\la]=0$. Thus if $L$ and $i$ denote the Lie derivative on differential forms and the interior product by vector fields, then $[L_X,L_{\la}]=0=[L_X,i_{\la}]$ for any $X\in\Lie\GG$ and $\la\in\gg$. We now introduce a BRS differential algebra \cite{D}. For any $n,k\in\nn$, let $C^{n,k}(P)$ be the space of $k$-cochains over $\Lie\GG$ with values on the differential forms of degree $n$ on $P$:
\be
C^{n,k}(P)=\hom( \Lie\GG^{\otimes k},\Om^n(P))\ .
\ee
The Lie derivative $L_X$ turns $\Om^n(P)$ into a $\Lie\GG$-module for any $n$. Thus we can endow $C^{*,*}(P)$ with the usual Lie algebra coboundary $s:C^{n,k}(P)\to C^{n,k+1}(P)$ which reads
\beq
(sf)(X_0,\ldots ,X_k) &=& \sum_{i=0}^k (-)^{n+i}L_{X_i}f(X_0,\ldots, \hat{X_i},\ldots ,X_k) +\\
&+& \sum_{0\le i<j\le k}(-)^{n+i+j}f([X_i,X_j], X_0,\ldots, \hat{X_i},\ldots ,\hat{X_j},\ldots ,X_k)\ ,\non
\eeq
for any $f\in C^{n,k}(P)$ and $X_i\in\Lie\GG$, where the hat $\hat{\cdot}$ stands for omission. Define also the de Rham coboundary $d:C^{n,k}(P)\to C^{n+1,k}(P)$:
\be
(df)(X_1,\ldots ,X_k)=d(f(X_1,\ldots ,X_k))\ .
\ee
One verifies readily that $s^2=sd+ds=d^2=0$, hence $(C^{*,*}(P),d,s)$ is a $\nn\times\nn$-graded bicomplex. Moreover it is naturally endowed with a graded commutative product: for any $f\in C^{n,k}(P)$ and $g\in C^{n',k'}(P)$, their product $fg\in C^{n+n',k+k'}(P)$ is
\beq
\lefteqn{(fg)(X_1,\ldots, X_{k+k'})=}\\
&& \frac{(-)^{kn'}}{k!\,k'!}\sum_{\si\in S_{k+k'}} \mathrm{sign}(\si)f(X_{\si(1)},\ldots,X_{\si(k)})\wedge g(X_{\si(k+1)},\ldots, X_{\si(k+k')})\ ,\non
\eeq
where $S_{k+k'}$ is the permutation group of $k+k'$ elements. $d$ and $s$ are odd derivations for the product, hence $C^{*,*}(P)$ is actually a bigraded commutative differential algebra.\\
We now define the Cartan operation of the Lie algebra $\gg$. The interior product $i_{\la}:C^{n,k}(P)\to C^{n-1,k}(P)$ and Lie derivative $L_{\la}=i_{\la}d+di_{\la}:C^{n,k}(P)\to C^{n,k}(P)$ associated to a fundamental vector field $\la\in\gg$ fulfill the relations
\be
i_{\la}s+si_{\la}=0\ ,\quad [L_{\la},s]=0\ ,
\ee
because $[L_{X},i_{\la}]=0$ for any $X\in\Lie\GG$. Consider the Lie superalgebra $\gg\otimes C^{*,*}(P)$, where $C^{*,*}(P)$ is a graded commutative algebra for the total degree of its components. There exists a canonical element $\om=e_{\al}\otimes \om^{\al}\in \gg\otimes C^{0,1}(P)$ defined by
\be
\om^{\al}(X)=X^{\al}\ ,\quad \forall X\in\Lie\GG\ .
\ee
$\om$ is the ``ghost field'' of Yang-Mills theory. On the other hand, any $G$-connection $A$ over the principal bundle $P$ may be viewed as an element $A=e_{\al}\otimes A^{\al}\in \gg\otimes C^{1,0}(P)$, with curvature $F=dA+\frac{1}{2}[A,A]$. Then $A$ and $\om$ verifiy the {\it BRS relations}
\be
sA=-d\om-[A,\om]\ ,\quad s\om=-\frac{1}{2}[\om,\om]\ ,\label{brs}
\ee
which imply the so-called {\it horizontality condition}
\be
(d+s)(A+\om)+\frac{1}{2}[A+\om,A+\om]=F\ .
\ee
The Cartan operation of $\gg$ acts on the connection $A$ and the ghost $\om$ as
\be
i_{\la}A=\la\ ,\quad L_{\la}A=-[\la,A]\ ,\quad i_{\la}\om=0\ ,\quad L_{\la}\om=-[\la,\om]\ ,
\ee
for any $\la\in\gg$ (to be precise, $\la$ appears as a vector field over $P$ in the l.h.s. and as an element of $\gg\otimes 1\subset \gg\otimes C^{*,*}(P)$ in the r.h.s.). The bigraded differential algebra $C^{*,*}(P)$ endowed with the Cartan operation and total connection $A+\om$ defines a {\it BRS $\gg$-operation} according to the terminology of \cite{D}.\\
The subalgebra of $\gg$-basic elements $f\in C^{*,*}(P)$, defined by $i_{\la}f=0=L_{\la}f$ for any $\la\in\gg$, is obviously stable by both $d$ and $s$, whence a bicomplex. Since we assume that $G$ is connected, the $\gg$-basic subalgebra identifies clearly with the bicomplex of $\Lie\GG$ cochains with values in the trivial module of differential forms over the base manifold $M$:
\be
C^{n,k}(M)=\hom( \Lie\GG^{\otimes k},\Om^n(M))\ .
\ee
We call $C^{*,*}(P)$ and $C^{*,*}(M)$ the BRS differential algebras of $P$ and $M$ respectively.\\

We are ready to introduce BRS cohomology, or cohomology ``$s$ modulo $d$''. Since $d$ anticommutes with $s$, the subspace $d C^{*,*}(P)$ is stable by $s$. By definition, the cohomology of the quotient complex $(C^{*,*}(P)/dC^{*,*}(P),s)$ is the BRS cohomology of $P$. Hence a cocycle $s$ mod $d$ of bidegree $(n,k)$ is represented by an element $\om^{n,k}\in C^{n,k}(P)$ such that there exists $\om^{n-1,k+1}\in C^{n-1,k+1}(P)$ with the relation
\be
s\om^{n,k}+d\om^{n-1,k+1}=0\ .
\ee
Two cocycles $\om^{n,k}$ and $\om'^{n,k}$ are cohomologous iff there exist two elements $\om^{n,k-1}$ and $\om^{n-1,k}$ such that
\be
\om'^{n,k}-\om^{n,k}=s\om^{n,k-1} + d\om^{n-1,k}\ .
\ee
We denote by $H^{n,k}(P,s\mod d)$ the BRS cohomology of bidegree $(n,k)$. The BRS cohomology $H^{n,k}(M,s\mod d)$ of the base manifold $M$ is defined in a similar way (note that the principal bundle $P$ enters indirectly in the definition of the BRS cohomology of $M$ through the gauge group $\GG$).\\
One sees that the definition of the cohomology $s$ mod $d$ of $M$ is somewhat reminiscent to the non-periodic cyclic homology of $\Ac=\cinf(M)$. In fact there is a well-defined pairing between the negative cyclic homology of $\Ac$ and the BRS cohomology of $M$, with values in the Lie algebra cohomology of $\GG$ with trivial coefficients. By composition with the Chern character on algebraic $K$-theory, we deduce the following 
\begin{lemma}\label{lcou}
Let $M$ be an oriented compact smooth manifold without boundary, of dimension $p$, $G$ a connected Lie group and $P\stackrel{G}{\longrightarrow}M$ a $G$-principal bundle. Let $\GG$ denote the gauge group of $P$. Then for any pair of integers $(n,k)$, with $0\le n\le p$, there is a well-defined pairing between the algebraic $K$-theory of $\Ac=\cinf(M)$ and the BRS cohomology of $M$,
\be
K_n(\Ac)\times H^{p-n,k}(M,s\mod d)\to H^k(\Lie\GG)\ .
\ee
For any $x\in K_n(\Ac)$ and a BRS cohomology class represented by a cocycle $\om^{p-n,k}$, the pairing reads
\be
\langle x,[\om^{p-n,k}] \rangle = \int_M \ch^-_n(x)\wedge \om^{p-n,k}\ ,
\ee
where $\ch^-_n:K_n(\Ac)\to HC^-_n(\Ac)$ is the negative Chern character composed with the projection $HC^-_n(\Ac)\to Z^n(M)$.
\end{lemma}
{\it Proof:} By taking the component of the negative Chern character $\ch^-_n(x)$ in the space of de Rham cocycles $Z^n(M)$, we only have to show that the pairing $Z^n(M)\times H^{p-n,k}(M,s\mod d)\to H^k(\Lie\GG)$ given by
$$
\int_M z_n\wedge \om^{p-n,k}\ ,\quad z_n\in Z^n(M)\ ,
$$
is well-defined. First of all, $\om^{p-n,k}$ is a cochain over $\Lie\GG$ with coefficients in the trivial module of differential forms over $M$, so that the above integral defines a cochain over $\Lie\GG$ with coefficients in $\cc$. Since $\om^{p-n,k}$ is a cocycle $s$ mod $d$, there is a cochain $\om^{p-n-1,k+1}$ such that $s\om^{p-n,k}+d\om^{p-n-1,k+1}=0$. Hence
$$
s\int_M z_n\wedge \om^{p-n,k}=(-)^{p-n} \int_M z_n\wedge s\om^{p-n,k}=(-)^{p}\int_M dz_n\wedge \om^{p-n-1,k+1}=0\ ,
$$
because $z_n$ is a cocycle. Moreover, if $\om'^{p-n,k}$ is cohomologous to $\om^{p-n,k}$, there exists $\om^{p-n,k-1}$ and $\om^{p-n-1,k}$ such that
$$
\om'^{p-n,k}=\om^{p-n,k}+ s\om^{p-n,k-1}+d\om^{p-n-1,k}
$$
and one has
$$
\int_M z_n\wedge \om'^{p-n,k}=\int_M z_n\wedge \om^{p-n,k}+(-)^{p-n} s\int_M z_n\wedge \om^{p-n,k-1}\ ,
$$
where the last term is a coboundary over $\Lie\GG$ with trivial coefficients. \cqfd\\

Let $A\in\gg\otimes C^{1,0}(P)$ be a $G$-connection on $P$. We are interested in the ($\zz_2$-graded) Lie subalgebra of $\gg\otimes C^{*,*}(P)$ generated by $A$, $dA$, $\om$, $d\om$, or equivalently by $A$, $F$, $\om$, $sA$. This Lie subalgebra is stable by $s$ and $d$. Let $\gg^*$ be the dual space of $\gg$, $S^n\gg^*$ the space of symmetric polynomials over $\gg$ of degree $n$, and $\Ic_S^n(\gg)\subset S^k\gg^*$ the subspace of invariant symmetric polynomials. Hence $I\in\Ic_S^n(\gg)$ if and only if 
\be
I(\la_1,\ldots,\la_n)=I(\la_{\si(1)},\ldots,\la_{\si(n)})\quad \mathrm{and}\quad \sum_{i=1}^nI(\la_1,\ldots,[\la,\la_i],\ldots,\la_n)=0\ ,
\ee
for any $\la,\la_i\in\gg$ and any permutation $\si$ of $n$ elements. Since one has $L_{\la}\chi=[-\la,\chi]$ for any $\chi$ in the Lie subalgebra generated by $A$, $F$, $\om$, $sA$, one deduces that $I(\chi_1,\ldots,\chi_n)$ is a $\gg$-invariant element of $C^{*,*}(P)$ for any such $\chi_i$'s, i.e. $L_{\la}I(\chi_1,\ldots,\chi_n)=0$. Moreover, since $i_{\la}F=i_{\la}\om=i_{\la}sA=0$, one has $i_{\la}I(\chi_1,\ldots,\chi_n)=0$ whenever the $\chi_i$'s are generated by $F$, $\om$, $sA$ alone, in which case $I(\chi_1,\ldots,\chi_n)$ is $\gg$-basic and defines an element of $C^{*,*}(M)$. There is a way of constructing BRS cocycles over $P$ and $M$ from the space of invariant symmetric polynomials $\Ic_S(\gg)$. The process is a generalization of the antitransgression map in ordinary Chern-Weil theory. This is provided by the {\it consistent} and {\it covariant} BRS descent equations.\\
Let $I\in\Ic_S^n(\gg)$ be an invariant symmetric polynomial of degree $n$, and $A$ a $G$-connection on $P$. On the tensor product of graded differential algebras $C^{*,*}(P)\otimes \Om[0,1]$, we consider the following connection (not a principal bundle connection):
\be
\nabla=s+d+d_t+t(A+\om)\ ,
\ee
where $d$ is the differential over $P$ and $d_t$ the differential over the interval $[0,1]$. Thus $\nabla$ interpolates between the BRS connection $s+d+A+\om$ for $t=1$ and the flat connection $s+d$ on $C^{*,*}(P)$ for $t=0$. Using the BRS relations (\ref{brs}), one computes the curvature:
\be
\nabla^2= dt(A+\om)+ (t^2-t)(\om^2+[A,\om])+tdA+t^2A^2\ .
\ee
Thus for $t=0$, one has $\nabla^2=dt(A+\om)$, and for $t=1$, $\nabla^2=dt(A+\om)+F$. The BRS cochain $I(\nabla^2,\ldots,\nabla^2)$ may be expanded in powers of $dt$:
\be
I(\nabla^2,\ldots,\nabla^2)=\Icons^0(A,\om)+dt\Icons^t(A,\om)\ ,
\ee
and the BRS transgression is defined to be
\be
\Qcons(A,\om)=\int_0^1 dt\Icons^t(A,\om)\ .
\ee
As usual, the Bianchi identity $[\nabla,\nabla^2]=0$ implies 
\be
(s+d+d_t)I(\nabla^2,\ldots,\nabla^2)=0\ ,\quad (s+d)(dt\Icons^t)+d_t\Icons^0=0\ .
\ee
Therefore
\be
(s+d)\Qcons(A,\om)=\int_0^1 dt(s+d)\Icons^1=\int_0^1 d_t\Icons^0=I(F,\ldots,F)
\ee
because $\Icons^0|_{t=1}=I(F,\ldots,F)$ and $\Icons^0|_{t=0}=0$. Expanding the cochain $\Qcons(A,\om)\in C^{*,*}(P)$ in bidegrees, one obtains the consistent BRS descent equations for the polynomial $I\in\Ic_S^n(\gg)$:
\beq
I(F,\ldots,F) &=& d\Qcons^{2n-1,0} \non\\
0 &=& d\Qcons^{2n-2,1} + s\Qcons^{2n-1,0}\non\\
0 &=& d\Qcons^{2n-3,2} + s\Qcons^{2n-2,1}\non\\
 &\vdots& \label{cons}\\
0 &=& d\Qcons^{0,2n-1} + s\Qcons^{1,2n-2}\non\\
0 &=& s\Qcons^{0,2n-1}\ .\non
\eeq
This shows that $\Qcons^{2n-k-1,k}$ defines a cocycle $s$ mod $d$ on the principal bundle $P$ for any $k\ge 0$. Another related construction is the set of covariant descent equations. The latter is obtained by a modification of the preceding connection and yields cocycles $s$ mod $d$ on the base manifold $M$. Given $I\in \Ic_S^n(\gg)$ and a $G$-connection $A$ on $P$, we form the following connection acting on $C^{*,*}(P)\otimes\Om[0,1]$:
\be
\nabla= s+d+d_t+A+t\om\ ,
\ee
with curvature
\be
\nabla^2= dt\,\om + F+(1-t)sA +(t^2-t)\om^2\ .
\ee
One sees that $\nabla^2$ involves only the horizontal and equivariant cochains $\chi=\om,F,sA,\om^2$, that is, $i_{\la}\chi=0$ and $L_{\la}\chi=-[\la,\chi]$ for any $\la\in\gg$. Thus $I(\nabla^2,\ldots,\nabla^2)$ is a $\gg$-basic element of $C^{*,*}(P)$, hence in $C^{*,*}(M)$. One expands it in powers of $dt$:
\be
I(\nabla^2,\ldots,\nabla^2)=\Icov^0(A,\om)+dt\Icov^t(A,\om)\ ,
\ee
and the covariant transgression $\Qcov\in C^{*,*}(M)$ is given by
\be
\Qcov(A,\om)=\int_0^1 dt\Icov^t(A,\om)\ .
\ee
The Bianchi identity implies
\beq
\lefteqn{(s+d)\Qcov(A,\om)=\Icov^0(A,\om)|_{t=1}-\Icov^0(A,\om)|_{t=0}}\\
&&=I(F,\ldots,F)-I(F+sA,\ldots,F+sA)\ .\non
\eeq
One sees that the component $\Qcov^{2n-1,0}(A,\om)$ vanishes because the term proportional to $dt$ in the curvature is $dt\,\om$, hence $\Icov^t(A,\om)$ contains at least one power of $\om$. Therefore, the expansion of $\Qcov(A,\om)$ in bidegrees yields the covariant descent equations
\beq
-C_n^1 I(sA,F,\ldots,F) &=& d\Qcov^{2n-2,1} \non\\
-C_n^2 I(sA,sA,F,\ldots,F) &=& d\Qcov^{2n-3,2} + s\Qcov^{2n-2,1}\non\\
 &\vdots& \non\\
-C_n^n I(sA,\ldots,sA) &=& d\Qcov^{n-1,n} + s\Qcov^{n,n-1}\non\\
0 &=& d\Qcov^{n-2,n+1} + s\Qcov^{n-1,n}\\
&\vdots& \non\\
0 &=& d\Qcov^{0,2n-1} + s\Qcov^{1,2n-2}\non\\
0 &=& s\Qcov^{0,2n-1}\ ,\non
\eeq
where $C_n^k=n!/k!(n-k)!$ are combinatorial constants. Hence we find that the component $\Qcov^{2n-k-1,k}\in C^{2n-k-1,k}(M)$ is a cocycle $s$ mod $d$ only for $k\ge n$. In fact, in appropriate degrees the BRS cohomology classes of $\Qcons$ and $\Qcov$ are independent of the choice of connection $A$ over $P$, and coincide in $H^{*,*}(P, s\mod d)$:
\begin{proposition}\label{pbrs}
Let $P\stackrel{G}{\longrightarrow}M$ be a $G$-principal bundle over $M$, and $A$ be a $G$-connection on $P$. Let $I\in\Ic_S^n(\gg)$ be an invariant symmetric polynomial over the Lie algebra $\gg=\Lie G$, and denote by $\Qcons(A,\om)\in C^{*,*}(P)$ and $\Qcov(A,\om)\in C^{*,*}(M)$ the consistent and covariant BRS transgressions of $I$. Then \\

\noindent i) The class of the consistent term $\Qcons^{2n-k-1,k}(A,\om)$ in the BRS cohomology $H^{2n-k-1,k}(P, s\mod d)$ is independent of the choice of connection $A$ for any $k\ge 1$.\\

\noindent ii) The class of the covariant term $\Qcov^{2n-k-1,k}(A,\om)$ in the BRS cohomology $H^{2n-k-1,k}(M, s\mod d)$ is independent of $A$ for any $k\ge n$.\\

\noindent iii) The image of $\Qcov^{2n-k-1,k}(A,\om)$ in $H^{2n-k-1,k}(P, s\mod d)$ coincides with the BRS cohomology class of $\Qcons^{2n-k-1,k}(A,\om)$ for any $k\ge n$.
\end{proposition}
{\it Proof:} This is proved as usual with a two-parameter transgression of $I$:\\
\noindent i) Consider a smooth family of $G$-connections $A_u$ on $P$ parametrized by $u\in [0,1]$. Let $t\in[0,1]$ be another parameter. Then we modify the construction of $\Qcons$ with the following connection
$$
\nabla= s+d+d_t+d_u +t(A_u+\om)\ ,
$$
whose curvature reads
$$
\nabla^2= dt(A+\om) + td_uA_u + (t^2-t)(\om^2+[A,\om])+tdA+t^2A^2\ .
$$
We develop the polynomial $I(\nabla^2,\ldots,\nabla^2)$ in powers of $dt, du$:
$$
I(\nabla^2,\ldots,\nabla^2)=I^0+dt I^t+du I^u+ dt\,du I^{t,u}\ ,
$$
and the Bianchi identity restricted to the $dt\,du$ component yields
$$
(s+d)dt\,du I^{t,u}+d_t(duI^u)+d_u(dtI^t)=0\ .
$$
After integration of $t,u$ over $[0,1]$, one has
$$
(s+d)\la + \int_0^1 du (I^u(t=1)-I^u(t=0))=\int_0^1 dt (I^t(u=1)-I^t(u=0))\ ,
$$
where $\la$ is the integration of $dt\,du I^{t,u}$. The right hand side is the difference $\Qcons(u=1)-\Qcons(u=0)$. The second term of the left hand side does not contain any ghost $\om$, because the pullback of $\nabla^2$ on the submanifolds $t=1$ and $t=0$ yield respectively $d_uA_u+F$ and $0$. Hence $\Qcons^{2n-k-1,k}(u=1)-\Qcons^{2n-k-1,k}(u=0)$ is a coboundary $s\mod d$ on $P$ whenever $k\ge 1$.\\
\noindent ii) One proceeds as in case i), with the connection
$$
\nabla=s+d+d_t+d_u +A_u+t\om
$$
and curvature
$$
\nabla^2= dt\,\om+d_uA_u+ F+ (1-t)sA+ (t^2-t)\om^2\ .
$$
All the terms in $\nabla^2$ are horizontal forms on $P$ (because the variation $d_uA_u$ is horizontal) and equivariant for the right action of $G$, hence in what follows all the differential forms are in fact basic, hence defined on $M$. One has as in i)
$$
(s+d)\la' + \int_0^1 du (I^u(t=1)-I^u(t=0))=\int_0^1 dt (I^t(u=1)-I^t(u=0))\ ,
$$
the r.h.s. being the difference $\Qcov(u=1)-\Qcov(u=0)$. The second term of the l.h.s. is zero whenever the ghost number is $\ge n$, because the bullbacks of the curvature $\nabla^2$ for $t=1$ and $t=0$ yield respectively $d_uA_u+F$ and $d_uA_u+F+sA$. Hence $\Qcov^{2n-k-1,k}(u=1)-\Qcov^{2n-k-1,k}(u=0)$ is a coboundary $s\mod d$ on $M$ whenever $k\ge n$.\\
\noindent iii) Now $A$ is fixed and we use the connection
$$
\nabla= s+d+d_t+d_u+u(A+t\om)\ .
$$
Of course the curvature is no longer horizontal and we work on $P$. From the expansion $I(\nabla^2,\ldots,\nabla^2)=I^0+dt I^t+du I^u+ dt\,du I^{t,u}$, we identify
$$
\Qcons=\int_0^1du I^u(t=1)\ ,\quad \Qcov=\int_0^1dt I^t(u=1)\ .
$$
The Bianchi identity and the fact that $\int_0^1dt I^t(u=0)=0$ imply
$$
\Qcov-\Qcons=-(s+d)\la''+\int_0^1du I^u(t=0)\ .
$$
But the bullback of the curvature for $t=0$ is $u(s+d)A+u^2A^2+duA$, hence the second term of the r.h.s. vanishes when the ghost number is $\ge n$. \cqfd\\

Let now $\Ac=\cinf(M)$ be the Fr\'echet algebra of smooth functions on $M$, and fix a connection $A$ on $P$. We return to the situation considered before and let $\Ec=(\Ac,\Hg,D_A)$ be the Dirac spectral triple associated to the twisted Dirac operator $D_A$. $\Ec$ is $p_+$-summable if $p$ is the dimension of $M$, and is $\Hc$-equivariant for the Hopf algebra $\Hc=\Uc(\Lie\GG)$. The secondary characteristic classes $\psi(\Ec): \Om^n(M)/Z^n(M)\to H^{p-n}(\Lie\GG)$ can be computed in terms of the BRS algebra $C^{*,*}(M)$:
\begin{proposition}
Let $M$ be a smooth compact riemannian spin manifold of dimension $p$, oriented by its spin structure, and $\Ac=\cinf(M)$. Let $G$ be a compact connected Lie group unitarily represented in a vector space $V$, and $\gg=\Lie G$. Let $P\stackrel{G}{\longrightarrow}M$ be a $G$-principal bundle over $M$, and $E=P\times_GV$ the hermitian vector bundle associated to the representation $V$. We denote by $\GG$ the gauge group of $P$, and $\Ec=(\Ac,\Hg,D_A)$ the $p_+$-summable $\GG$-equivariant Dirac spectral triple associated to $E$ and a choice of $G$-connection $A$ on $P$. Then the secondary characteristic classes $\psi(\Ec): \Om^n(M)/Z^n(M)\to H^{p-n}(\Lie\GG)$, $0\le n<p$, are represented by the following maps:
\be
\Om^n(M)\ni\ \om_n\mapsto \frac{1}{(2\pi i)^{p/2}}\int_M \om_n\wedge \Tr(e^{-sA})\ \in C^{p-n}(\Lie\GG)\ ,\label{lim}
\ee
where $sA\in \gg\otimes C^{1,1}(P)$ is the BRS variation of $A$ and $\Tr$ is the trace of endomorphisms of $E$. Remark that the differential form $\Tr(\exp(-sA))$ is $\gg$-basic, hence defined on $M$.
\end{proposition}
{\it Proof:} This is a direct but lengthy computation of the relevant components of the chain map $\psi(\rho,D_A)=\Pf \chi(\rho,tD_A)$ discussed in section \ref{ssec}. One uses well-known asymptotic expansions of the heat kernel $\exp(-t^2D_A^2)$. This is not too difficult since $M$ is a smooth manifold, and only few terms remain after taking the limit $t\to 0$. We don't give the details here. \cqfd
\begin{remark}\textup{
Observe that the Dirac $\hat{A}$-genus familiar to topology does not appear in the above formula. This is due to the truncature of the chain map $\psi(\rho,D_A)$ used to exhibit the secondary characteristic classes. In some sense, the topological content of $(\Ac,\Hg,D_A)$ is dropped, and only the BRS part is keeped intact. By the way, this also shows the independence of $\psi(\Ec)$ with respect to the metric on $M$. }
\end{remark}

In order to show that the map (\ref{lim}) is well-defined in cohomology, we use the set of covariant descent equations associated to the invariant symmetric polynomial $I\in\Ic_S^{p-n}(\gg)$:
\be
I(\la_1,\ldots,\la_{p-n})=\frac{1}{(p-n)!}\sum_{\si\in S_{p-n}}\frac{1}{(p-n)!}\Tr(\la_{\si(1)}\ldots\la_{\si(p-n)})\ ,
\ee
for any $\la_i\in\gg$, where $S_{p-n}$ is the permutation group of $p-n$ elements. Indeed the $\gg$-basic differential form
\be
I(sA,\ldots,sA)=\frac{1}{(p-n)!}\,\Tr\big((sA)^{p-n}\big)
\ee
is, up to a sign, the component of $\Tr(\exp(-sA))$ entering in formula (\ref{lim}). Then from the covariant descent equations, there exists BRS cochains $\Qcov^{p-n-1,p-n}(A,\om)$ and $\Qcov^{p-n,p-n-1}(A,\om)$ such that
\be
-I(sA,\ldots,sA)= d\Qcov^{p-n-1,p-n}(A,\om) + s\Qcov^{p-n,p-n-1}(A,\om)\ .
\ee
Therefore, the image of a $n$-form $\om_n\in \Om^n(M)$ under $\psi(\Ec)$ is represented in the cohomology $H^{p-n}(\Lie\GG)$ by
\be
\frac{(-)^{p-n}}{(2\pi i)^{p/2}}\int_M \om_n\wedge I(sA,\ldots,sA)\equiv\frac{(-)^{p}}{(2\pi i)^{p/2}}\int_M d\om_n \wedge \Qcov^{p-n-1,p-n}(A,\om)\ .
\ee
Hence $\psi(\Ec)$ vanishes on $Z^n(M)$ and factors through the boundary map $d:\Om^n(M)/Z^n(M)\to B^{n+1}(M)\subset HC^-_{n+1}(\Ac)$. This combined with the commutativity of the diagram relating the long exact sequences of $K$-theory and cyclic homology
\be
\xymatrix{
 \Kt_{n+2}(\Ac) \ar[d]^{\cht_{n+2}} \ar[r] & \Kr_{n+1}(\Ac) \ar[d]^{\chr_{n+1}} \ar[r] & K_{n+1}(\Ac) \ar[d]^{\ch^-_{n+1}} \ar[r] & \Kt_{n+1}(\Ac) \ar[d]^{\cht_{n+1}} \ar[r] &  \Kr_{n}(\Ac) \ar[d]^{\chr_{n}}  \\
 HP_{n+2}(\Ac) \ar[r]^{S} & HC_{n}(\Ac) \ar[r]^{B} & HC^-_{n+1}(\Ac) \ar[r]^{I} & HP_{n+1}(\Ac) \ar[r]^{S} & HC_{n-1}(\Ac)  }
\ee
shows that the composite map
\be
\begin{CD}
\Kr_{n+1}(\Ac) @>{\chr_{n+1}}>> HC_n(\Ac) @>{\psi(\Ec)}>> H^{p-n}(\Lie\GG)
\end{CD}
\ee
factors through the image of the homomorphism $\Kr_{n+1}(\Ac)\to K_{n+1}(\Ac)$, or equivalently by exactness, the kernel $\Ka_{n+1}=\ker(K_{n+1}(\Ac)\to \Kt_{n+1}(\Ac))$:
\be
\begin{CD}
\Ka_{n+1}(\Ac) @>{\ch^-_{n+1}}>> B^{n+1}(M) @>{\cap \Qcov}>> H^{p-n}(\Lie\GG)\ .
\end{CD}
\ee
Lemma \ref{lcou} shows that this map amounts to compute the pairing between the kernel $\Ka_{n+1}(\Ac)\subset K_{n+1}(\Ac)$ and the class of $\Qcov^{p-n-1,p-n}(A,\om)$ in the BRS cohomology $H^{p-n-1,p-n}(M,s\mod d)$. \\
It is instructive to determine the image of the negative Chern character $\ch^-_n:\Ka_n(\Ac)\to B^n(M)$. For $n=1$, the homomorphism $K_1(\Ac)\to \Kt_1(\Ac)$ is clearly surjective. From Milnor \cite{M}, one knows that the algebraic $K_1$ of any commutative (unital) Banach algebra $\Asf$ is given by the direct sum
\be
K_1(\Asf)= \Asf^{\times}\oplus \pi_0(SL(\Asf))
\ee
of the abelian multiplicative group $\Asf^{\times}$ of invertible elements of $\Asf$, and the homotopy group $\pi_0$ of the invertible matrices over $\Asf$ with determinant 1. This is true in particular for the Banach algebra of continuous functions over $M$. On the other hand, the topological $K$-theory group of $\Asf$ is by definition $\Kt_1(\Asf)=\pi_0(GL(\Asf))$, hence we deduce that the kernel of the homomorphism $K_1(\Asf)\to \Kt_1(\Asf)$ is isomorphic to the quotient of abelian groups
\be
\Ka_1(\Asf)=\Asf^{\times}/\pi_0(\Asf^{\times})\ .
\ee
By \cite{R}, the latter equality is true also for the dense Fr\'echet algebra $\Ac=\cinf(M)\subset\Asf$, thus $\Ka_1(\Ac)$ is the connected component of 1 in the abelian group $\Ac^{\times}$. The negative Chern character $\ch^-_1:\Ka_1(\Ac)\to B^1(M)$ is described as follows. An element $g\in\Ka_1(\Ac)$ is a complex-valued invertible function over $M$ homotopic to the constant function 1, hence it has a logarithm $\log g$. As an element of $K_1(\Ac)$, its Chern character in $HC^-_1(\Ac)=Z^1(M)\oplus H^3(M)\oplus H^5(M)\ldots$ is \cite{L}
\be
\ch^-_1(g)=g^{-1}dg + \sum_{k\ge 1} (-)^k\frac{k!}{(2k+1)!}g^{-1}dg(dg^{-1}dg)^k\ ,
\ee
and since $g$ has a logarithm, the components in de Rham cohomology vanish, hence
\be
\ch^-_1(g)=g^{-1}dg= d\log g\ \in B^1(M)\ .
\ee
This shows that the Chern character $\ch^-_1:\Ka_1(\Ac)\to B^1(M)$ is {\it surjective}. By a cup-product argument in algebraic $K$-theory and cyclic homology \cite{L}, one shows more generally that the image of the abelian group $\Ka_n(\Ac)$ under $\ch^-_n$ is a {\it dense} subset of the locally convex space $B^n(M)$. For example when $n=2$, there is a graded commutative cup-product \cite{L}
\be
\cup : K_1(\Ac)\times K_1(\Ac)\to K_2(\Ac)\ ,
\ee
which carries two invertible elements $x,y\in\Ac^{\times}$ to their image $x\cup y=-y\cup x$ (Steinberg symbol). If $x$ and $y$ are homotopic to 1, the cup-product of their Chern characters in cyclic homology is given by
\be
\ch^-_2(x\cup y)=d(\log x \, d\log y)\ \in B^2(M)\ ,
\ee
which shows that the image of $\Ka_2(\Ac)$ spans a dense additive group in $B^2(M)$. Hence, for any $n\ge 0$ the pairing between $\Ka_{n+1}(\Ac)$ and the BRS cohomology class of the covariant term $\Qcov^{p-n-1,p-n}(A,\om)$ is optimal. We introduce the following 
\begin{definition}
Let $M$ be a smooth compact manifold, $G$ a compact connected Lie group and $P\stackrel{G}{\longrightarrow}M$ a $G$-principal bundle over $M$. We consider that $G$ is unitarily represented in a vector space $V$, and let $E\to M$ be the associated vector bundle. For any $n\ge 0$, let $I^n_V\in\Ic_S^n(\gg)$ be the invariant symmetric polynomial over $\gg=\Lie G$ corresponding to the degree $n$ of the development of $\Tr\exp(\la)$, for $\la\in\gg$ represented in $V$. The \emph{Chern character of $E$} in the BRS cohomology of bidegree $(2n-k-1,k)$, for any $k\ge n$, is the class 
\be
\ch^{2n-k-1,k}(E)=[\Qcov^{2n-k-1,k}(A,\om)]\ \in H^{2n-k-1,k}(M,s\mod d)
\ee
associated to the covariant transgression of the invariant polynomial $I_V^n$, for any choice of $G$-connection $A$ on $P$.
\end{definition}
Then collecting all the results of this section together, we relate the pairing of theorem \ref{t3} with BRS cohomology in the case $\Ac=\cinf(M)$:
\begin{theorem}\label{t4}
Let $M$ be a smooth compact riemannian spin manifold of dimension $p$, oriented by its spin structure, and $\Ac=\cinf(M)$. Let $G$ be a compact connected Lie group unitarily represented in a vector space $V$, $P\stackrel{G}{\longrightarrow}M$ a $G$-principal bundle over $M$, and $E=P\times_GV$ the hermitian vector bundle associated to the representation $V$. We denote by $\GG$ the gauge group of $P$, and $\Ec=(\Ac,\Hg,D_A)$ the $p_+$-summable $\GG$-equivariant Dirac spectral triple associated to $E$ and a choice of $G$-connection $A$ on $P$. Then the pairing of theorem \ref{t3}
\be
\Ka_{n+1}(\Ac)\times \Esf^{\GG}_p(\Ac,\cc)\to H^{p-n}(\Lie\GG)\ ,\quad 0\le n<p\ ,
\ee
between a $K$-theory class $x\in \Ka_{n+1}(\Ac)$ and the Dirac spectral triple is given by
\be
\langle x,\Ec\rangle= \frac{(-)^{p}}{(2\pi i)^{p/2}}\langle x, \ch^{p-n-1,p-n}(E)\rangle
\ee
where $\ch^{*,*}(E)\in H^{*,*}(M,s\mod d)$ is the Chern character of $E$ in BRS cohomology. This pairing does not depend on the connection $A$ nor on the metric on $M$ (homotopy invariance for the Dirac operator). \cqfd
\end{theorem}
\begin{remark}\textup{
The image of algebraic $K$-theory under the Chern character $\ch^-_{n}:\Ka_{n}(\Ac)\to B^{n}(M)$ acts as {\it test functions} for the BRS cohomology class of $E$. This allows to reconstruct the class $s\mod d$ of $\ch(E)$ {\it locally} on the manifold $M$. The process is independent of the topology of $M$, which is dropped by restriction to the kernel of the homomorphism $K_n(\Ac)\to\Kt_n(\Ac)$. }
\end{remark}
When the dimension of $M$ is even, the BRS descent equations giving rise to the representatives $\Qcov^{2n-k-1,k}(A,\om)$, $k\ge n$, of the Chern character $\ch(E)$, also contain the chiral anomaly and Schwinger term of quantum Yang-Mills theory \cite{MSZ}. For example when $M$ has dimension $p=2$, one gets characteristic maps for $\Ka_1(\Ac)$ and $\Ka_2(\Ac)$ only. In the first case, the covariant descent equations correspond to the invariant polynomial $I(\la_1,\la_2)=\frac{1}{4}\Tr(\la_1\la_2+\la_2\la_1)$:
\beq
-\Tr(sAF) &=& d\Tr(\om F)\ =\ 0\non\\
-\frac{1}{2}\Tr(sAsA) &=& d \Tr(\frac{1}{2}\om sA)+s\Tr(\om F)\\
0 &=& d\Tr(\frac{-1}{6}\om^3) +s \Tr(\frac{1}{2}\om sA)\non\\
0 &=& s\Tr(\frac{-1}{6}\om^3)\ ,\non
\eeq
where $\Qcov^{2,1}=\Tr(\om F)$ is the covariant anomaly, $\Qcov^{1,2}=\Tr(\frac{1}{2}\om sA)$ is the covariant Schwinger term, and $\Qcov^{0,3}=\Tr(\frac{-1}{6}\om^3)$ is the Maurer-Cartan cocycle. The pairing with a class $[g]\in\Ka_1(\Ac)$ involves the Schwinger term:
\be
\langle [g], \Ec \rangle = \frac{1}{2\pi i}\int_M d(\log g)\wedge \Tr(\frac{1}{2}\om sA)\ .
\ee
This amounts to integrate the Schwinger term on all contractible loops $S^1\to M$ (boundary term). In the second case $\Ka_2(\Ac)$, the invariant polynomial is $I(\la)=\Tr\la$:
\beq
-\Tr(sA) &=& d\Tr\om \non\\
0  &=& s\Tr\om\ .
\eeq
Hence $\Qcov^{0,1}=\Tr\om$, and the pairing with a Steinberg symbol $x\cup y \in \Ka_2(\Ac)$ reads
\be
\langle x\cup y,\Ec\rangle = \frac{1}{2\pi i}\int_M x^{-1}dx\wedge y^{-1}dy\, \Tr \om\ .
\ee
When $M$ has higher (even) dimension $p$, the algebraic $K$-theory of middle dimension $\Ka_{p/2}(\Ac)$ pairs with the covariant term
\be
[\Qcov^{p/2,p/2+1}(A,\om)]\in H^{p/2,p/2+1}(M,s\mod d)\ . 
\ee
By proposition \ref{pbrs}, the image of $\Qcov^{p/2,p/2+1}(A,\om)$ in the BRS cohomology of $P$ corresponds to the {\it consistent} term 
\be
[\Qcons^{p/2,p/2+1}(A,\om)]\in H^{p/2,p/2+1}(P,s\mod d)\ .
\ee 
It follows that the pairing with algebraic $K$-theory detects also the homological information contained in the consistent anomaly and Schwinger term
\be
[\Qcons^{p,1}(A,\om)]\in H^{p,1}(P,s\mod d)\ ,\quad [\Qcons^{p-1,2}(A,\om)]\in H^{p-1,2}(P,s\mod d)
\ee 
transiting through the consistent descent equations (\ref{cons}) until the position of $\Qcons^{p/2,p/2+1}$.

\begin{exercise}\textup{
Compute and interpret the secondary characteristic classes associated to non-commutative examples of equivariant spectral triples as stated at the beginning of this section.}
\end{exercise}

\end{document}